# Omics-scale polymer computational database transferable to real-world artificial intelligence applications

The RadonPy consortium

Ryo Yoshida[1,2,3*†], Yoshihiro Hayashi[1,2,3*†], Hidemine Furuya[4], Ryohei Hosoya[4], Kazuyoshi Kaneko[5], Hiroki Sugisawa[6], Yu Kaneko[7], Aiko Takahashi[1], Yoh Noguchi[1,8], Shun Nanjo[3], Keiko Shinoda[1], Tomu Hamakawa[1], Mitsuru Ohno[7], Takuya Kitamura[9], Misaki Yonekawa[9], Stephen Wu[1,3], Masato Ohnishi[1], Chang Liu[1,2], Teruki Tsurimoto[10], Arifin[11], Araki Wakiuchi[11], Kohei Noda[11], Junko Morikawa[4], Teruaki Hayakawa[4], Junichiro Shiomi[1,2,12,13,14], Masanobu Naito[2,15], Kazuya Shiratori[6], Tomoki Nagai[11], Norio Tomotsu[16], Hiroto Inoue[17], Ryuichi Sakashita[17], Masashi Ishii[18], Isao Kuwajima[18], Kenji Furuichi[19], Norihiko Hiroi[19], Yuki Takemoto[19], Takahiro Ohkuma[20], Keita Yamamoto[21], Naoya Kowatari[5], Masato Suzuki[5], Naoya Matsumoto[5], Seiryu Umetani[22], Hisaki Ikebata[23], Yasuyuki Shudo[24], Mayu Nagao[24], Shinya Kamada[24], Kazunori Kamio[22], Taichi Shomura[22], Kensaku Nakamura[22], Yudai Iwamizu[22], Atsutoshi Abe[22], Koki Yoshitomi[22], Yuki Horie[22], Katsuhiko Koike[22], Koichi Iwakabe[22], Shinya Gima[7], Kota Usui[6], Gikyo Usuki[25], Takuro Tsutsumi[25], Keitaro Matsuoka[25], Kazuki Sada[25], Masahiro Kitabata[26], Takuma Kikutsuji[26], Akitaka Kamauchi[27], Yusuke Iijima[28], Tsubasa Suzuki[28], Takenori Goda[28], Yuki Takabayashi[28], Kazuko Imai[28], Yuji Mochizuki[29], Hideo Doi[29], Koji Okuwaki[30], Hiroya Nitta[30], Taku Ozawa[30], Hitoshi Kamijima[31], Toshiaki Shintani[31], Takuma Mitamura[31], Massimiliano Zamengo[4], Yuitsu Sugami[10], Seiji Akiyama[2], Yoshinari Murakami[32], Atsushi Betto[33], Naoya Matsuo[33], Satoru Kagao[33], Tetsuya Kobayashi[34], Norie Matsubara[34], Shosei Kubo[34], Yuki Ishiyama[35], Yuri Ichioka[35], Mamoru Usami[36], Satoru Yoshizaki[37], Seigo Mizutani[37], Yosuke Hanawa[38], Shogo Kunieda[38], Mitsuru Yambe[38], Takeru Nakamura[38], Hiromori Murashima[38], Kenji Takahashi[39], Naoki Wada[39], Masahiro Kawano[40], Yosuke Harada[40], Takehiro Fujita[41], Erina Fujita[1], Ryoji Himeno[3], Hiori Kino[1,3], Kenji Fukumizu[1,3]

[1] The Institute of Statistical Mathematics, Research Organization of Information and Systems, Tachikawa, Tokyo 190-8562, Japan
[2] Advanced General Intelligence for Science Program (AGIS), TRIP Headquarters, RIKEN, Wako, Saitama 351-0198, Japan
[3] Graduate Institute for Advanced Studies, SOKENDAI, Tachikawa, Tokyo 190-8562, Japan
[4] School of Materials and Chemical Technology, Institute of Science Tokyo, Meguro-ku, Tokyo 152-8550, Japan
[5] Research & Advanced Development Division, The Yokohama Rubber Co., Ltd., Hiratsuka, Kanagawa, 254-8601, Japan
[6] Science & Innovation Center, Mitsubishi Chemical Corporation, Yokohama 227-8502, Japan
[7] Business Development Center, R&D Headquarters, Daicel Corporation, Himeji, Hyogo 671-1283, Japan
[8] School of Life Sciences, Tokyo University of Pharmacy and Life Sciences, Hachioji, 192-0392 Tokyo, Japan
[9] Imaging & Informatics Laboratories, FUJIFILM Corporation, Ashigarakami-gun, Kanagawa 258-8577, Japan
[10] R&D Center, Corporate, Sekisui Chemical Co., Ltd., Mishima-gun, Osaka 618-0021, Japan
[11] Materials Informatics Initiative, JSR Corporation, Kawasaki, Kanagawa 210-0821, Japan
[12] Institute of Engineering Innovation, The University of Tokyo, Bunkyo-ku, Tokyo 113-0032, Japan
[13] Department of Mechanical Engineering, The University of Tokyo, Bunkyo, Tokyo 113-8656, Japan
[14] Center for Advanced Intelligence Project, RIKEN, Wako, Saitama, 351-0198, Japan
[15] Research Center for Macromolecules and Biomaterials, National Institute for Materials Science, Tsukuba, Ibaraki 305-0047, Japan
[16] Department of Applied Chemistry and Biotechnology, Chiba University, Inage-ku, Chiba 263-8522, Japan
[17] Advanced Materials Research Laboratory, TOSOH Corporation, Yokkaichi, Mie 510-8540 Japan
[18] Research and Service Division of Materials Data and Integrated System, National Institute for Materials Science, Tsukuba, Ibaraki 305-0047, Japan
[19] Corporate Research Center, TOYOBO Co., Ltd., Otsu, Shiga 520-0292, Japan
[20] Fundamental Analytical Research Department, Bridgestone Corporation, Kodaira, Tokyo 187-8531, Japan
[21] Data Science & Informatics Promotion Department, NIPPON SHOKUBAI Co., Ltd., Suita, Osaka 564-0034, Japan
[22] Research Center, Mitsui Chemicals, Inc., Sodegaura, Chiba 299-0265, Japan
[23] CrowdChem, Inc., Shinagawa-ku, Tokyo 140-0013, Japan
[24] Corporate Engineering Center, Sumitomo Bakelite Co., Ltd., Fujieda, Shizuoka 426-0041, Japan
[25] Faculty of Science, Hokkaido University, Sapporo, Hokkaido 060-0810 Japan
[26] Advanced Materials Research Laboratories, Toray Industries, Inc., Otsu, Shiga 520-8558, Japan
[27] Green Innovation Center, Panasonic Holdings Corporation, Kadoma, Osaka 571-8501, Japan
[28] Toppan Technical Research Institute, TOPPAN Holdings Inc., Sugito-machi, Saitama 345-8508, Japan
[29] Faculty of Science, Rikkyo University, Toshima-ku, Tokyo 171-8501, Japan
[30] Engineering Technology Business Unit, JSOL Corporation, Chiyoda-ku, Tokyo 102-0074, Japan
[31] Business Division, CLI Business Unit, Research Institute of Systems Planning, Inc., Shibuya-ku, Tokyo 150-0031, Japan

[32] Data Science Center, DIC Corporation, Sakura, Chiba 285-8668, Japan
[33] Group R&D Division, artience Co.,Ltd., Chiyoda Sakado, Saitama 350-0214, Japan
[34] Research & Development Division, NIPPON STEEL Chemical & Material Co., Ltd., Kitakyushu-shi, Fukuoka 805-8503, Japan
[35] Corporate Research & Development, Asahi Kasei Corporation, Chiyoda-ku, Tokyo 100-0006, Japan
[36] ASMS Co., Ltd., Shinagawa, Tokyo 141-0022, Japan
[37] Technology and Innovation Center, Daikin Industries, Ltd., Settsu, Osaka 566-8585, Japan
[38] R&D Strategy Division, SCREEN Holdings. Co. Ltd., Settsu, Fushimi-ku, Kyoto 612-8486, Japan
[39] Institute of Science and Technology, Kanazawa University, Kakuma-machi, Kanazawa 920-1192, Japan
[40] Innovation Center, Idemitsu Kosan Co., Ltd, Sodegaura, Chiba 299-0293, Japan
[41] Advanced Materials Company, Idemitsu Kosan Co., Ltd, Ichihara, Chiba 299-0193, Japan

Corresponding authors: yoshidar@ism.ac.jp; yhayashi@ism.ac.jp
[†]These authors contributed equally to this work.

**Developing large-scale foundational datasets is a critical milestone in advancing artificial intelligence (AI)-driven scientific innovation. However, unlike AI-mature fields such as natural language processing, materials science, particularly polymer research, has significantly lagged in developing extensive open datasets. This lag is primarily due to the high costs of polymer synthesis and property measurements, along with the vastness and complexity of the chemical space. This study presents PolyOmics, an omics-scale computational database generated through fully automated molecular dynamics simulation pipelines that provide diverse physical properties for over $10^5$ polymeric materials. The PolyOmics database is collaboratively developed by approximately 260 researchers from 48 institutions to bridge the gap between academia and industry. Machine learning models pretrained on PolyOmics can be efficiently fine-tuned for a wide range of real-world downstream tasks, even when only limited experimental data are available. Notably, the generalisation capability of these simulation-to-real transfer models improve significantly as the size of the PolyOmics database increases, exhibiting power-law scaling. The emergence of scaling laws supports the "more is better" principle, highlighting the significance of ultralarge-scale computational materials data for improving real-world prediction performance. This unprecedented omics-scale database reveals vast unexplored regions of polymer materials, providing a foundation for AI-driven polymer science.**

## Main

Large-scale foundational datasets are essential for artificial intelligence (AI)-driven scientific advancements. However, unlike advanced AI fields such as computer vision, natural language processing, biology, and medicine, materials science continues to face a significant shortage of open, large-scale databases. This gap largely arises from the high costs associated with data generation, including synthesis, sample preparation, compositional and structural characterisation, and property measurements, along with the vastness, hierarchy, and complexity of the material space. To address these challenges, omics-scale databases generated through high-throughput first-principles density functional theory (DFT) computations have played a crucial role in advancing AI in materials science. Notable examples include the pioneering Materials Project[1] for inorganic crystalline materials, followed by AFLOWLIB[2], OQMD[3], GNoME[4], and QM9 for small organic molecules[5]. These efforts have established a strong foundation for transformative AI technologies, including foundational generative models for crystal structures[6] and molecules[7], and universal machine learning potentials[8].

Despite these advances, the development of extensive databases for polymer materials has lagged significantly (Table S1 for the existing polymer databases). Experimental databases remain limited in both coverage and scale, and typically contain only a few property types for several thousand or fewer polymers[9,10]. Although several large-scale virtual polymer databases have emerged in recent years[11–13], they lack comprehensive property data mainly due to the technical barriers inherent in high-throughput computational experiments. Polymers, with their large-scale non-periodic structures, cannot be handled directly using ordinary periodic-system calculations based on first principles. Furthermore, conducting molecular dynamics (MD) simulations to handle large-scale, long-term stochastic dynamics of polymer systems incurs extremely high computational costs. Fully automating tractable multistep modelling processes for diverse structures, such as amorphous, branched, and mechanically stretched structures, pose significant technical challenges.

To address these issues, we have developed RadonPy, a Python-based pipeline software for fully automating all-atom classical MD simulations of various polymeric materials[14]. Given polymer constitutional repeat units with additional simulation conditions, RadonPy automatically executes the entire simulation workflow, including charge calculations, force field parameter assignment, initial structure generation, and equilibrium and non-equilibrium MD (NEMD) simulations. The latest version (version 1.0), released in this study, implements automated simulation pipelines for 62 distinct polymer properties and structural and dynamic features, including thermal, optical, dielectric, and mechanical properties, across a wide array of polymer systems such as homopolymers, copolymers, and crosslinked polymers with various higher-order structures.

In this study, RadonPy is used to develop PolyOmics, the world's largest polymer materials database, which encompasses a diverse set of physical properties of more than $10^5$ distinct polymeric materials. The database is constructed using the Fugaku supercomputer, which has been operated at approximately 100 million node-hours for four years. The project is conducted by an industry–academia consortium of more than 260 researchers from 3 national research institutes, 8 universities, and 37 companies (of which approximately 100 researchers from 41 organisations have contributed as co-authors of this paper). PolyOmics provides critical insights into previously unexplored regions of the polymer material space, such as the Pareto fronts formed by trade-offs between multiple properties. The database comprises a diverse collection of specialised datasets, including 20 distinct classes of general polymers, cellulose derivatives, biodegradable candidate polymers, uniaxially oriented structures, solubility parameters, and dielectric properties. Notably, PolyOmics is primarily designed for use in simulation-to-reality (Sim2Real) machine learning[15–17]. Models pretrained on PolyOmics are fine-tuned for downstream real-world tasks, where only limited experimental data are available. Remarkably, these Sim2Real transferred models exhibit outstanding transferability with power-law scaling across diverse real-world property prediction tasks as the PolyOmics dataset size increases. This unprecedented omics-scale database opens access to vast, previously unexplored regions of polymer material research, establishing a new foundation for AI-driven polymer research.

Simulations cannot fully capture the complexities and uncertainties inherent in real-world systems. In recent years, several new inorganic crystalline compounds have been identified using ultralarge-scale brute-force DFT calculations. However, the scientific significance of these findings has been criticised[18]. One of the key concerns lies in the discrepancy between DFT calculations, which assume perfectly periodic systems at absolute zero temperature, and the behaviour of real-world materials. However, this criticism overlooks the intrinsic value of computational data. The PolyOmics database is not intended for the individual retrieval of data in a dictionary-like manner. Instead, its fundamental value emerges when it is used collectively for AI model training. The scaling laws that emerge when bridging the gap between imperfect simulations and complex real-world systems through Sim2Real transfer learning demonstrate that ultralarge-scale computational materials data possess practical utility, validating the "more is better" principle.

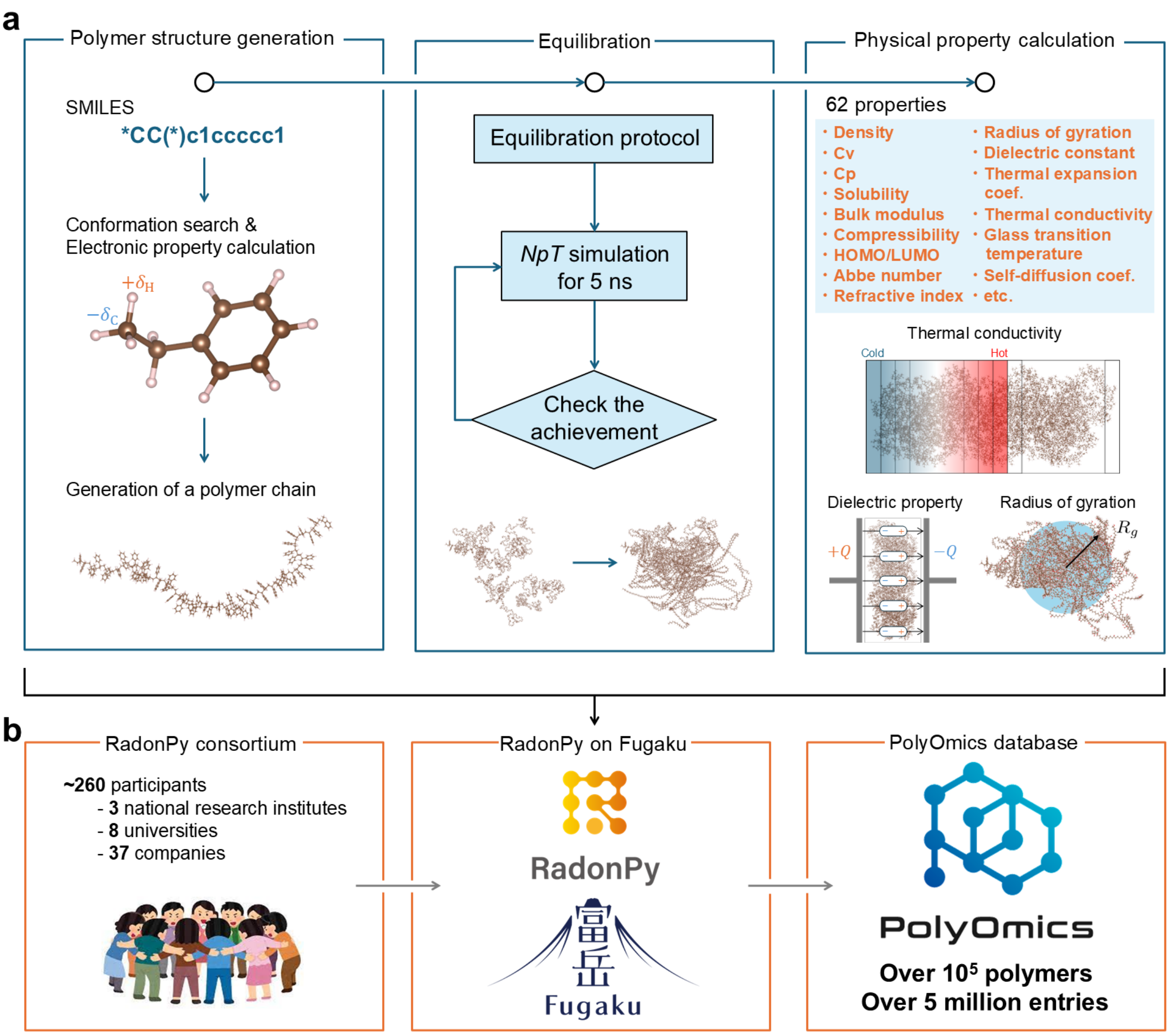

**Fig. 1.** Creation of PolyOmics, an omics-scale polymer materials database using the automated computational experiment software RadonPy. **a.** RadonPy's automated MD simulation workflows. In the latest version, automated calculation algorithms for 62 distinct properties have been implemented (Table S3). **b.** Collaborative database development through an industry–academia consortium using the supercomputer Fugaku.

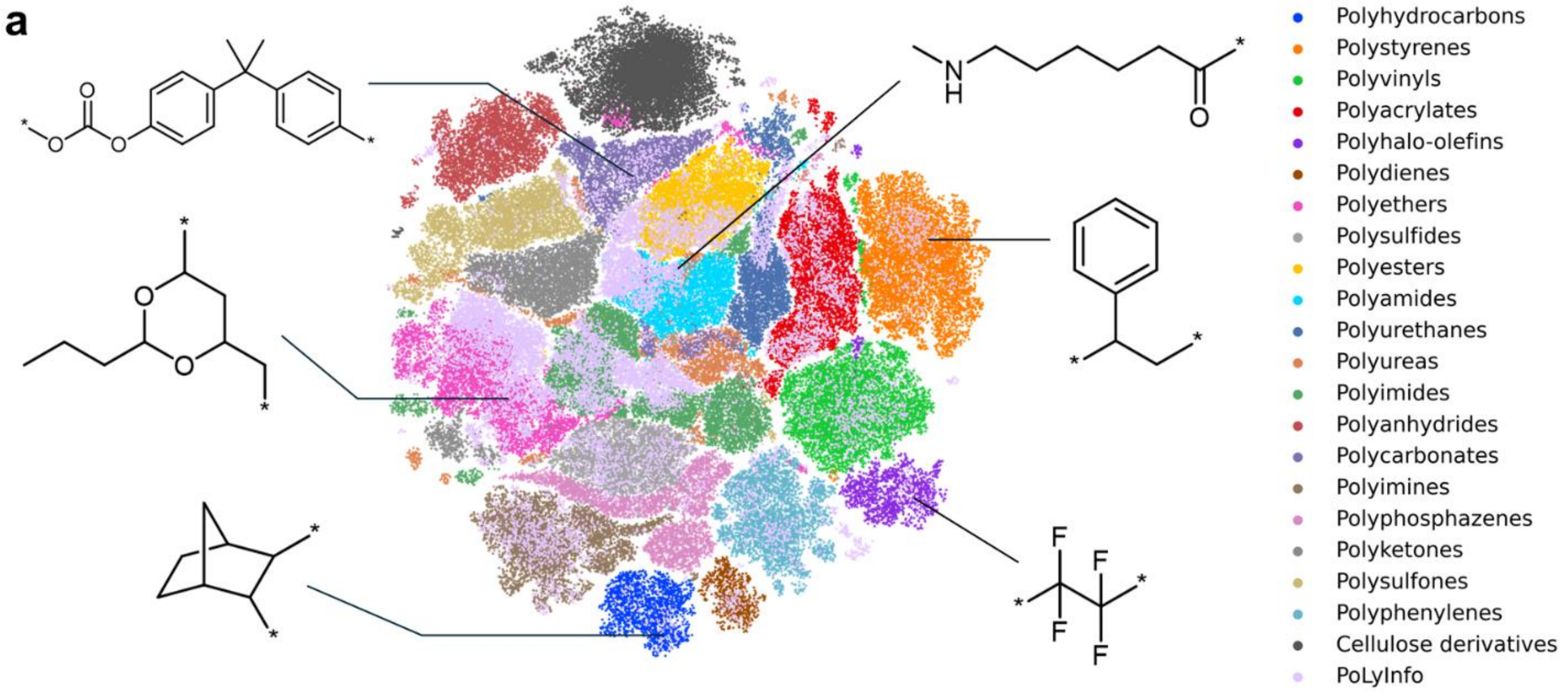

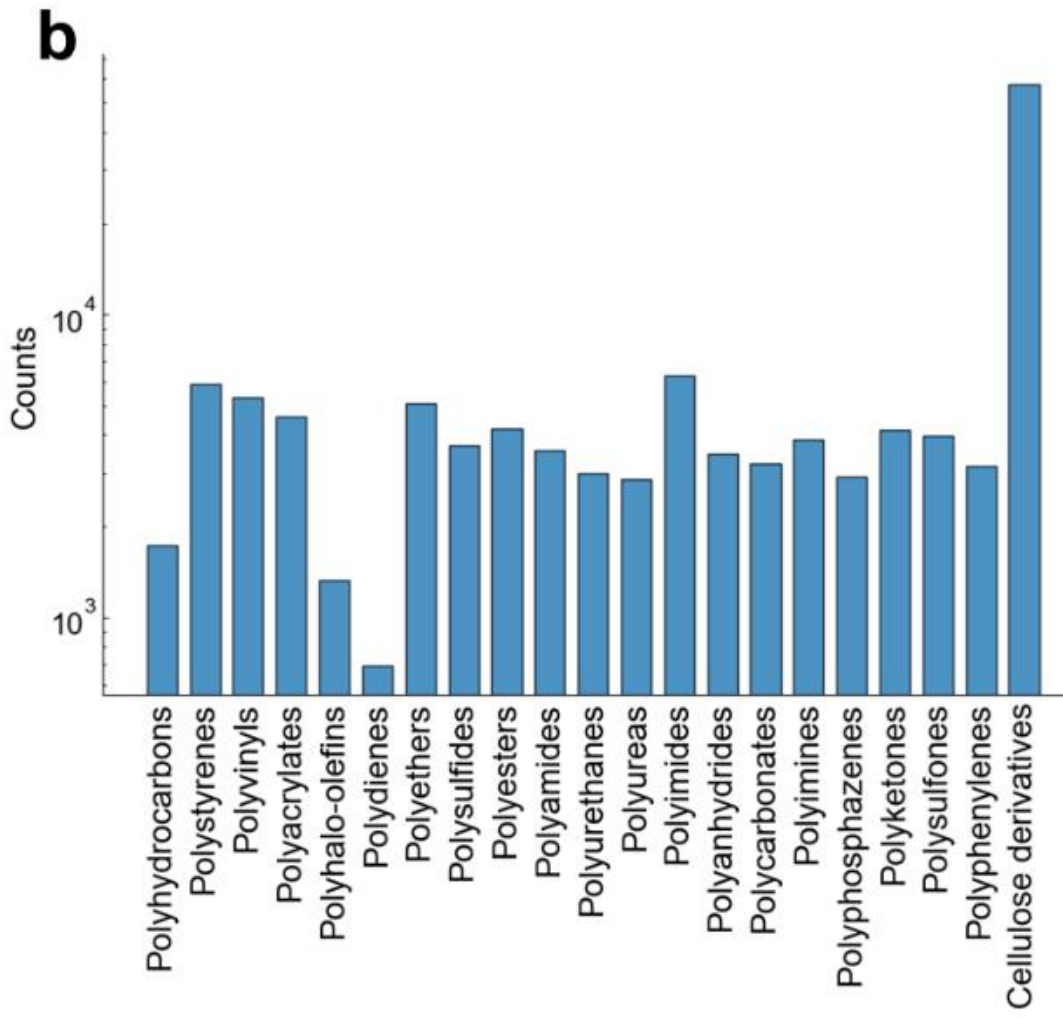

**c**

| Dataset name | Subset name | Description | # of entries |
|---|---|---|---|
| Isotropic amorphous polymers | All data | 62 properties for 125,658 isotropic amorphous polymers | 4,873,198 |
| | General polymers | 62 properties for 73,045 general polymers classified into 20 classes, generated by probabilistic language models and rule-based generators | 3,020,351 |
| | Cellulose derivatives | 50,661 cellulose derivatives | 1,811,440 |
| | PFASs | 3,821 PFASs | 157,661 |
| | Biodegradable candidates | 357 polyesters and 26 polycarbonates that were synthesized for the biodegradability screening (Fransen et al.) | 6,511 |
| | Ladder polymers | 519 ladder polymers and their corresponding non-ladder polymers | 34,513 |
| | Additional optical properties | Abbe number, wave-number dependent permittivity and refractive index for 13,240 polymers | 92,680 |
| | Dynamic dielectric properties | Dynamic dielectric constant and loss tangent for 1,084 polymers | 2,168 |
| | Solubility parameters | Solubility parameters for 58,834 polymers | 294,170 |
| Uniaxial stretched polymers | | Thermal conductivity profiles for 1,379 uniaxial stretched polymers | 7,316 |
| Small molecules | | 23,361 small molecules | 654,820 |
| $\chi$ parameters | | $\chi$ parameters of 93,331 polymers with respect to 19 commonly used organic solvents that are estimated using machine learning models | 1,773,289 |

**Fig.** 2. Data summary of PolyOmics. **a.** UMAP projection[19] showing the distribution of polymer species in the general polymers and the cellulose derivatives dataset of the PolyOmics (color-coded according to their polymer classes), alongside 15,323 currently synthesized polymers from PoLyInfo (in light purple). **b.** Number of polymers in PolyOmics, the general polymer dataset classified into 20 structural classes (Table S2), and the cellulose derivatives. **c.** Selected data collections made available separately on the PolyOmics website https://huggingface.co/datasets/yhayashi1986/PolyOmics.

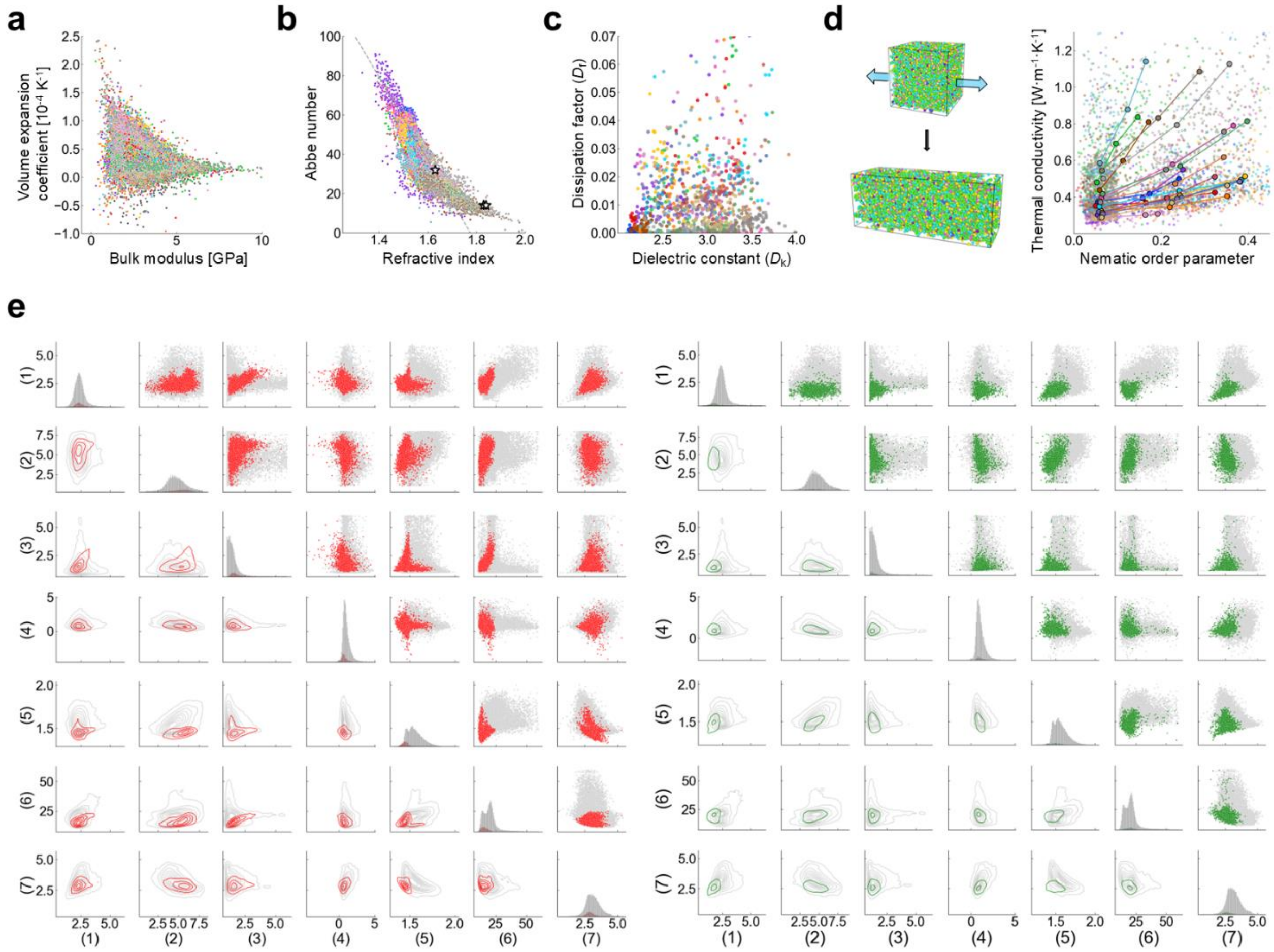

**Fig. 3.** Data collections in PolyOmics. **a.** Joint distribution of volume expansion coefficient versus bulk modulus across 79,041 polymers. Fluoropolymers are highlighted in green. **b.** Joint distribution of refractive index versus Abbe number, where the property values are linearly calibrated with reference to experimental values, following the procedure in a prior work[19]. The solid line indicates an empirically known Pareto frontier. The star markers represent three polymers that are predicted and discovered using the machine learning polymer design system SPACIER[20] in combination with RadonPy. **c.** Joint distribution of dielectric constant and dielectric dissipation factor at 10 GHz for 1,084 polymers. **d.** Dependence of thermal conductivity on molecular orientation in 1,379 uniaxially stretched polymers. **e.** Joint distribution of following seven properties for 3,887 cellulose derivatives (left) and 1,616 PFAS compounds (right), compared with other 35,713 general polymers (light gray): (1) thermal conductivity [$10^{-1}$ W·m$^{-1}$·K$^{-1}$], (2) glass transition temperature [$10^{2}$ K], (3) static dielectric constant, (4) linear expansion coefficient [$10^{-4}$ K$^{-1}$], (5) refractive index, (6) radius of gyration [Å], and (7) $C_P$ [$10^{3}$ J·kg$^{-1}$·K$^{-1}$].

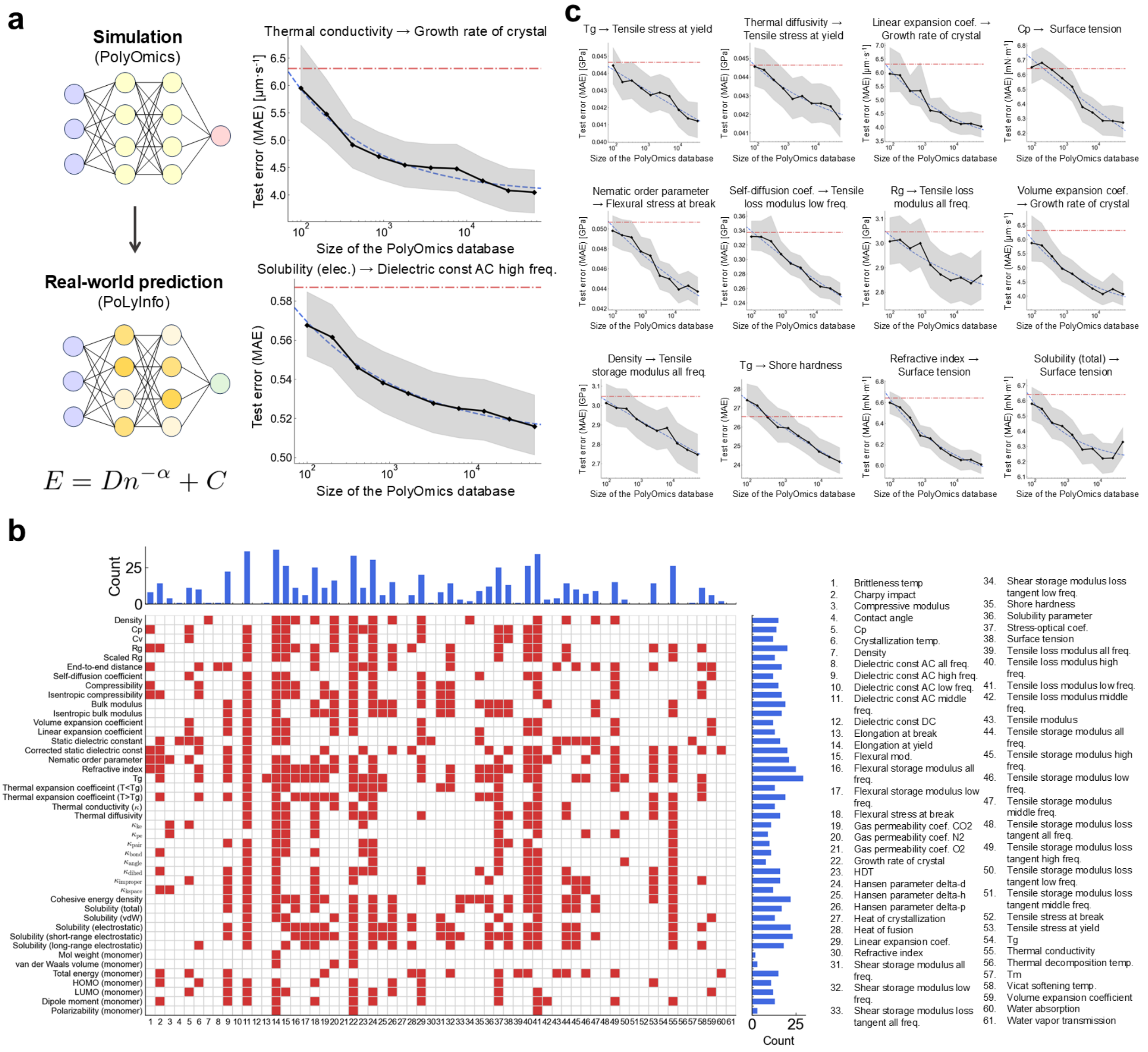

**Fig. 4.** Scaling law of Sim2Real transfer learning using PolyOmics. **a.** Scaling law of Sim2Real transfer learning. **b.** Transfer learning and presence/absence of significant scaling across all combinations of the 61 property datasets from PoLyInfo and the 43 property datasets from PolyOmics. The bar plots arranged in each row and column represent the number of cases where significant scaling was observed. **c.** Example of observed scaling behaviours.

## RadonPy

RadonPy is an open-source Python software developed to automate a wide range of computational experiments on polymer materials, with a primary focus on all-atom classical MD simulations[14]. Once the chemical structure of the polymer chain repeating units with other simulation conditions is specified, RadonPy streamlines the entire simulation workflow, including the conformational search, charge calculations via DFT, force field parameter assignment, simulation cell construction, equilibrium and NEMD simulations, and property calculations (Fig. 1a, Methods, and Section 1 in Supplementary Notes). The main simulation engines are the open-source software LAMMPS[21] for MD simulations and Psi4[22] for DFT calculations. Currently, automated simulation pipelines for 62 different properties and polymer structural and dynamic features (Table S3 lists these properties) have been implemented for homopolymers, random copolymers, block copolymers, alternating copolymers, and crosslinked polymers with various higher-order structures, including isotropic amorphous and uniaxially oriented structures, thermal transport properties via NEMD simulations, glass transition temperatures, and dielectric properties under applied alternating electric fields (Methods and Section 1 in Supplementary Notes).

**PolyOmics database**

Using RadonPy, the PolyOmics database is constructed, comprising multiple properties of over $10^5$ polymeric materials, with the total number of data entries currently exceeding 7,301,307. As of the current release, the database contains 62 properties: 43 properties derived from RadonPy (Table S3), for which no data are currently available for the remaining 19 properties, and machine learning–predicted $\chi$ parameters (Section 3 in Supplementary Notes) that quantify the miscibility between polymers and 19 widely used organic solvents and plasticisers (Table S4). The chemical spaces of these polymers are shown in Fig. 2a, along with those of the 15,323 currently available polymers from PoLyInfo[23,24]. By running independent jobs for different polymers in parallel across numerous nodes on Fugaku, the unprecedented-scale database has been developed over approximately four years, consuming approximately 100 million node-hours (Fig. 1b). Within the consortium, multiple working groups have independently defined objectives and collaboratively generated a wide range of datasets. The resulting data comprise various subsets (Fig. 2c), including a property dataset for over $10^5$ isotropic amorphous polymers; ~73,000 general polymers classified into 20 classes; ~50,000 cellulose derivatives; ~13,000 candidates for optical polymers; 3,821 per and polyfluoroalkyl substances (PFAS); 383 polyesters and polycarbonates targeting biodegradable polymer development; 519 ladder polymers; dynamic dielectric properties for 1,084 polymers; thermal transport properties under uniaxial stretching for 1,379 polymers; $\chi$ parameters for ~93,000 polymers mixed with the 19 organic solvents; solubility parameters for ~59,000 polymers, among others. Each polymer in the general polymer dataset typically belongs to one of the 20 classes defined by PolyOmics (Table S2). Examples selected from these datasets are further discussed.

The isotropic amorphous polymer dataset is the largest subset in the current PolyOmics collection and includes 43 properties for 125,658 isotropic amorphous polymers, all obtained under nearly identical calculation conditions (Methods). Most of the polymers are generated using custom molecule generators based on SMILES linguistic learning[25] and SMiPoly, a rule-based polymerisation reaction algorithm[13] (Methods and Section 2 in Supplementary Notes). For the simulation conditions, the degree of polymerisation is adjusted such that each polymer chain contains approximately 1,000 atoms, and an initial isotropic amorphous structure is generated by replicating ten polymer chains arranged based on a self-avoiding random-walk algorithm. In the MD simulations, potential energy is evaluated using the general Amber force field version 2 (GAFF2)[26]. To ensure the reliability of the simulation settings, comparisons are made with experimentally measured values reported in PoLyInfo and other sources (Fig. S2).

This dataset provides a comprehensive overview of the expansive design space of amorphous polymers. To date, no previous studies have elucidated the joint distribution of such a large number of properties and polymer species at this scale (Fig. S1 shows the joint distribution of 19 selected properties for approximately 70,000 isotropic amorphous polymers from the general polymer and cellulose derivative datasets, while Fig. S7 shows the marginal distribution of each of the 43 properties across the 20 polymer classes of the general polymer and cellulose derivative datasets). By leveraging this material map, Pareto fronts shaped by trade-offs are identified among multiple properties, along with previously unrecognised classes of polymers that constitute or even surpass empirically known Pareto boundaries.

For example, we analysed the trade-off between increasing heat resistance and flexibility. The coefficient of linear thermal expansion serves as an indicator of heat resistance, while the elastic modulus represents flexibility. Owing to this inherent trade-off, Pareto fronts emerged in the lower regions of the corresponding scatter plots (Fig. 3a). Moreover, various polymers located near the boundaries of the joint property space were systematically identified. In the development of optical polymers, the primary objective is to enhance both the refractive index and the Abbe number. The Abbe number is a dimensionless parameter that quantifies the chromatic dispersion, specifically representing the ability of light to refract across different wavelengths. The simultaneous increase in both properties presented a trade-off, which resulted in an empirically known upper boundary of achievable performance[27] (Fig. 3b). PolyOmics provided a substantial number of polymers that exceeded this upper boundary. Many of these polymers contained multiple sulfonyl groups (-$SO_2$-) or other sulfur-containing substructures. Previous studies have reported that the incorporation of sulfonyl groups into molecules is an effective strategy for designing high-performance optical polymers that can reach the empirical boundary[28,29]. In a previous research[20], SPACIER, a polymer design platform that integrates RadonPy's automated computer experiments with Bayesian optimisation, was used to successfully synthesise three novel polymers that surpassed the currently known performance limits (star markers in Fig. 3b). Furthermore, PolyOmics provided more polymers than the benchmarks.

**Data collections**

PolyOmics contains diverse datasets (Fig. 2c). The solubility parameter dataset provides key quantities related to polymer–polymer or polymer–solvent miscibility and phase separation, such as the electrostatic interaction energy and dispersion interaction energy, across 58,834 polymers (Methods). Machine learning–predicted $\chi$ parameters, which quantify the miscibility of each polymer with 15 commonly used organic solvents (such as acetone, toluene, *N*,*N*-dimethylformamide) and four plasticisers (such as di(2-ethylhexyl)phthalate), are included in the database[30] (Table S4). These parameters are essential for a wide range of fundamental research tasks, such as assessing solubility in organic solvents crucial for polymer synthesis, predicting phase separation in polymer blends and composites, gas permeability, dispersibility, and crystallinity.

RadonPy implements an automated simulation workflow to evaluate the dielectric properties of polymers under an applied alternating current (AC) electric field (Methods and Section 1 in Supplementary Notes). The time-dependent AC electric field is modelled as $E(t) = E_0 \sin(\omega t)$, where $E_0$ represents the amplitude of the electric field, $\omega$ is the angular frequency, and $t$ is time. While applying an electric field with an arbitrary frequency in a specific direction to an equilibrated isotropic amorphous structure, the dielectric constant and dielectric dispersion factors are calculated by analysing the electric displacements obtained from MD simulations with a time step of 1 fs, along with the dielectric loss. PolyOmics records the dielectric properties of several thousand polymers. The joint distributions of the dielectric constant and dielectric dispersion factors at 10 GHz for 1,084 isotropic amorphous polymers selected from the 20 polymer classes are shown in Fig. 3c. Polymers exhibiting both low dielectric constants and dissipation factors typically feature rigid aromatic rings containing fluorine atoms or low concentrations of polar functional groups, as observed for polyhydrocarbons. These findings provide valuable insights for the rational design of polymers with favourable dielectric properties.

Polymer stretching significantly affects thermal conductivity. When molecular chains align, phonons (acoustic vibrations) and intramolecular vibrations propagate more efficiently along the stretching direction, leading to an increase in the thermal conductivity along that direction. For example, highly stretched polyethylene fibres exhibit extremely low thermal conductivity in the perpendicular direction (~0.1–0.5 $W{\cdot}m^{-1}{\cdot}K^{-1}$) while achieving exceptionally high thermal conductivity (>100 $W{\cdot}m^{-1}{\cdot}K^{-1}$) along the direction of stretch[31,32]. PolyOmics contains thermal conductivity profiles for 1,379 uniaxially stretched polymers. Isotropic amorphous structures are generated using equilibrium simulations. A uniaxial orientation is induced by iteratively stretching and compressing the simulation box in a single direction at a stretching speed of 25 m/s and target strains of 0.5 and 1.0. Following the stretching process, the resulting structures are equilibrated for an additional 500 ps (Methods and Section 1 in Supplementary Notes). This dataset enables the analysis of variations in thermal conductivity with the degree of polymer chain alignment, quantified by the orientation order parameter (Fig. 3d).

PolyOmics contains a property dataset for 519 distinct types of ladder polymers (Methods). Ladder polymers are a class of macromolecules characterised by a molecular structure resembling a ladder. Their fundamental framework comprises adjacent backbone units connected via dual linkages—either double or covalent bonds—forming a rigid integrated molecular framework (Fig. S3). Owing to these dual-bond or covalent linkage structures, ladder polymers exhibit enhanced resistance to thermal degradation and chemical reactions, along with outstanding optical, electronic, thermal, and mechanical properties. However, synthesis is difficult. Consequently, experimental data are extremely limited. PolyOmics provides ladder polymers along with a property dataset of their corresponding non-ladder polymers, in which one of the double linkages has been removed, serving as a control set.

Furthermore, the database contains a candidate set of biodegradable polymers. Fransen et al.[33] conducted high-throughput biodegradability screening of 642 distinct polyesters and polycarbonates using their proprietary experimental platform[33]. Exploring the relationship between the MD-calculated properties and biodegradability of these polymers, many of which are included in PolyOmics, offers important and intriguing applications (Fig. S4). Additionally, the database includes cellulose derivatives, a group of compounds obtained by chemically modifying natural cellulose. These derivatives have a linear structure comprising glucose units connected via β-1,4 linkages, with each unit containing three hydroxyl groups (–OH). By introducing ether or ester groups to these hydroxyls, various properties such as biodegradability, solubility, viscosity, thermal and mechanical characteristics, and chemical stability can be tailored for specific applications. In this study, ~50,000 cellulose derivatives were generated by substituting the hydroxyl sites with five different degrees of substitution using ~10,000 types of substituents extracted from a commercially available reagent database[34] (Methods and Fig. S5), and their isotropic amorphous properties were computed (Fig. 3e).

Additionally, PolyOmics offers a diverse collection of data. For example, the database contains a property dataset of 23,361 small molecules. By comparing the properties of the polymers with those of their repeating units, the effects of the polymerisation degree and molecular weight on the polymeric properties can be investigated (Fig. S6). Furthermore, PolyOmics includes a dedicated dataset of 3,821 PFAS resins (Fig. 3e). Notably, the physical properties of numerous PFAS are distributed near the tails of the joint property distributions, indicating their distinctive characteristics within the broader polymer space. This dataset offers significant potential to guide the identification of promising seed compounds for the development of alternative materials for PFASs.

**Scaling law of Sim2Real transfer learning**

The PolyOmics database was developed primarily as a data resource for Sim2Real transfer learning. Models pre-trained on PolyOmic were fine-tuned with limited experimental data to achieve superior generalisation in real-world tasks beyond the capabilities of models trained from scratch. Here, the scaling law underlying Sim2Real transfer learning is discussed, which is the core design principle of the PolyOmics database. Mikami et al.[35] theoretically elucidated the scaling law and empirically confirmed it using computer vision tasks. AccordinglyClick or tap here to enter text., our previous study demonstrated that the scaling law holds consistently across a wide range of Sim2Real transfer scenarios in materials science[17]. These prior studies showed that the generalisation error $E$ of a Sim2Real transferred model with respect to experimental properties decreases monotonically with the size $n$ of the computational database, following a power-law relationship (Fig. 4a).

$$E = Dn^{-\alpha} + C \qquad (1)$$

A desirable database should exhibit a large decay exponent $\alpha$ with respect to increasing $n$, and a small transfer gap $C$. The term $C$ represents the asymptotic limit of achievable performance as the computational data expand, serving as a key indicator of the expected value of the computational property database.

We performed Sim2Real transfer learning using PolyOmics and PoLyInfo, a comprehensive experimental property database compiled from literature. A conventional fully connected neural network was employed to predict each property based on the compositional and structural features of polymer repeating units (Methods). From PoLyInfo, we extracted datasets for 61 distinct experimental properties of amorphous polymers, in which only the data corresponding to temperatures at room temperature (0–50 °C) were considered for temperature-dependent properties (Table S5 lists the properties and dataset sizes). For PolyOmics, 43 isotropic amorphous property datasets were used separately to generate pretrained source models. Sim2Real transfer was conducted across all combinations of 43 source and 61 target tasks, and the presence or absence of a scaling law was evaluated. When the area under the estimated scaling curve was sufficiently smaller than the baseline for scratch learning, the observed scaling behaviour was considered significant (Methods). Consequently, notable scaling was identified in 635 of the 43×61 transfer scenarios (Fig. 4b).

For 13 property prediction tasks, where both experimental and computational data were available, Sim2Real consistently outperformed scratch learning, showing significantly strong scaling. In particular, for density, refractive index, and glass transition temperature—with experimental dataset sizes of 607, 234, and 4,936, respectively—Sim2Real transfer exhibited pronounced scalability (Fig. 4c) with the estimated decay rates of $\alpha \approx$ 0.011–0.438. These properties exhibited a high consistency between the RadonPy-derived computational and experimental values (Fig. S2), indicating that expanding the computational dataset almost directly translated to an improved performance in real-world prediction tasks.

For other Sim2Real transfers between identical properties, the specific heat capacity at constant pressure ($C_\mathrm{p}$) exhibited comparatively strong scaling (Fig. 4c), although the $C_\mathrm{p}$ values obtained from classical MD simulations inherently neglected quantum effects, resulting in a systematic bias from the experimental values (Fig. S2). The thermal conductivity and the coefficient of linear thermal expansion converged in their scaling curves around a dataset size of $n = 10^4$ (Fig. 4c). When such convergence was attained, a strategy can be devised to halt further data generation and reallocate surplus computational resources to other projects. Alternatively, modification of the data-generation protocol to improve scalability could be considered. Notably, the observed scaling behaviours reflected generalisation performance only within a finite subset of polymers across a vast chemical space—particularly those with limited experimental data. Even when the generalisation performance improved outside of the data distribution, measurements are not possible. Furthermore, transferability is dependent not only on the quantity of computational data but also on the amount and quality of the experimental data. For example, the number of experimental thermal conductivity values was only 39. In addition, the coefficient of linear thermal expansion exhibited substantial measurement variability (Fig. S2). Thus, even when the computational dataset is expanded, it is infeasible to surpass the intrinsic performance limits imposed by the available experimental data.

During transfer learning between different physical properties, several cases exhibited significant scaling. Among all task pairs, cases with clear negative transfers accounted for less than 7%. As illustrated in Fig. 4c, Sim2Real transfer learning exhibited strong scaling for a wide variety of real-world systems in which direct computational evaluation is technically challenging or even infeasible due to the spatiotemporal limitations of MD simulations, such as crystal growth rate, tensile stress at yield, fracture bending stress, and surface tension. Notably, the radius of gyration $R_g$, glass transition temperature, nematic order parameter, and solubility parameters showed significant scaling in 20, 29, 21, and 22 of the 61 target tasks, respectively. $R_g$, calculated as the root-mean-square distance from the centre of mass to each repeating unit, quantitatively represented the spatial extent and folding of a polymer molecule. In particular, $R_g$ exhibited high transferability to mechanical properties such as yield stress and dynamic viscoelasticity. The yield stress of a polymer depends on its molecular entanglements and structural characteristics. $R_g$, which represented the expansion of the polymer chains, indirectly affected the yield stress through its influence on intermolecular interactions and entanglements.

**Conclusions and future perspectives**

The PolyOmics database, which encompasses the computed properties of polymers on a scale of over $10^5$ polymeric materials, is a landmark achievement in data-driven polymer material research. Similar to the DFT-based material databases that have revolutionised AI for crystalline materials, PolyOmics is expected to stimulate a similar transformation in polymer science. This study illuminates a substantial portion of the vast polymer material space; however, many regions remain unexplored. In particular, the properties of polymeric materials are significantly influenced by higher-order structural features such as crystallinity, amorphous regions, liquid crystallinity, and microphase separation in the polymer blends. Automating MD simulations for such complex heterogeneous systems at the atomic scale remains technically and computationally challenging. Future database development will require multiscale and hierarchical approaches that incorporate automated coarse-grained simulations alongside atomistic models.

Unlike prior DFT databases, PolyOmics was specifically developed as a foundational data resource for Sim2Real machine learning. MD simulations of polymer materials are subject to inherent uncertainties and imperfections stemming from limitations in polymer chains, polymerisation degree, spatiotemporal scales, and the stochasticity of initial configurations. These limitations reduce the utility of dictionary-style analyses of individual polymers, which have historically hindered the development of high-throughput computational databases for polymers. To overcome this problem, we re-conceptualised computational databases as foundational training resources for Sim2Real machine learning, redefining their value in terms of transferability and scalability to real-world tasks.

Placing transfer learning scaling laws at the core provides a roadmap for mitigating data scarcity in materials science. In several cutting-edge fields, collecting sufficient experimental data for AI model training is infeasible. The solution lies in identifying domains, such as computer simulations, where large-scale data can be systematically generated, and then bridging the domain gap through machine learning. Such a foundational source database should be designed such that with its expansion, the model's capability to generalise across diverse downstream task scales accordingly. To broaden the transferable domains and enhance transferability, both the quantity and quality of data should be enhanced. Thus, we are focused on developing RadonPy and PolyOmics through cross-institutional collaborative efforts.

## Methods

## RadonPy

The RadonPy project was initiated to fully automate a wide range of computational experiments on polymer materials, with the focus on automating all-atom classical MD simulations. The input parameters included the chemical structure of one or more polymer repeat units given as SMILES strings, degree of polymerisation, number of polymer chains, temperature, pressure, and so on. The basic workflow included: (1) conformational search for the given repeat units; (2) calculation of electronic properties, including atomic charges, using DFT; (3) generation of initial polymer configurations; (4) assignment of force field parameters; (5) equilibrium MD simulation; (6) NEMD simulation; and (7) property calculation from molecular trajectories. RadonPy integrated the open-source software Psi4 and LAMMPS for the DFT and MD calculations, respectively, with all steps orchestrated via a Python interface. Optimisation for high-throughput computation across numerous nodes on supercomputers was achieved, including built-in mechanisms to automatically detect equilibration, handle failures in non-equilibrium simulations, and manage rescheduling and restarts, which are essential for accommodating the wall-time limits set by supercomputer usage policies. The details of the RadonPy software are provided in Section 1 of Supplementary Notes.

In the latest version (version 1.0), automated calculation algorithms for 62 distinct properties were implemented for homopolymers, alternating/random/block copolymers, crosslinked polymers, and polymer blends. Higher-order structures that can be handled included isotropic amorphous structures with uniaxial stretching orientations and crystalline polymers. Software development was conducted on a consortium basis, with ongoing efforts to expand the range of target properties and polymer systems. For each simulation workflow, standardised protocols, including simulation box settings such as the degree of polymerisation and number of polymer chains, were defined by computational chemistry experts. The primary criterion prioritised in the parameter settings was consistent with the experimental properties. For example, in isotropic amorphous polymers, computational properties such as the density, refractive index, linear expansion coefficient, volume expansion coefficient, thermal conductivity, glass transition temperature, and solubility parameters have demonstrated quantitative predictive accuracy relative to experimental values (Fig. S2). For properties with unavailable comprehensive experimental datasets, simulation conditions were established based on expert knowledge and insights from computational chemistry.

Notably, computational data do not necessarily match the experimental values quantitatively. In particular, to prioritise throughput in data production, the calculation conditions were set more loosely compared with conventional polymer material MD simulations. Validity is entrusted to the performance of machine learning transferred to real-world systems, specifically in terms of transferability and scalability. For instance, classical MD simulations for deriving specific heat capacity do not account for quantum effects, resulting in an upward bias in the MD-derived values compared with experimental observations (Fig. S2c). Similarly, property calculations for the linear expansion coefficient exhibited high variability owing to their significant dependence on the initial structure randomness, leading to weak correlations with the experimental data (Fig. S2g). Nevertheless, the transferred models demonstrated strong generalisation performance with respect to these experimental properties (Fig. 4). Sim2Real machine learning bridged the gap between PolyOmics' vast yet imperfect computational datasets and the complex and uncertain realities of experimental systems.

## Virtual polymer generation

In PolyOmics, virtually generated polymers were automatically classified into 20 polymer classes, including polyesters, polyimides, polyamides, and polyacrylates (Fig. 2b and Table S2) and assigned unique identifiers before registration in the database. Most of these polymers were generated using in-house polymer generators based on SMILES probabilistic language models[25] and the rule-based polymerisation reaction algorithm SMiPoly[13]. These polymers were recorded as a general polymer dataset in PolyOmics.

For each polymer class, the probabilistic language model was trained on the SMILES strings of previously synthesised polymers. Specifically, a generator was constructed by feeding synthetic polymers from each class in the PoLyInfo database, resulting in the creation of a virtual polymer library. The repeating unit of each existing polymer was encoded as a SMILES string, and using a language model, a structure generator was constructed to mimic the frequently occurring patterns found in existing molecules, such as molecular fragments and bonding patterns. The training and generation processes were performed using XenonPy, the in-house materials informatics platform[36]. The number of polymers generated for each class is summarised in Table S2.

SMiPoly implements 22 polymerisation reaction rules, which are categorised into four major reaction types: addition chain polymerisation, ring-opening chain polymerisation, polycondensation, and polyaddition. Each rule is defined by a combination of reaction centres and transformation rules encoded by Reaction SMARTS[37]. When a starting molecule is provided, SMiPoly checks for the presence of these reaction centres and performs *in silico* polymerisation according to applicable transformation rules. Using ~1,000 starting readily available molecules, SMiPoly generated ~160,000 unique polymers, which were classified into 20 polymer classes defined by the PolyOmics taxonomy (Table S2). Ohno et al.[13] reported that the resulting polymers spanned approximately 50% of the structural space occupied by previously synthesised polymers, whereas approximately 50% of the generated polymers were within previously unexplored regionsClick or tap here to enter text.. Currently, MD calculations for a subset of these virtual polymers have been completed.

## Equilibration simulation of isotropic amorphous polymers

Hayashi et al.[14] developed an automated workflow for calculating the physical properties of isotropic amorphous polymersClick or tap here to enter text.. A large portion of PolyOmics data was generated under the same condition. The calculation process is summarised as follows (Section 1 in Supplementary Notes).

***Conformation search of repeating units*****:** For a given polymer repeat unit with the SMILES string, where the asterisk symbols representing the connecting points of the repeating unit were capped with hydrogen atoms, 1,000 different initial conformations were generated using the ETKDG version 2 method implemented in the Python library RDKit[38]. The generated structures were further relaxed by performing geometrical optimisation using DFT calculations with Psi4. The most stable conformation was selected from 1,000 different conformations based on total DFT energies.

***Electronic property calculation of repeating units*****:** The atomic charges of the stable conformation were calculated using the restrained electrostatic potential (RESP) charge model[39] based on a Hartree–Fock single-point calculation[40]. Additional properties, including total energy, the highest occupied molecular orbital (HOMO) and the lowest unoccupied molecular orbital (LUMO) energy levels, dipole moment, and dipole polarizability tensor, were computed using the ωB97M-D3BJ functional combined with the appropriate basis sets[41,42].

***Generation of polymer chains*****:** Polymer chains were constructed by connecting repeating units using a self-avoiding random-walk algorithm. Each chain was created to include ~1,000 atoms. During chain growth, the bond between the head and capped atoms of the growing chain, along with the bond between the tail and capped atoms of the next repeat unit, was arranged coaxially and antiparallel to prevent unwanted chiral inversions or cis/trans conversions. The generated polymers were atactic (non-ordered).

***Force field parameter assignment*****:** The GAFF2 force-field parameters were automatically assigned to each polymer chain in RadonPy. When any parameters were missing for certain atomic groups, these values were empirically estimated in the same manner as for GAFF2.

***Generation of simulation cells*****:** A simulation cell containing ~10,000 atoms was created by randomly arranging and rotating ten polymer chains such that they did not overlap. The initial cell density was set to 0.05 $\mathrm{g \cdot cm^{-3}}$ and was subsequently increased through a packing simulation.

***Packing simulation*****:** The packing simulation was performed using *NVT* simulations, where the temperature was increased from 300 to 700 K, and the density was increased to 0.8 $\mathrm{g \cdot cm^{-3}}$. During this process, the Coulomb interaction was disabled to prevent the polymer chains from aggregating into globule-like structures, ensuring that the chains remained in random coil form and did not pass through one another.

***Equilibration simulation*****:** The system underwent Larsen's 21-step equilibration protocol to achieve equilibrium[43]. This process involved temperature cycling from 300 to 700 K over ~1.5 ns, pressure cycling from 50,000 to 1 atm, and repeated compression and decompression by combining the *NVT* and *NPT* simulations with a Nosé–Hoover thermostat and barostat. After completing the 21-step equilibration, *NPT* simulations were performed for over 5 ns at 300 K and 1 atm until equilibrium was achieved. Equilibrium was determined when fluctuations in energy, density, and other parameters fell below the specified thresholds, ensuring system stability throughout the 50 ns equilibration period.

***Property calculation*****:** In the latest version, the 62 distinct properties can be automatically calculated from the equilibrium MD simulations (Table S3): density, radius of gyration, scaled radius of gyration, end-to-end distance, specific heat capacities at constant pressure/volume, isothermal/isentropic compressibility, isothermal/isentropic bulk modulus, volume expansion coefficient, linear expansion coefficient, self-diffusion coefficient, refractive index, static dielectric constant, corrected static dielectric constant, and nematic order parameter.

### Calculation of thermal conductivity

To calculate thermal conductivity, we performed reverse NEMD simulations, specifically employing the Müller–Plathe method, which imposes a temperature gradient and measures the resulting heat flux[44]. The simulation box was created by triplicating an equilibrated isotropic amorphous cell along the *x*-axis under periodic boundary conditions. The box was divided into slabs, and a temperature gradient was induced by exchanging velocities between the coldest atom in the middle slab and the hottest atom in the first slab. This created a temperature gradient, with the hottest slab in the centre and the temperature decreasing towards the ends. A preheating step at 300 K for 2 ps using the $NVT$ ensemble was performed before running the reverse NEMD simulation for 1 ns using the NVE ensemble. The number of slabs was set to 20, with velocity swapping occurring every 200 fs. The time step was 0.2 fs. Non-bonded interactions were calculated using the twin-range cutoff method with cutoffs of 8 and 12 Å, and long-range Coulomb interactions were controlled using the PPPM method. The adequacy of the reverse NEMD calculations was validated by confirming a linear temperature gradient, with poor results removed from the data (Section 1 of Supplementary Notes).

In RadonPy, thermal conductivity was decomposed into six components over a 100-ps NEMD simulation, based on their contribution to energy flux through convection and interatomic interactions. These components included convection, bond, angle, dihedral, improper, and nonbonded. The non-bonded contributions were further divided into pairwise and K-space components, referred to as TC_ke, TC_pe, TC_pair, TC_bond, TC_angle, TC_dihed, TC_improper, and TC_kspace, respectively, in the PolyOmics database (Table S3)[45–47].

### Calculation of glass transition temperature

The system was initially equilibrated in the $NVT$ ensemble at high temperature, followed by a controlled cooling process at a constant rate from the liquid state to the glassy state. Cooling was performed at a rate of 800 K/ns using a Nosé–Hoover thermostat with a timestep of 1 fs. The initial temperature was determined to be the temperature at which the density was ~0.5 $\mathrm{g \cdot cm^{-3}}$, and the system was subsequently cooled to 50 K. The simulations were conducted for several nanoseconds to allow proper equilibration and transition. Glass transition was identified by a notable change in the slope of the observed density values or a sharp drop in diffusion, signalling the onset of vitrification.

### Calculation of dynamic dielectric properties

The dynamic dielectric constants ($\varepsilon'$) and dielectric loss tangent ($\tan\delta$) of isotropic amorphous polymers were evaluated over a frequency range of 10–500 GHz by simulating time-dependent dielectric polarisations under an AC electric field. The dielectric properties were calculated by performing MD simulations under the $NPT$ ensemble at 1 atm. The external AC electric field $E(t) = E_0 \sin(\omega t)$ with an arbitrary frequency was applied in the $x$, $y$, and $z$ directions to induce polarisation, defined as $P = \mu/V$, where $V$ is the volume of a simulation cell and $\mu$ is the total dipole moment. The corresponding force, depending on the charge of an atom, $q$, $F = qE(t)$ was applied via the LAMMPS `efield` command, with the field amplitude kept low to avoid non-linear effects. The field frequency was set to 10–500 GHz. The electric displacement was calculated as

$$D(t) = \varepsilon_0 E(t) + P(t) \tag{2}$$

with the phase difference $\delta$ between $D(t)$ and $E(t)$ determined via fitting. Here, $\varepsilon_0$ denotes the dielectric constant of free space. The dynamic dielectric constant $\varepsilon'$ and dielectric loss tangent $\tan\delta$ were derived as:

$$\varepsilon' = 1 + \frac{P_o}{\varepsilon_0 E_0}\cos\delta + \frac{P_e}{\varepsilon_0 E} \tag{3}$$

$$\tan\delta = \frac{P_o}{\varepsilon_0 E_0}\sin\delta \times \frac{1}{\varepsilon'} \quad (4)$$

where $P_e$ represents the electronic polarisation obtained from DFT calculations and $P_o$ denotes the orientational dipolar polarisation. The calculated values were averaged over five independent initial configurations (Section 1 of Supplementary Notes).

The calculated $\varepsilon'$ and $\tan\delta$ values exhibited good agreement with the experimental values of 13 polyimides at 0% relative humidity (RH) (Fig. S2). The experimental values, originally measured at multiple RH levels by Sawada et al.[48], were extrapolated to 0% RH for comparison.

## Calculation of thermal conductivity under uniaxial stretching

The thermal conductivities of polymers can be significantly enhanced by uniaxial stretching (Section 1 in Supplementary Notes). After generating the isotropic amorphous structures through equilibrium simulations, an iterative stretching and relaxation process was performed by extending the simulation box in one direction. Stretching was performed under *NPT* conditions at 300 K with a stretching speed of 25 m/s and strain of 0.5 and 1.0. Following deformation, the structures were equilibrated for 500 ps. The thermal conductivity along the stretching direction was evaluated using NEMD simulations implemented in RadonPy. Furthermore, the degree of chain alignment in the stretched structures was quantified by calculating the order parameter based on the moments of inertia of the individual repeating units.

## Calculation of solubility parameters

The Hansen solubility parameter is commonly expressed as three types of cohesive energy densities, corresponding to the intermolecular interaction energies derived from the dispersion term $\delta_{\mathrm{d}}$, the polar interaction term $\delta_{\mathrm{p}}$, and the hydrogen bonding term $\delta_{\mathrm{h}}$. However, when calculating the solubility parameter using the GAFF2 force field, the non-bonded interactions comprise van der Waals and electrostatic interactions. Thus, the solubility parameter comprises three components: the interaction originating from van der Waals interactions $\delta_{\mathrm{vdw}}$, the interaction from electrostatic interactions $\delta_{\mathrm{e}}$, and their total sum $\delta_{\mathrm{total}}$. In this study, we considered that $\delta_{\mathrm{vdw}}$ corresponds to $\delta_{\mathrm{d}}$ and $\delta_{\mathrm{e}}$ corresponds to $\delta_{\mathrm{e}} = \sqrt{\delta_{\mathrm{p}}^2 + \delta_{\mathrm{h}}^2}$, and we evaluated the consistency between the MD-calculated values and experimental values (Fig. S2). The solubility parameters in the PolyOmics database followed this definition.

Generally, to calculate the solubility parameter, computing the cohesive energy and molar volume for a single molecule in the gas phase (with no intermolecular interactions) and for the bulk system is necessary. Therefore, to compute the energy of the former, we extracted a single molecule from an equilibrated multiparticle bulk system in an MD simulation. The cohesive energy density was calculated from the interaction energy within the molecule and molar volume, and the difference from the bulk system energy was used to calculate the solubility parameter (Section 1 in the Supplementary Notes).

## Data collections

In this study, the following datasets from PolyOmics were analysed and illustrated as examples.

***Small molecules***: The database contains the MD-derived physical property data of 23,361 small molecules. By comparing the properties of the polymers with those of their isolated repeating units, it is possible to systematically analyse how the degree of polymerisation and molecular weight influence polymeric properties (Fig. S6).

***PFAS resins***: The database includes a dataset comprising 3,821 PFAS resins. Notably, the physical properties of these PFAS are located near the tails of the joint property distributions, indicating their distinctive characteristics within a broader polymer space (Fig. 3e). The PolyOmics PFAS dataset is expected to facilitate identification of seed compounds in the search for alternatives to PFAS.

***Cellulose derivatives***: Cellulose derivatives are a group of polymers obtained by chemically modifying natural cellulose. Cellulose, the primary component of plant cell walls, is a polysaccharide comprising numerous linked glucose units. Its structure consists of glucose molecules connected in a linear chain via $\beta$-1,4-glycosidic bonds with each glucose unit containing three hydroxyl groups (–OH). By introducing ether or ester groups to these hydroxyl groups through chemical modification, the properties of cellulose, such as biodegradability, solubility, viscosity, thermal and mechanical properties, and chemical stability, can be tailored to suit specific applications. In this study, using a set of ~300,000 substituents extracted from the commercially available reagent database ZINC[34], ~50,000 cellulose derivatives were randomly generated with degrees of substitution ranging from 0 to 3 (Fig. S5). The isotropic amorphous properties of these derivatives were calculated using RadonPy, which covered 43 different physical properties (Fig. 3).

***Biodegradable polymer candidates***: The database includes a collection of candidate materials for biodegradable polymer alternatives to cellulose derivatives. Fransen et al.[33] conducted biodegradability assessments of 642 distinct types of polyesters and polycarbonates using their proprietary high-throughput experimental platform. Exploring the relationship between the material properties of these polymers, as archived in PolyOmics, and their biodegradability metrics is an important and interesting application of the database.

***Uniaxially stretched oriented polymers***: Currently, PolyOmics contains thermal conductivity profiles for 1,379 uniaxially stretched polymers. The relationship between the nematic order parameter and thermal conductivity for 1,379 polymers at strains of 0, 0.5, and 1.0 is shown in Fig. 3d. The orientation order ranged from 0 to 0.6, while thermal conductivity values spanned from 0.2 to 2.5 $W{\cdot}m^{-1}{\cdot}K^{-1}$. The thermal conductivity generally increased with a higher orientation order.

Some polymers exhibited relatively high thermal conductivities (1.5–2.0 $W{\cdot}m^{-1}{\cdot}K^{-1}$) despite having orientation order values below 0.2. The correlation between the average thermal conductivity and the average orientation order parameter at each strain level for each polymer class is shown in Fig. 3d.

Notably, polysulfides, polyphenylenes, and polyimines demonstrate high thermal conductivities at a strain of 1.0, whereas polyacrylates, polyphosphazenes, and polystyrenes exhibited minimal changes in conductivity upon stretching. These results suggest that polymers with rigid backbones, such as those containing multiple bonds and aromatic rings, exhibit significant enhancement in thermal conductivity through orientation. Conversely, polymers with bulky side chains showed limited improvement in conductivity even when their orientation order increased.

***Ladder polymers***: A structure generator in RadonPy was used to systematically construct ladder polymer structures. As depicted in Fig. S3, 170 distinct ring structures (R) were defined, each possessing four polymerisable sites, and eight types of alkyl linker groups (L) containing two polymerisable sites. By exhaustively combining these components within an R-R-L framework, 1, 360 ladder polymers (170 × 8 combinations) were generated. A total of 519 ladder polymers and their corresponding non-ladder polymers were calculated and included in PolyOmics. Furthermore, as a control, non-ladder polymers were constructed by removing one of the two polymerisable sites from the R-R framework. The isotropic amorphous properties of the polymers were evaluated thoroughly. The results indicate that compared to non-ladder polymers, ladder polymers exhibit a significant and simultaneous reduction in both thermal expansion and volumetric expansion coefficients. Furthermore, the analysis of the isotropic amorphous structures demonstrated that ladder polymers tend to adopt a zigzag-like higher-order conformation at the junctions of the building blocks (R-R-L) and exhibit a lower density than non-ladder polymers. This reduction in density is attributed to the increased rigidity of the dual linkages, which hindered efficient packing. The ladder architecture significantly reduces molecular mobility, as evidenced by a decrease in the self-diffusion coefficient. Elongation of the alkyl linker chains promotes densification and increases the self-diffusion coefficient.

## Transfer learning workflow and scaling analysis

For transfer learning between 43 and 61 datasets from PolyOmics and PoLyInfo, respectively, we employed a conventional multilayer perceptron. The model input was defined as a 190-dimensional kernel-mean force field descriptor vector that represents the compositional and structural features of a polymer repeating unit, as detailed later. The mapping from the vectorised polymer to each property was modelled using a neural network with five hidden layers, where each layer applied a rectified linear unit (ReLU) activation function to a fully connected linear layer. The numbers of neurons in the hidden layers were fixed at 512, 256, 128, 64, and 32. The output regression layer was placed on the head to calculate the scalar value of the MD-calculated or experimental properties. During model training, we used the mean squared error as the loss function and optimised the model using the Adam optimiser. The learning rate was set to 0.001, and the batch size was set to 32. Pre-training and fine-tuning were performed under the same conditions. Of the training dataset, 10% and 20% were masked as the validation set in pre-training and fine-tuning, respectively, and training was stopped when there was no improvement in the validation loss for 50 and 20 consecutive steps in pre-training and fine-tuning, respectively. For further details, refer to the distributed PyTorch implementation.

To determine the presence or absence of transfer learning scaling, we used the area ratio between the estimated scaling curve and the baseline of the scratch learning. The estimated power-law function was integrated over the range $n \in [10^2, 10^5]$ to compute the area $S_1$. As the baseline, we used a constant function corresponding to the mean MAE of models trained solely on experimental data, and calculated the area $S_0$ over the same range of $n \in [10^2, 10^5]$. Cases in which the area ratio $S_1/S_0$ was less than or equal to 0.95 were judged to exhibit scalable transfer learning.

**Kernel mean embedding of force field parameters**

The chemical structure of a given polymer was described using force field parameters defined by the AMBER (GAFF2) potential. The potential of the all-atom classical MD simulation consists of seven distinct terms representing van der Waals interaction energy, electrostatic interactions, bond stretching, bond angle bending, dihedral angle rotation, and improper torsion. The atoms constituting the polymer were classified according to their element species, and predefined parameter values were assigned to each atom or atomic group based on the element species. Table S6 summarises the GAFF2 parameters. Ten distinct parameters were assigned to their corresponding components, with four distinct parameters applied to individual atoms, three parameters to the two-body atom species forming a bond, two parameters to the three-body atom species forming a bond angle, and one parameter to the three-body atom species forming a dihedral angle.

As the number of force field parameters varies depending on the polymer species, these parameters must be converted into a fixed-length descriptor vector for input into the neural network. Therefore, we employed a machine learning technique called kernel mean embedding[49] to encode the distribution of force field parameters into a fixed-length vector[50]. This procedure involved smoothing the histogram of the parameter component values using kernel density estimation with a Gaussian radial basis function kernel. The discretised points were ten points corresponding to ten different element species, such as hydrogen and carbon, for mass, and 20 equally spaced grid points for the other nine parameters. Therefore, each polymer was encoded as a 190-dimensional descriptor.

## Data availability

All data in the PolyOmics database, including polymer repeat units, MD-calculated properties, machine learning–predicted miscibility with organic solvents and plasticizers, MD simulation conditions, and JSON files of relaxed structures obtained from equilibrated simulations, are available at Hugging Face (https://huggingface.co/datasets/yhayashi1986/PolyOmics). For convenience, a curated CSV file containing the specific datasets highlighted in this paper is provided as a supplementary resource.

Upon publication of this paper, all PolyOmics data will be accessible through a dedicated web-based database (https://radonpydb.org/). This online platform offers user-friendly functionalities, such as graphical interface–based data search and browsing, as well as programmatic access and data downloading through an application programming interface (API).

Owing to licensing restrictions, the experimental datasets from the PoLyInfo database could not be redistributed. Access to the PoLyInfo database requires contacting the designated office of the National Institute for Materials Science via the PoLyInfo website (https://polymer.nims.go.jp).

## Code availability

The latest version of RadonPy (version 1.0) is available at https://github.com/RadonPy/RadonPy. Source code for transfer learning and a pretrained machine learning model for predicting polymer–solvent miscibility are also provided.

## Acknowledgements

We express our sincere gratitude to all members of the RadonPy consortium for their valuable contributions to the advancement of their respective projects and the overall operation of the consortium. This research was primarily supported by the Ministry of Education, Culture, Sports, Science and Technology (MEXT) through the “Program for Promoting Researches on the Supercomputer Fugaku” (JPMXP1020200314). Additional support was provided by the Japan Science and Technology Agency (JST) (JPMJCR19I3, JPMJCR22O3, JPMJCR2332, JPMJCR2546, and

JPMJPF2102) and the Japan Society for the Promotion of Science (JSPS) (25H01126, 19H01132, 22K11949, and 25K00147). Computational resources were provided by Fugaku at the RIKEN Center for Computational Science, Kobe, Japan (hp210264, hp220179, hp230190, hp240216, and hp250235) and a supercomputer at the Research Center for Computational Science, Okazaki, Japan (21-IMS-C126, 22-IMS-C125, 23-IMS-C113, 24-IMS-C107, and 25-IMS-C107).

## Contributions

R.Y. conceived and led the project, formulated the design principles of the PolyOmics database, and developed a core conceptual framework for Sim2Real transfer learning. He also established the RadonPy consortium and oversaw the project's overall direction.

Y.H. developed and implemented both the RadonPy software and the PolyOmics database, as well as the experimental validation of scaling laws. Most of the functionalities in RadonPy were designed and implemented by Y.H..

H.F. and R.H. developed an automated property calculation system and generated data for dielectric properties and thermal conductivity under uniaxial stretching. H.S. implemented an automated system for calculating glass transition temperatures. K.K. developed an automated system for calculating solubility parameters. S.N. implemented the automated system and generated the refractive index and Abbe number data. T.K. contributed to the automated generation and data production of the ladder polymers. K.S. generated a dataset of cellulose derivatives. Y.K. generated the dataset for small molecules. S.W. and K.S. developed a machine learning model to predict $\chi$ parameters and applied it to estimate the miscibility between virtual polymers and organic solvents in PolyOmics. R.S. contributed to the implementation of the automated property calculation system for the rheological properties. Y.N. generated virtual polymers using a chemical language model. M.O. developed the SMiPoly software and generated the virtual polymers. Additionally, many other consortium members voluntarily defined the research tasks and contributed to data generation. M.I. and I.K. provided the experimental data from the PoLyInfo database. C.L., Y.H., and A.T. developed a web-based version of the PolyOmics database that includes a set of APIs.

As the consortium's administrative manager, A.T. coordinated the overall project operations and support, including the collection and organization of data produced within the consortium. J.S., T.N., N.T., K.S., M.O., H.S., H.F., H.I., Y.M., H.I., M.Z., Y.H., and A.T. served as members of the steering committee under the leadership of R.Y. and contributed to project management and coordination.

R.Y. drafted the initial manuscript and Y.H. and H.F. revised the technical sections. T.H. and M.O. contributed to figure preparation and final proofreading of the manuscript.

Supplementary Information

# Omics-scale polymer computational database transferable to real-world artificial intelligence applications

The RadonPy consortium*

* The authors and their affiliations are listed at the end of the main text.

## Table of Contents

### Supplementary Tables

### Supplementary Figures

### Supplementary Notes

**Table S1.** List of currently available polymer databases: P = presence; A = absence; E = experimental data; C = computational data.

| Database | Description | Property data | Data type | Bulk download | URL | Ref |
|---|---|---|---|---|---|---|
| PoLyInfo | ~100 distinct properties compiled for ~19,000 homopolymers and several thousand copolymers and polymer blends, extracted from published literature | P | E | A | https://polymer.nims.go.jp/ | [1,2] |
| PI1M | ~1M polymers generated from deep generative models; no property data is included. | A | C | P | https://github.com/RUIMINMA1996/PI1M | [3] |
| Polymer Scholar | 24 properties of ~100k polymers, extracted from published literature using LLM | P | E | A | https://polymerscholar.org/ | [4] |
| polyVERSE | A compilation of several datasets gathered by a research group at Georgia Institute of Technology | P | E & C | P | https://github.com/Ramprasad-Group/polyVERSE | [5] |
| OMG | A dataset of ~12M polymers generated by applying 17 polymerisation rules to purchasable monomers; no property data is included. | A | C | P | https://zenodo.org/records/7556992 | [6] |
| SMiPoly | A dataset of ~180k polymers generated by applying 22 polymerisation rules to purchasable monomers; no property data is included. | A | C | P | https://github.com/PEJpOhno/SMiPoly | [7] |
| PolyOne | ~100M hypothetical polymers with 29 predicted properties using machine learning models | P | C | P | https://zenodo.org/records/7766806 | [8] |
| RadonPy | ~1k polymers with 15 properties calculated using RadonPy | P | C | P | https://github.com/RadonPy/RadonPy | [9] |
| PolyOmics | ~130k polymers with 43 properties calculated using RadonPy and machine learning χ parameters for 19 solvents and plasticisers | P | C | P | https://huggingface.co/datasets/yhayashi1986/PolyOmics | This article |

**Table S2.** 20 polymer classes with the number of data entries in the general polymers dataset of the PolyOmics database.

| Polymer class name | Class label | # of entries | # of unique chemical structures | # of entries from XenonPy | # of entries from SMiPoly |
|---|---|---|---|---|---|
| Polyhydrocarbons | PHYC | 3,596 | 1,724 | 1,708 | 107 |
| Polystyrenes | PSTR | 7,435 | 5,925 | 20,233 | 32,994 |
| Polyvinyl, polyvinylidene | PVNL | 9,967 | 5,320 | 23,824 | 16,160 |
| Polyacrylate | PACR | 7,194 | 4,613 | 95,855 | 88,614 |
| Halogen-containing polymers | PHAL | 1,387 | 1,331 | 1,614 | 2,112 |
| Polyene | PDIE | 1,239 | 693 | 778 | 3 |
| Polyether | POXI | 6,675 | 5,081 | 91,035 | 61 |
| Polythioether | PSUL | 3,794 | 3,693 | 30,631 | 0 |
| Polyester | PEST | 4,446 | 4,190 | 458,961 | 9,632 |
| Polyamide | PAMD | 3,740 | 3,564 | 370,460 | 7,946 |
| Polyurethane | PURT | 3,095 | 2,991 | 146,887 | 2,772 |
| Polyurea | PURA | 2,876 | 2,857 | 55,968 | 0 |
| Polyimide | PIMD | 8,237 | 6,268 | 176,614 | 6,553 |
| Polyanhydrides | PANH | 3,879 | 3,472 | 61,776 | 1,434 |
| Polycarbonate | PCBN | 3,388 | 3,235 | 38,891 | 138 |
| Polyimines and nitrogen-containing polymers other than polyamines and polyamides | PIMN | 4,615 | 3,875 | 34,658 | 0 |
| Polyphosphazene | PPHS | 2,921 | 2,915 | 13,762 | 0 |
| Polyketones | PKTN | 4,218 | 4,166 | 81,027 | 130 |
| Polysulfon | PSFO | 4,064 | 3,974 | 49,147 | 332 |
| Polyphenylene | PPNL | 3,200 | 3,159 | 24,210 | 359 |

**Table S3.** 62 properties with automated computation pipelines implemented in the current version of RadonPy, along with the number of data entries registered in the PolyOmics database.

| Property name (this article) | Abbreviation (this article) | Notation in the database | Unit in the database | Remarks | # of entries |
|---|---|---|---|---|---|
| Density | $P$ | Density | g·cm$^{-3}$ | | 125,658 |
| Specific heat capacity at constant pressure | $C_p$ | Cp | J·mol$^{-1}$·K$^{-1}$ | | 125,654 |
| Specific heat capacity at constant volume | $C_v$ | Cv | J·mol$^{-1}$·K$^{-1}$ | | 125,654 |
| Compressibility | $\beta_T$ | compressibility | Pa$^{-1}$ | | 125,654 |
| Isentropic compressibility | $\beta_S$ | isentropic_compressibility | Pa$^{-1}$ | | 125,654 |
| Bulk modulus (isothermal) | $K_T$ | bulk_modulus | Pa | | 125,654 |
| Isentropic bulk modulus | $K_S$ | isentropic_bulk_modulus | Pa | | 125,654 |
| Self-diffusion coefficient | $D$ | self-diffusion | m$^2$·s$^{-1}$ | | 125,653 |
| Linear expansion coefficient | CLTE | linear_expansion | K$^{-1}$ | | 125,654 |
| Volume expansion coefficient | CVTE | volume_expansion | K$^{-1}$ | | 125,654 |
| Radius of gyration | $R_g$ | Rg | Å | | 125,658 |
| Scaled radius of gyration | Scaled $R_g$ | scaled_Rg | Å | The scaled $R_g$ was defined as $R_g$ scaled by $1/M_w^{0.6}$ to remove the dependency on the molecular weight of the polymer ($M_w$). | 125,658 |
| Mean square end-to-end distance | $R_e$ | end-to-end_distance | $Å^2$ | | 125,259 |
| Nematic order parameter | $S_n$ | nematic_order_parameter | dimensionless | | 125,275 |
| Free volume | $V_{free}$ | free_volume | $Å^3$ | | 0 |
| Fractional free volume | $V_{FFV}$ | fractional_free_volume | $Å^3$ | | 0 |
| Refractive index | RI | refractive_index | dimensionless | | 123,015 |
| Refractive index (486 nm) | $RI_D$ | refractive_index_486 | dimensionless | Refractive index at 486 nm | 13,240 |
| Refractive index (589 nm) | $RI_F$ | refractive_index_589 | dimensionless | Refractive index at 589 nm | 13,240 |
| Refractive index (656 nm) | $RI_C$ | refractive_index_656 | dimensionless | Refractive index at 656 nm | 13,240 |
| Abbe's number | $\upsilon$ | abbe_number | dimensionless | | 13,240 |
| Cohesive energy density | CED | sp_cep | $J \cdot cm^{-3}$ | | 58,834 |
| Solubility parameter (total) | $\delta_{\mathrm{total}}$ | sp_total | $MPa^{0.5}$ | | 58,834 |
| Solubility parameter (vdw) | $\delta_{\mathrm{vdw}}$ | sp_vdw | $MPa^{0.5}$ | vdW (Lennard-Jones potential) term in solubility parameter | 58,834 |
| Solubility parameter (electric) | $\delta_{\mathrm{ele}}$ | sp_ele | $MPa^{0.5}$ | Electrostatic potential term in the solubility parameter | 58,834 |
| Solubility parameter (short-range electric) | $\delta_{\mathrm{ele-sr}}$ | sp_ele_short | $MPa^{0.5}$ | Pairwise coulombic potential term in solubility parameter | 58,834 |
| Solubility parameter (long-range electric) | $\delta_{\mathrm{ele-lr}}$ | sp_ele_long | $MPa^{0.5}$ | Particle-Particle Particle-Mesh (PPPM) method term (kspace term in LAMMPS) in the solubility parameter | 58,834 |
| Dielectric constant (static) | $\varepsilon(0)$ | static_dielectric_const | dimensionless | | 132,828 |
| Corrected dielectric constant (static) | $\varepsilon_{\mathrm{corr}}(0)$ | dielectric_const_dc | dimensionless | Defined as (static dielectric constant) – 1 + (refractive index)$^2$ | 123,015 |
| Dynamic dielectric constant | $\varepsilon'$ | efdp_permittivity_real | dimensionless | Dynamic dielectric constant with a frequency range from 10 to 500 GHz | 1,084 |
| Dielectric loss | $\varepsilon''$ | efdp_permittivity_imaginary | dimensionless | Frequency range from 10 to 500 GHz | 1,084 |
| Dielectric loss tangent | $\tan\delta$ | efdp_dielectric_loss_tan | dimensionless | Frequency range from 10 to 500 GHz | 1,084 |
| Molecular weight (repeating unit) | $M$ | mol_weight_monomer | | | 125,258 |
| vdw volume (repeating unit) | $V_{vdw}$ | vdw_volume_monomer | $Å^3$ | | 125,258 |
| Dipole moment | $\mu$ | qm_dipole_moment_monomer | debye | DFT calculated value | 125,225 |
| Polarizability | $\alpha$ | qm_polarizability_monomer | $Å^3$ | DFT calculated value | 124,274 |
| Polarizability (486 nm) | $\alpha_D$ | qm_polarizability_sos_486_monomer | $Å^3$ | Dynamic polarizability at 486 nm by sum-over-state (SOS) approach using TD-DFT | 13,240 |
| Polarizability (589 nm) | $\alpha_F$ | qm_polarizability_sos_589_monomer | $Å^3$ | Dynamic polarizability at 589 nm by the SOS approach using TD-DFT | 13,240 |
| Polarizability (656 nm) | $\alpha_C$ | qm_polarizability_sos_656_monomer | $Å^3$ | Dynamic polarizability at 656 nm by the SOS approach using TD-DFT | 13,240 |
| Total energy | $E_{etot}$ | qm_total_energy_monomer | hartree | DFT calculated value | 125,225 |
| HOMO | HOMO | qm_homo_monomer | eV | DFT calculated value | 125,225 |
| LUMO | LUMO | qm_lumo_monomer | eV | DFT calculated value | 125,225 |
| Thermal conductivity | K | thermal_conductivity | $W \cdot m^{-1} \cdot K^{-1}$ | | 116,922 |
| Thermal diffusivity | Λ | thermal_diffusivity | $m^2 \cdot s^{-1}$ | | 116,919 |
| Thermal conductivity (kinetic energy) | $K_{ke}$ | TC_ke | $W \cdot m^{-1} \cdot K^{-1}$ | Kinetic energy term in the thermal conductivity by decomposition analysis | 116,539 |
| Thermal conductivity (potential energy) | $K_{pe}$ | TC_pe | $W \cdot m^{-1} \cdot K^{-1}$ | Potential energy term in the thermal conductivity by decomposition analysis | 116,539 |
| Thermal conductivity (pairwise interaction) | $K_{pair}$ | TC_pair | $W \cdot m^{-1} \cdot K^{-1}$ | Pairwise coulombic interaction term in the thermal conductivity by decomposition analysis | 116,539 |
| Thermal conductivity (bond) | $K_{bond}$ | TC_bond | $W \cdot m^{-1} \cdot K^{-1}$ | Bond stretching term in the thermal conductivity by decomposition analysis | 116,539 |
| Thermal conductivity (angle) | $K_{angle}$ | TC_angle | $W \cdot m^{-1} \cdot K^{-1}$ | Bond bending term in the thermal conductivity by decomposition analysis | 116,539 |
| Thermal conductivity (dihedral angle) | $K_{dihed}$ | TC_dihed | $W \cdot m^{-1} \cdot K^{-1}$ | Bond rotation term in the thermal conductivity by decomposition analysis | 116,539 |
| Thermal conductivity (improper angle) | $K_{improper}$ | TC_improper | $W \cdot m^{-1} \cdot K^{-1}$ | Improper angle term in the thermal conductivity by decomposition analysis | 116,539 |
| Thermal conductivity (K-space) | $K_{kspace}$ | TC_kspace | $W \cdot m^{-1} \cdot K^{-1}$ | K-space term in the thermal conductivity by decomposition analysis | 116,539 |
| Glass transition temperature | $T_g$ | tg | K | | 84,368 |
| Thermal expansion coefficient (below $T_g$) | $CTE_b$ | tg_thermal_expansion_coef(below_tg) | $K^{-1}$ | Calculated from the gradient of density and temperature | 83,985 |
| Thermal expansion coefficient (upper $T_g$) | $CTE_u$ | tg_thermal_expansion_coef(upper_tg) | $K^{-1}$ | Calculated from gradient of density and temperature | 83,985 |
| Tensile modulus | $Y$ | tem_tensile_modulus | GPa | | 0 |
| Poisson ratio | $\nu$ | tem_poisson_ratio | dimensionless | | 0 |
| Bulk modulus (calculated by uniaxial stretching) | $K_{tensile}$ | tem_bulk_modulus | GPa | Calculated from Young modulus and Poisson ratio by uniaxial stretching simulation | 0 |
| Shear modulus | $G_{tensile}$ | tem_shear_modulus | GPa | Calculated from Young modulus and Poisson ratio by uniaxial stretching simulation | 0 |
| Lame constant | $\Lambda_{tensile}$ | tem_lame_constant | GPa | Calculated from Young modulus and Poisson ratio by uniaxial stretching simulation | 0 |
| Tensile viscosity | $\eta_{tensile}$ | tem_tensile_viscosity | Pa·s | Calculated from Young modulus by uniaxial stretching simulation | 0 |
| Speed of sound | $C$ | tem_speed_of_sound | $m \cdot s^{-1}$ | Calculated from Young modulus and shear modulus by uniaxial stretching simulation | 0 |

**Table S4.** 19 organic solvents and plasticisers for $\chi$ parameter prediction

| Name | Type | SMILES | CAS number | Notes |
|---|---|---|---|---|
| Hexane | Solvent | `CCCCCC` | 110-54-3 | Non-polar solvent, used for polyolefins and cleaning. |
| Benzene | Solvent | `c1ccccc1` | 71-43-2 | Aromatic non-polar solvent, toxic and regulated. |
| Acetone | Solvent | `CC(=O)C` | 67-64-1 | Polar aprotic, low boiling point, used for cleaning and dissolution. |
| Ethanol | Solvent | `CCO` | 64-17-5 | Polar aprotic, low boiling point, used for cleaning and dissolution. |
| DMSO (dimethyl sulfoxide) | Solvent | `CS(=O)C` | 67-68-5 | Strongly polar aprotic solvent with high solvation power. |
| Water | Solvent | `O` | 7732-18-5 | Polar protic solvent, essential for hydrophilic systems. |
| Chloroform | Solvent | `ClC(Cl)Cl` | 67-66-3 | Non-polar halogenated solvent, toxic. |
| DMF (dimethylformamide) | Solvent | `O=CN(C)C` | 68-12-2 | Polar aprotic solvent, used for polyamide and polyurethane. |
| THF (tetrahydrofuran) | Solvent | `C1CCOC1` | 109-99-9 | Medium polarity, widely used in polymerisation and dissolution. |
| Diethyl ether | Solvent | `CCOCC` | 60-29-7 | Low boiling, highly flammable, used in extraction and washing. |
| Methanol | Solvent | `CO` | 67-56-1 | Polar protic solvent, common reaction medium. |
| Ethyl acetate | Solvent | `CC(=O)OCC` | 141-78-6 | Medium polarity ester, used in coatings and adhesives. |
| Toluene | Solvent | `Cc1ccccc1` | 108-88-3 | Aromatic non-polar solvent, used for polystyrene, etc. |
| Cyclohexane | Solvent | `C1CCCCC1` | 110-82-7 | Non-polar solvent, solubility testing. |
| 1,4-Dioxane | Solvent | `O1CCOCC1` | 123-91-1 | Ether-type solvent with high solvation power, regulated in some regions. |
| DEHP (di(2-ethylhexyl) phthalate) | Plasticizer | `CCCCC(CC)COC(=O)c1ccccc1C(=O)OCC(CC)CCCC` | 117-81-7 | Phthalate plasticizer (PVC), regulated in many regions. |
| DEHA (di(2-ethylhexyl) adipate) | Plasticizer | `CCCCC(CC)COC(=O)CCCCC(=O)OCC(CC)CCCC` | 103-23-1 | Adipate-type plasticizer, improves low-temperature flexibility. |
| ATBC (acetyl tributyl citrate) | Plasticizer | `CCCCOC(=O)CC(CC(=O)OCCCC)(C(=O)OCCCC)OC(=O)C` | 77-90-7 | Citrate-type non-phthalate plasticizer. |
| DEGDB (diethylene glycol dibenzoate) | Plasticizer | `O=C(OCCOCCOC(=O)c1ccccc1)c2ccccc2` | 120-55-8 | Benzoate-type non-phthalate plasticizer. |

**Table S5.** 61 property datasets in the PoLyInfo database used for Sim2Real transfer from 43 source datasets in PolyOmics.

| Property names | Remarks on data extraction protocol | # of entries |
|---|---|---|
| Brittleness temperature | | 61 |
| Charpy impact | | 17 |
| Compressive modulus | | 14 |
| Contact angle | | 132 |
| $C_p$ | Data measured near $T_g$ is excluded | 104 |
| Crystallization temperature | | 151 |
| Deflection temperature under load | | 44 |
| Density | | 607 |
| Dielectric constant (under AC; all data) | | 296 |
| Dielectric constant (under AC; high-frequency) | Frequency range >= 1 GHz | 80 |
| Dielectric constant (under AC; middle-frequency) | Frequency range >= 1 MHz, < 1 GHz | 78 |
| Dielectric constant (under AC; low-frequency) | Frequency range < 1 MHz | 193 |
| Dielectric constant (under DC) | | 34 |
| Elongation at break | | 485 |
| Elongation at yield | | 51 |
| Flexural modulus | | 34 |
| Flexural storage modulus (all data) | | 31 |
| Flexural storage modulus (low-frequency) | Frequency range <= 1 Hz | 27 |
| Flexural stress at break | | 24 |
| Gas permeability coefficient ($CO_2$) | | 276 |
| Gas permeability coefficient ($N_2$) | | 244 |
| Gas permeability coefficient ($O_2$) | | 321 |
| Growth rate of crystal | | 26 |
| Hansen solubility parameter (dispersion) | | 56 |
| Hansen solubility parameter (hydrogen bond) | | 55 |
| Hansen solubility parameter (polar) | | 55 |
| Heat of crystallization | | 55 |
| Heat of fusion | | 258 |
| Linear expansion coefficient | | 215 |
| Refractive index | | 234 |
| Shear storage modulus (all data) | | 51 |
| Shear storage modulus (low-frequency) | Frequency range <= 1 Hz | 43 |
| Shear storage modulus loss tangent (all data) | | 46 |
| Shear storage modulus loss tangent (low-frequency) | Frequency range <= 1 Hz | 37 |
| Shore hardness | | 28 |
| Solubility parameter (Hildebrand) | | 292 |
| Stress-optical coefficient | | 36 |
| Surface tension | | 130 |
| Tensile loss modulus (all-data) | | 82 |
| Tensile loss modulus (high-frequency) | Frequency range >= 100 Hz | 40 |
| Tensile loss modulus (middle-frequency) | Frequency range > 1 Hz, < 100 Hz | 29 |
| Tensile loss modulus (low-frequency) | Frequency range <= 1 Hz | 18 |
| Tensile modulus | | 293 |
| Tensile storage modulus (all data) | | 164 |
| Tensile storage modulus (high-frequency) | Frequency range >= 100 Hz | 50 |
| Tensile storage modulus (middle-frequency) | Frequency range > 1 Hz, < 100 Hz | 87 |
| Tensile storage modulus (low-frequency) | Frequency range <= 1 Hz | 41 |
| Tensile storage modulus loss tangent (all data) | | 107 |
| Tensile storage modulus loss tangent (high-frequency) | Frequency range >= 100 Hz | 48 |
| Tensile storage modulus loss tangent (middle-frequency) | Frequency range > 1 Hz, < 100 Hz | 45 |
| Tensile storage modulus loss tangent (low-frequency) | Frequency range <= 1 Hz | 25 |
| Tensile stress at break | | 534 |
| Tensile stress at yield | | 138 |
| $T_g$ (glass transition temperature) | | 4,936 |
| Thermal conductivity | | 39 |
| Thermal decomposition temperature | | 2,747 |
| $T_m$ (melting point) | | 1,655 |
| Vicat softening temperature | | 63 |
| Volume expansion coefficient | | 166 |
| Water absorption | | 321 |
| Water vapor transmission | | 57 |

**Table S6.** 10 different parameters constituting the kernel mean force field descriptor. Each parameter value is defined on the elemental types of an atomic group forming an atom, bond, bond angle, and dihedral angle. For a given polymer, the descriptor is constructed by embedding the occurrence frequency of force field parameters into a feature space.

| Atomic group | Parameter | Description |
|---|---|---|
| Atom | Mass | Atomic mass |
| | $\sigma_{\mathrm{LJ}}$ | Determining the equilibrium distance of vdW interactions |
| | $\varepsilon_{\mathrm{LJ}}$ | Depth of the potential well of vdW interactions |
| | Charge | Atomic charge of the Gasteiger charge model |
| Bond | $r_0$ | Equilibrium length of chemical bonds |
| | $K_{\mathrm{bond}}$ | Force constant of bond stretching |
| | Polar | Bond polarisation defined by the absolute value of charge difference between atoms in a bond |
| Bond angle | $\theta_0$ | Equilibrium angle of bond angles |
| | $K_{\mathrm{angle}}$ | Force constant of bond bending |
| Dihedral angle | $K_{\mathrm{dihedral}}$ | Rotation barrier height of dihedral angles |

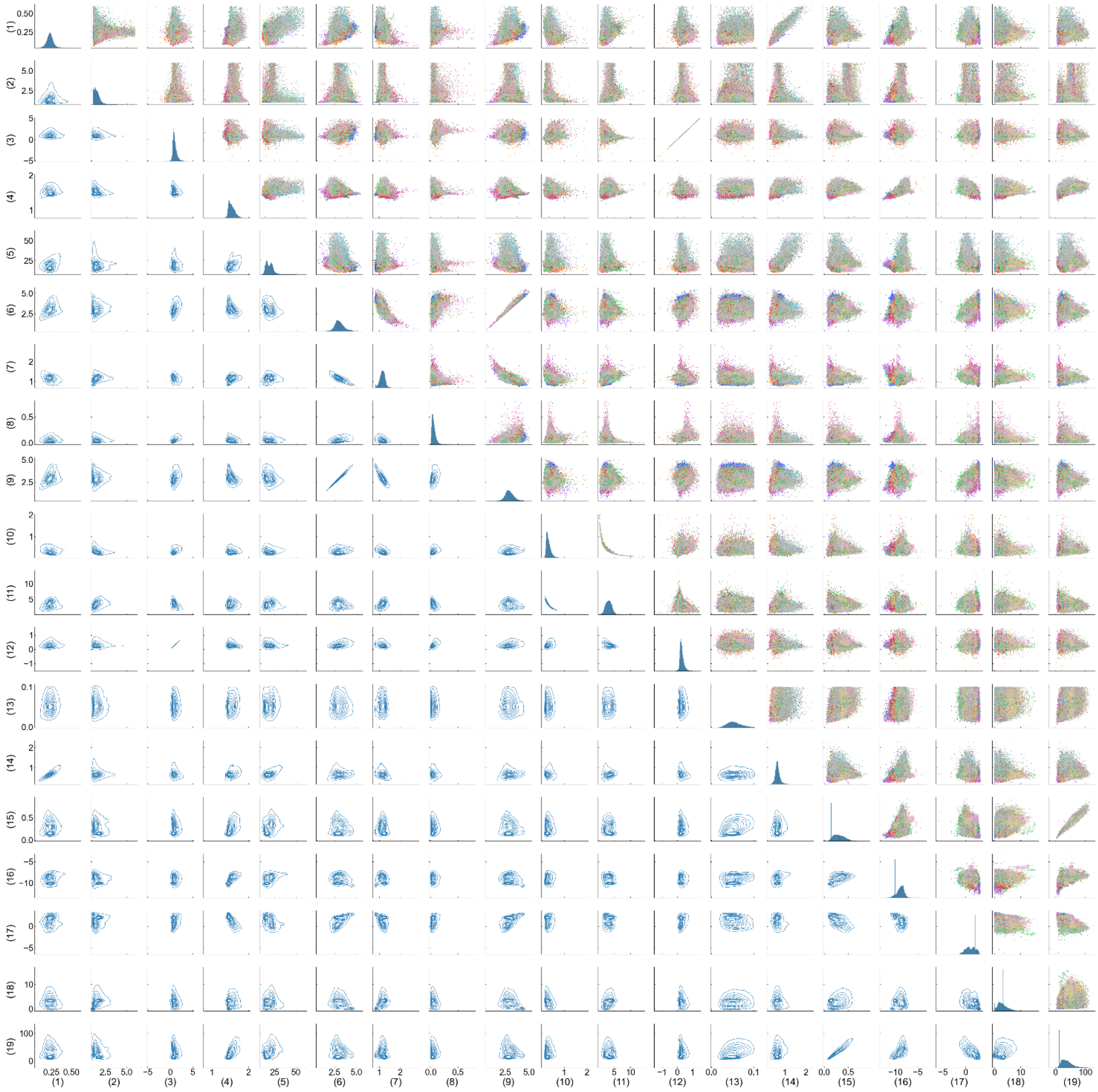

**Fig. S1.** Joint distribution of 19 different properties across over 69,532 isotropic amorphous polymers in the general polymer dataset and the cellulose derivatives dataset: (1) thermal conductivity [$W{\cdot}m^{-1}{\cdot}K^{-1}$], (2) static dielectric constant, (3) linear expansion coefficient [$10^{-4}$ $K^{-1}$], (4) refractive index, (5) radius of gyration [Å], (6) $C_p$ [$10^3$ $J{\cdot}kg^{-1}{\cdot}K^{-1}$], (7) density [$g{\cdot}cm^3$], (8) self-diffusion coefficient [$10^{-11}$ $m^2{\cdot}s^{-1}$], (9) $C_v$ [$10^3$ $J{\cdot}kg^{-1}{\cdot}K^{-1}$], (10) compressibility [$GPa^{-1}$], (11) bulk modulus [GPa], (12) volume expansion coefficient [$10^{-3}$ $K^{-1}$], (13) nematic order parameter, (14) thermal diffusivity [$10^{-7}$ $m^2{\cdot}s^{-1}$], (15) van der Waals volume (constitutional repeating unit; CRU) [$10^3$ Å$^3$], (16) HOMO (CRU) [eV], (17) LUMO (CRU) [eV], (18) dipole moment (CRU) [Debye], and (19) polarizability (CRU) [Å$^3$].

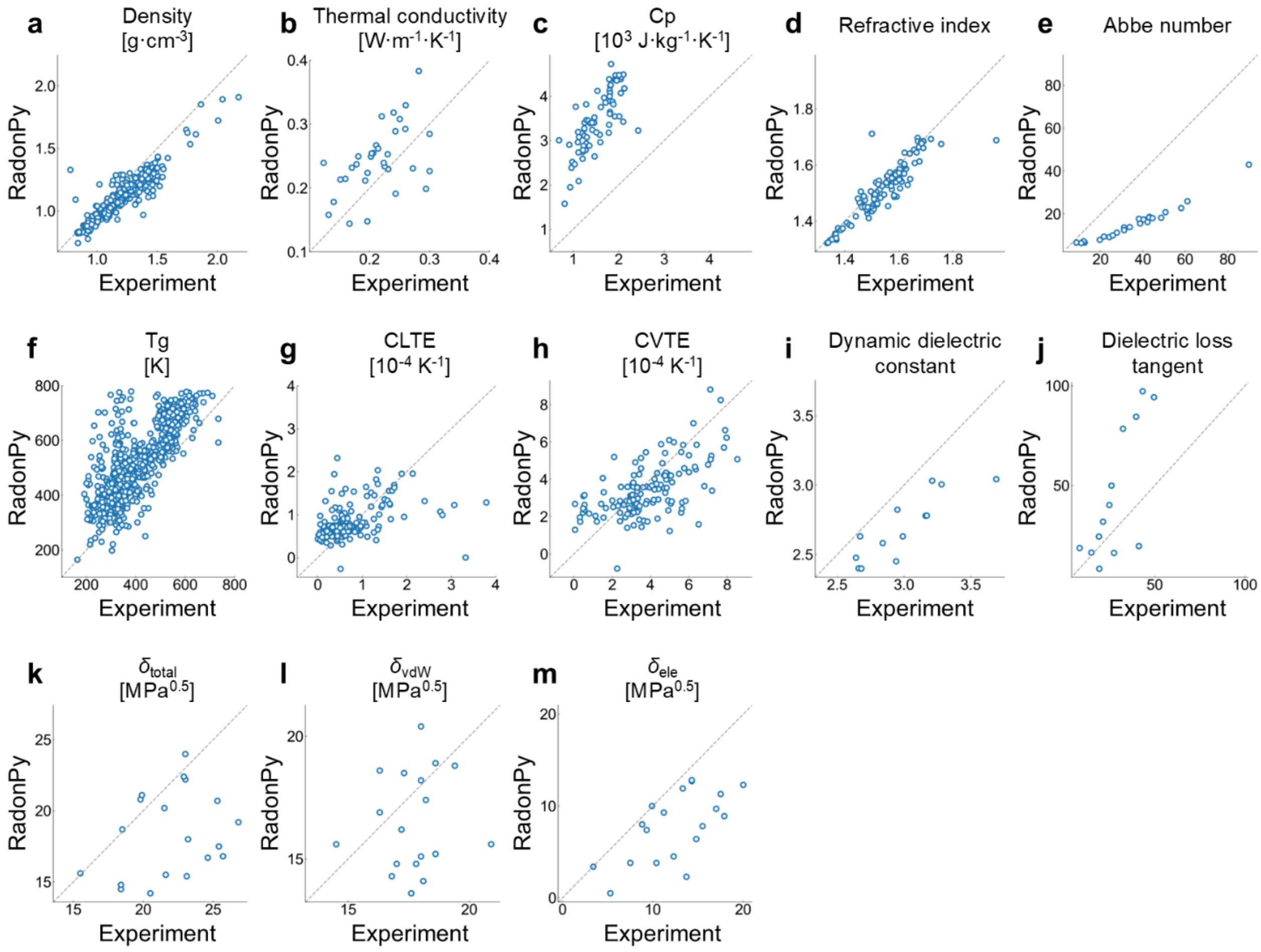

**Fig. S2.** Comparison between RadonPy-calculated properties and experimental data: **a.** density, **b.** thermal conductivity ($\kappa$, rt), **c.** specific heat capacity at constant pressure ($C_p$, rt), **d.** refractive index, **e.** Abbe number, **f**. glass transition temperature ($T_g$), **g**. linear thermal expansion coefficient (CLTE, rt), **h**. volumetric thermal expansion coefficient (CVTE, rt), **i**. dynamic dielectric constant ($\varepsilon'$, 10 GHz), **j**. dielectric loss tangent (tan $\delta$, 10 GHz), **k**. solubility parameter ($\delta_{\mathrm{total}}$), **l**. solubility parameter ($\delta_{\mathrm{vdw}}$), and **m**. solubility parameter ($\delta_{\mathrm{e}}$).

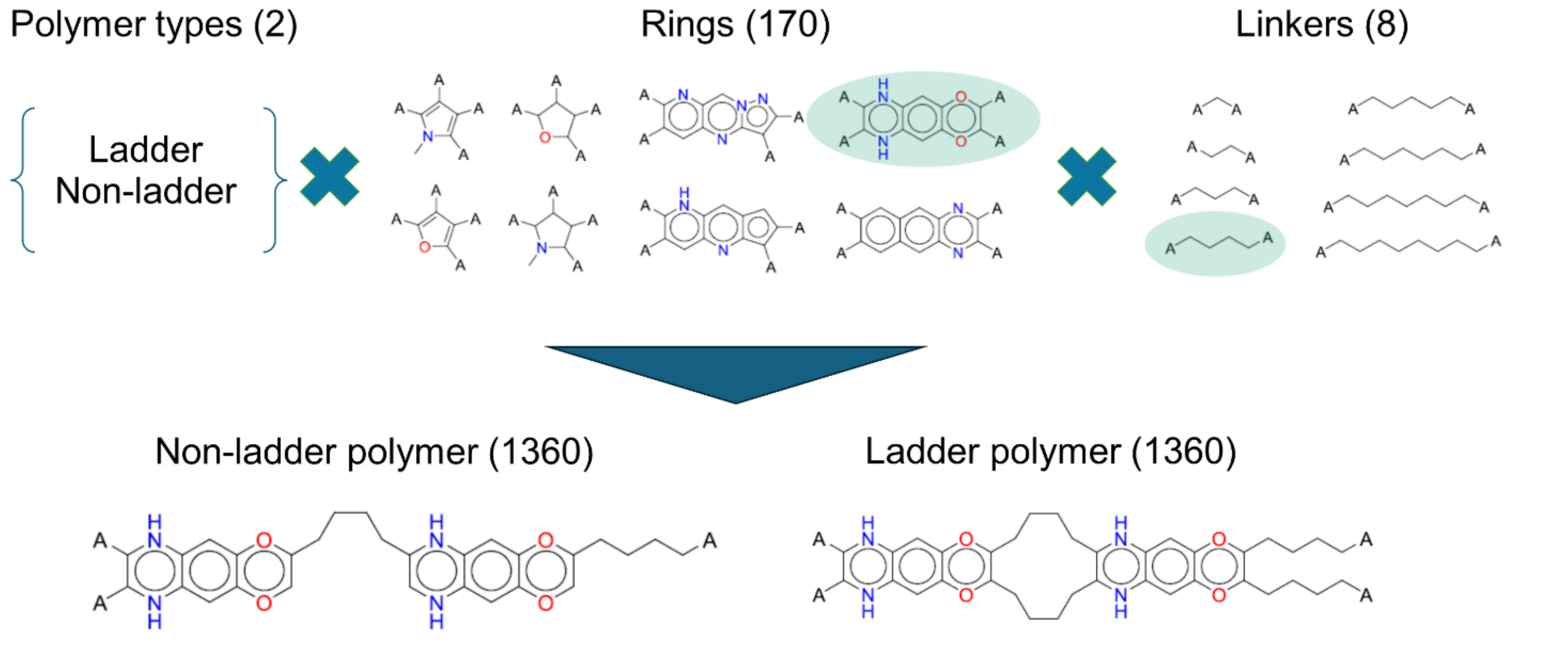

**Fig. S3.** Generation of virtual ladder polymers. We use 170 distinct ring structures (R), each having four polymerisable sites, in conjunction with eight types of alkyl linker groups (L), each containing two polymerisable sites. By combining these components within an R-R-L framework, we generate a total of 1,360 distinct ladder polymers (170 × 8 combinations) and their corresponding non-ladder polymers. Selected 519 ladder polymers and their corresponding non-ladder polymers are calculated and included in the PolyOmics.

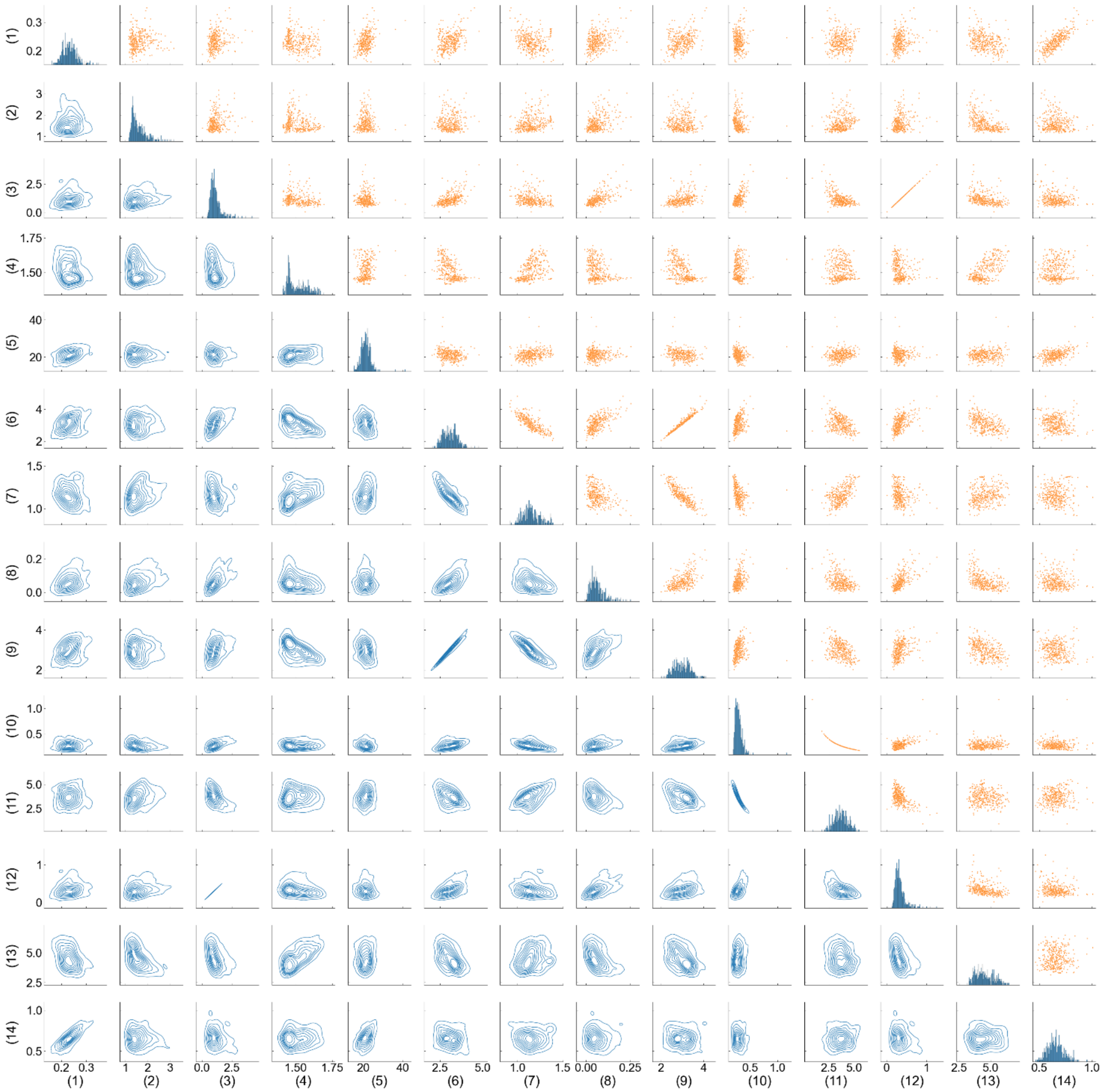

**Fig. S4.** Joint distributions of 14 different properties across 357 polyesters and 26 polycarbonates that are synthesised during biodegradability screening[10]: (1) thermal conductivity [$W \cdot m^{-1} \cdot K^{-1}$], (2) static dielectric constant, (3) linear expansion coefficient [$10^{-4}$ $K^{-1}$], (4) refractive index, (5) radius of gyration [Å], (6) $C_P$ [$10^3$ $J \cdot kg^{-1} \cdot K^{-1}$], (7) density [$g \cdot cm^{-3}$], (8) self-diffusion coefficient [$10^{-11}$ $m^2 \cdot s^{-1}$], (9) $C_V$ [$10^3$ $J \cdot kg^{-1} \cdot K^{-1}$], (10) compressibility [$GPa^{-1}$], (11) bulk modulus [GPa], (12) volume expansion coefficient [$10^{-3}$ $K^{-1}$], (13) glass transition temperature [$10^2$ K], and (14) thermal diffusivity [$10^{-7}$ $m^2 \cdot s^{-1}$].

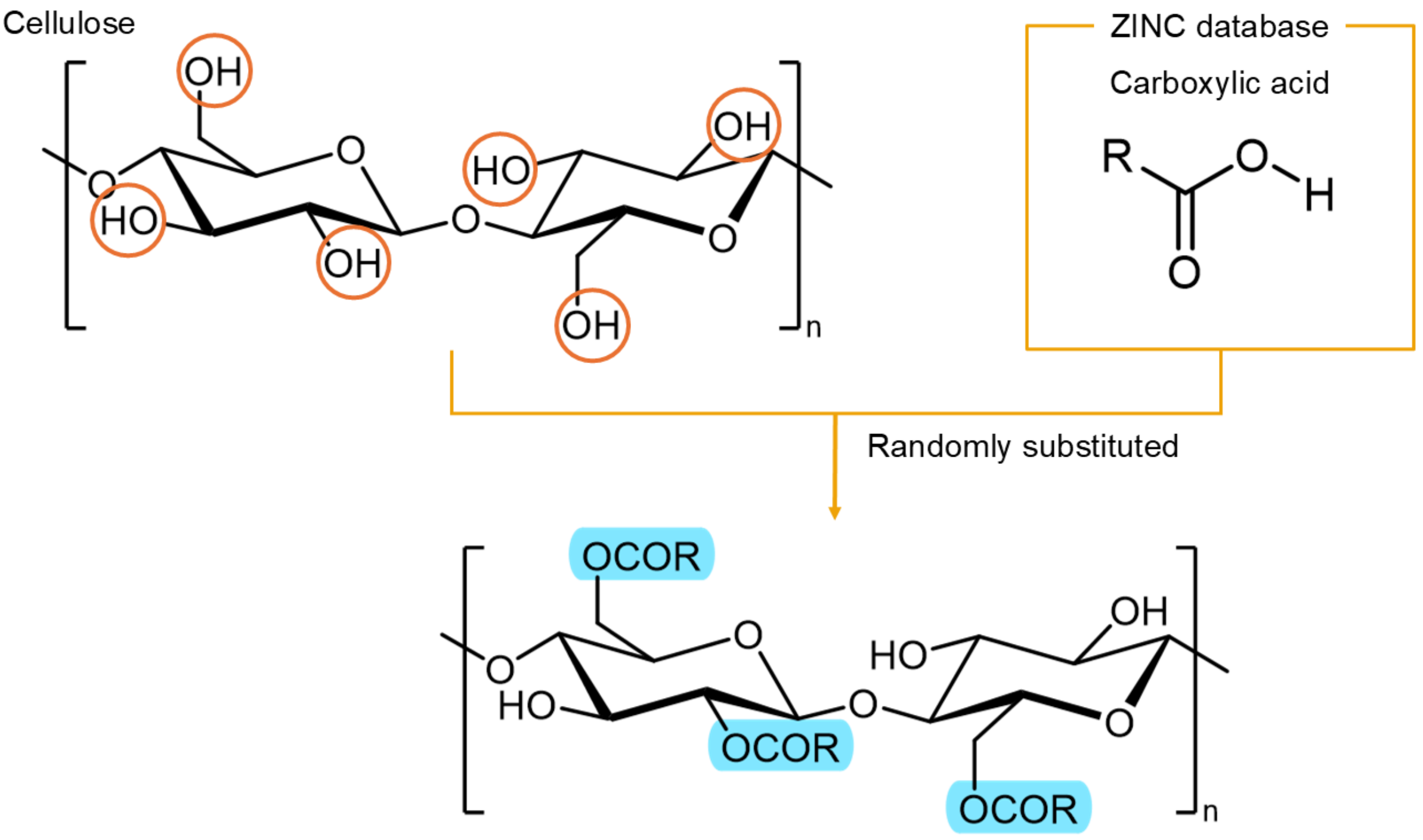

**Fig. S5.** Generation of 50,661 virtual cellulose derivatives. Purchasable fragments from the ZINC database are inserted into one or more of the three substitution sites on the cellulose molecule.

**a**

| PHYC | PSTR | PVNL | PACR | PHAL | PDIE | POXI | PSUL | PEST | PAMD | PURT | PURA | PIMD | PANH | PCBN | PIMN | PPHS | PKTN | PSFO | PPNL |
|---|---|---|---|---|---|---|---|---|---|---|---|---|---|---|---|---|---|---|---|
| 998 | 1295 | 1299 | 1294 | 1145 | 773 | 1300 | 1302 | 1297 | 1233 | 1193 | 1161 | 1160 | 1094 | 1080 | 1132 | 1095 | 990 | 1180 | 1107 |

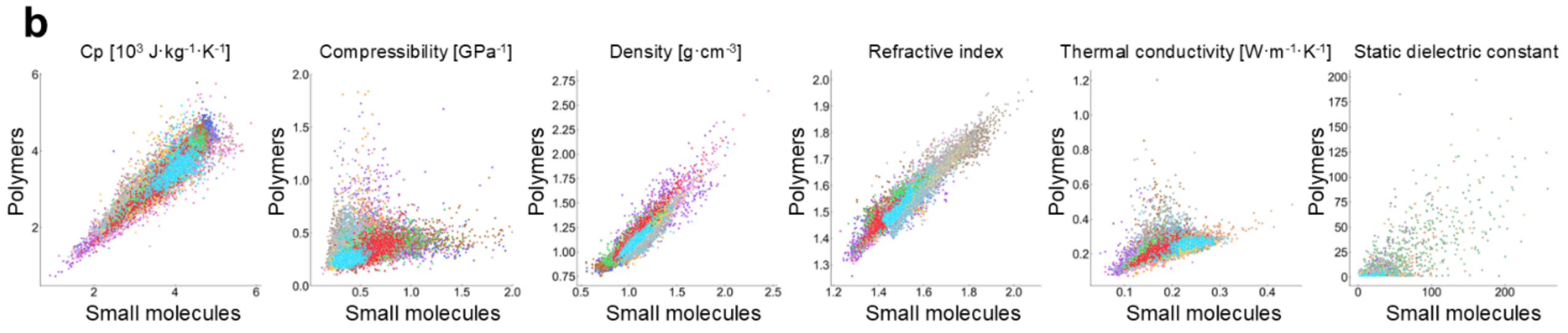

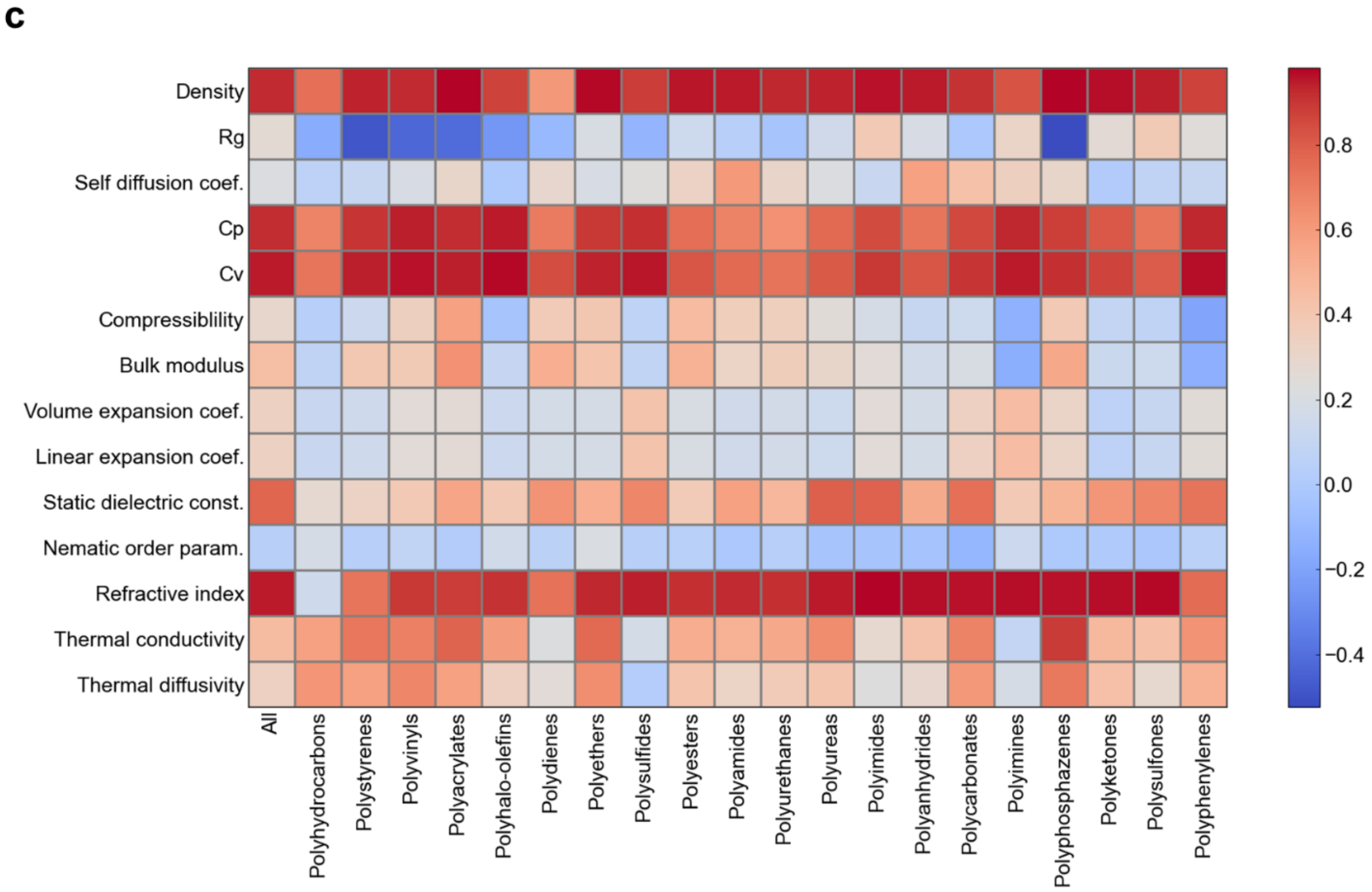

**Fig. S6.** Comparison of properties between polymers and small molecules, which correspond to repeating units. **a**. Number of small molecule data entries for each of the 20 polymer classes. **b**. Comparison of small molecules and polymer properties for 14 properties. **c**. Pearson's correlation coefficients between small molecules and polymer properties, organised by polymer class (rows) and property (columns).

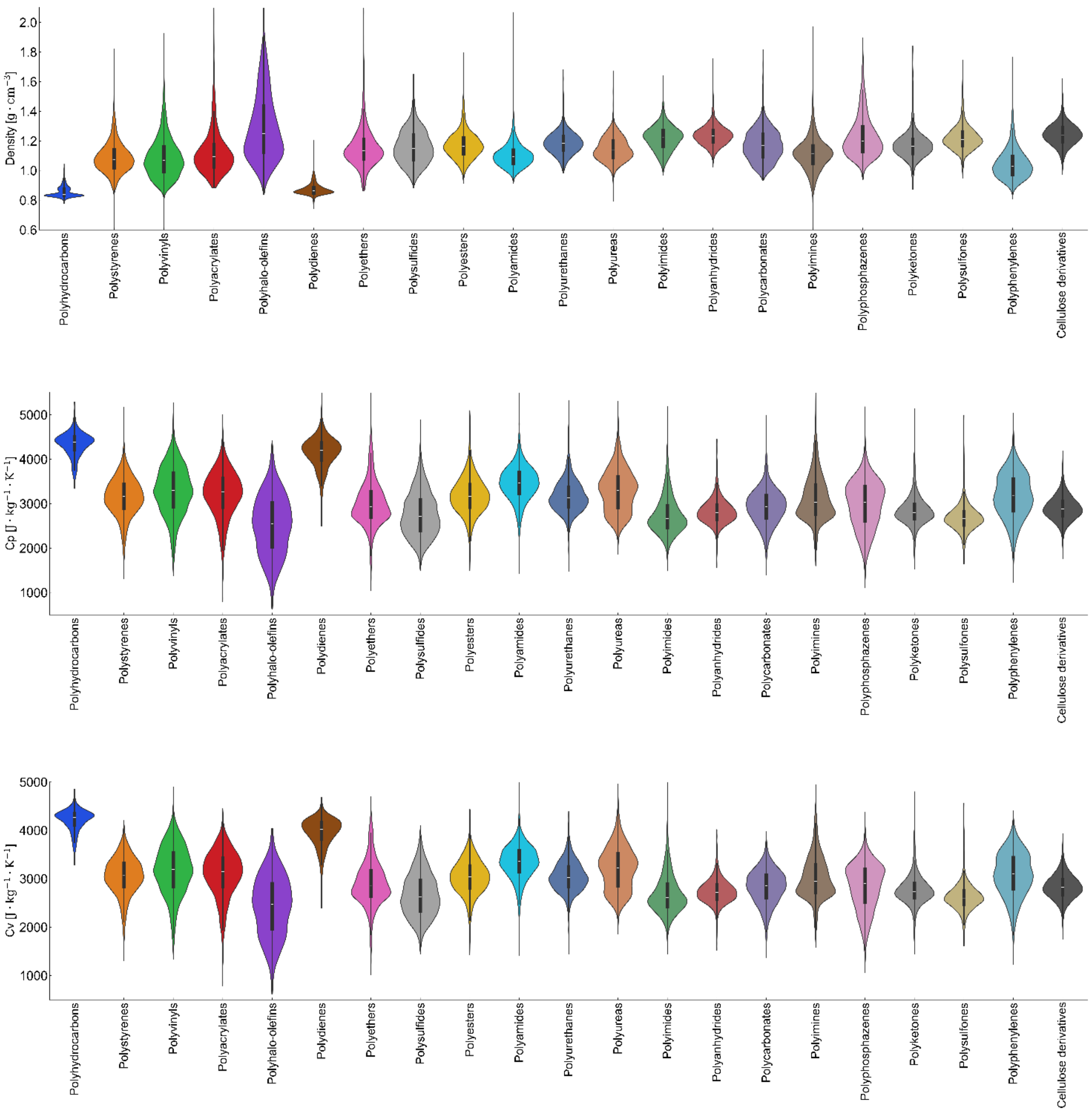

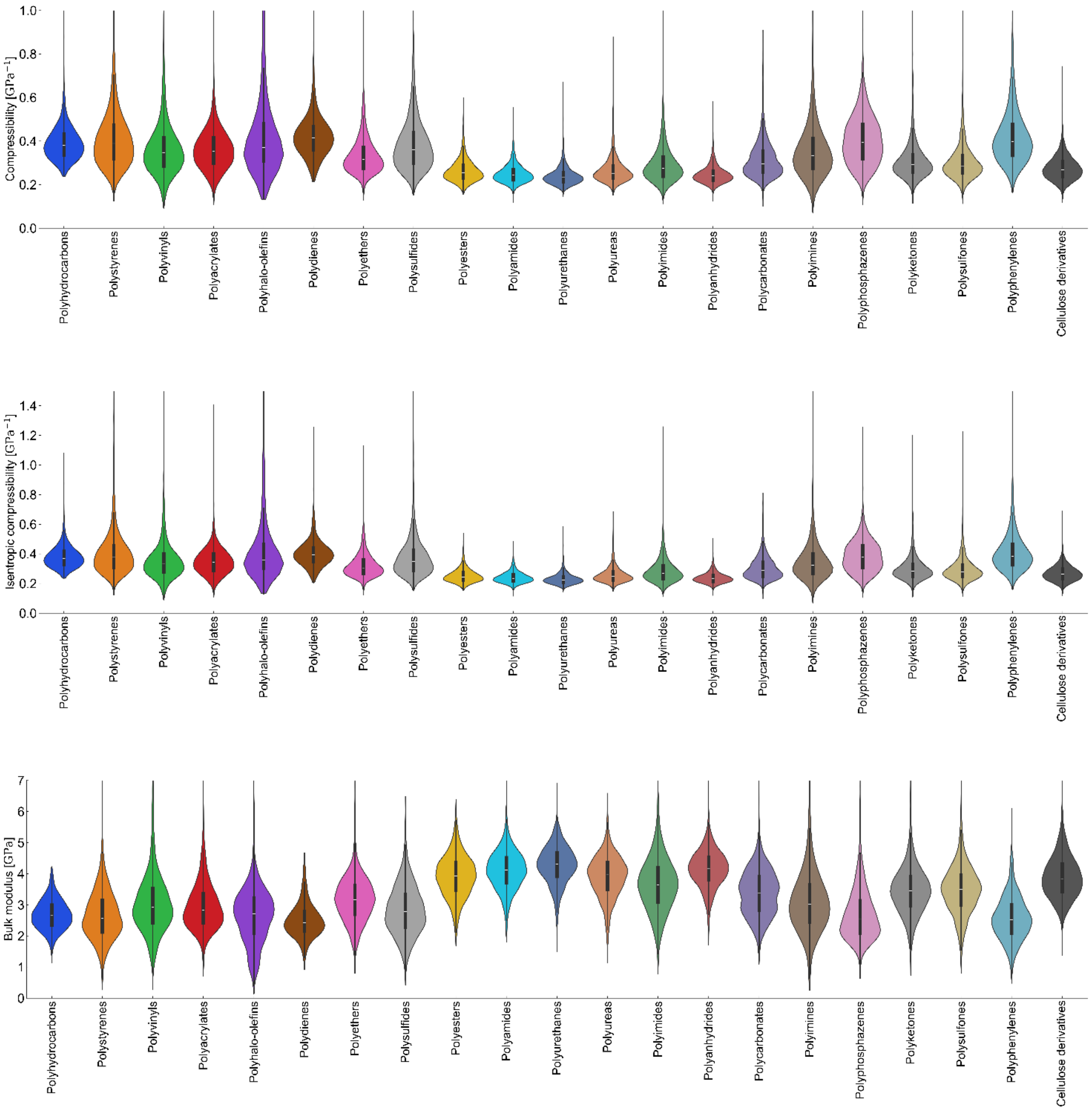

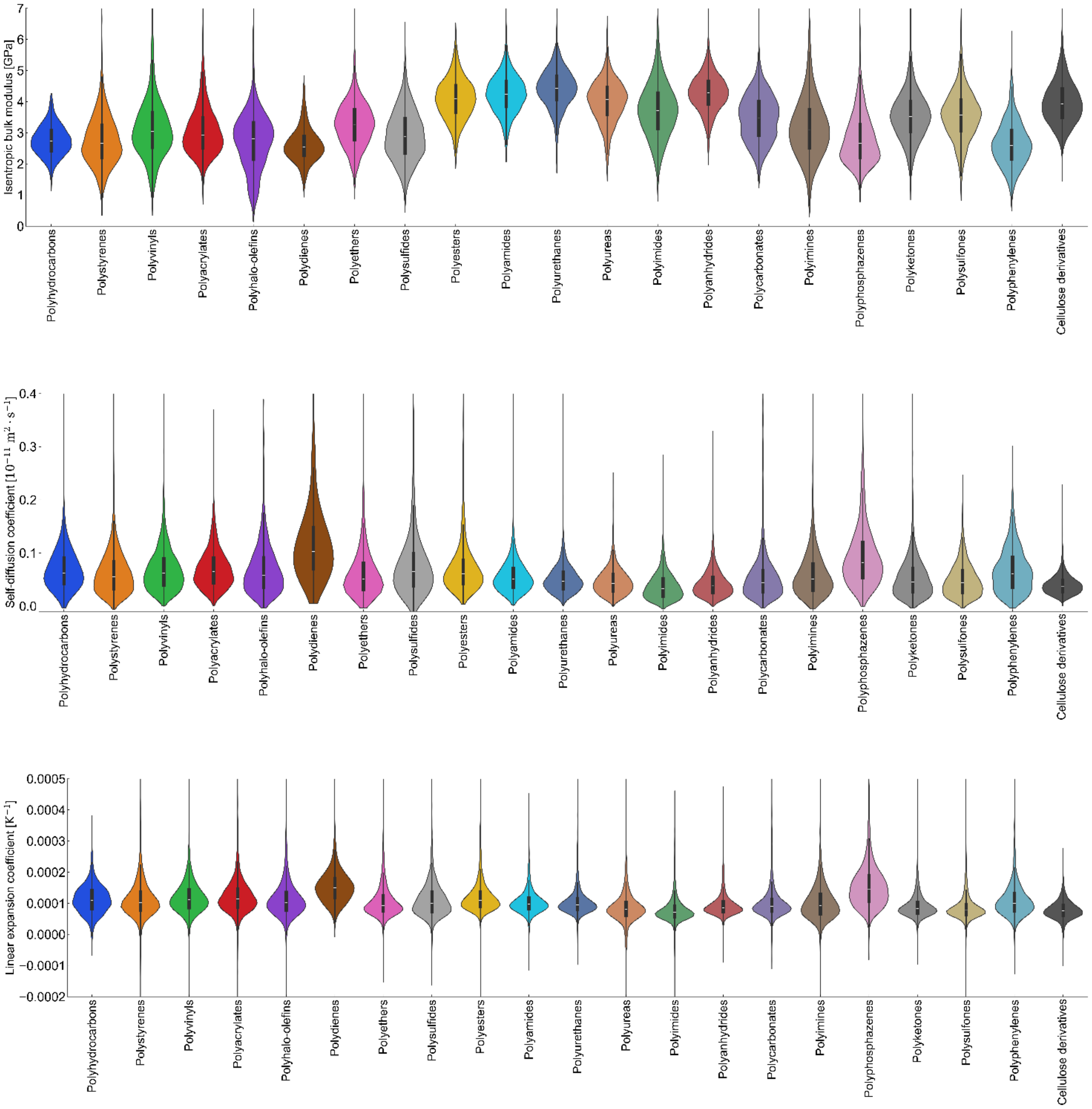

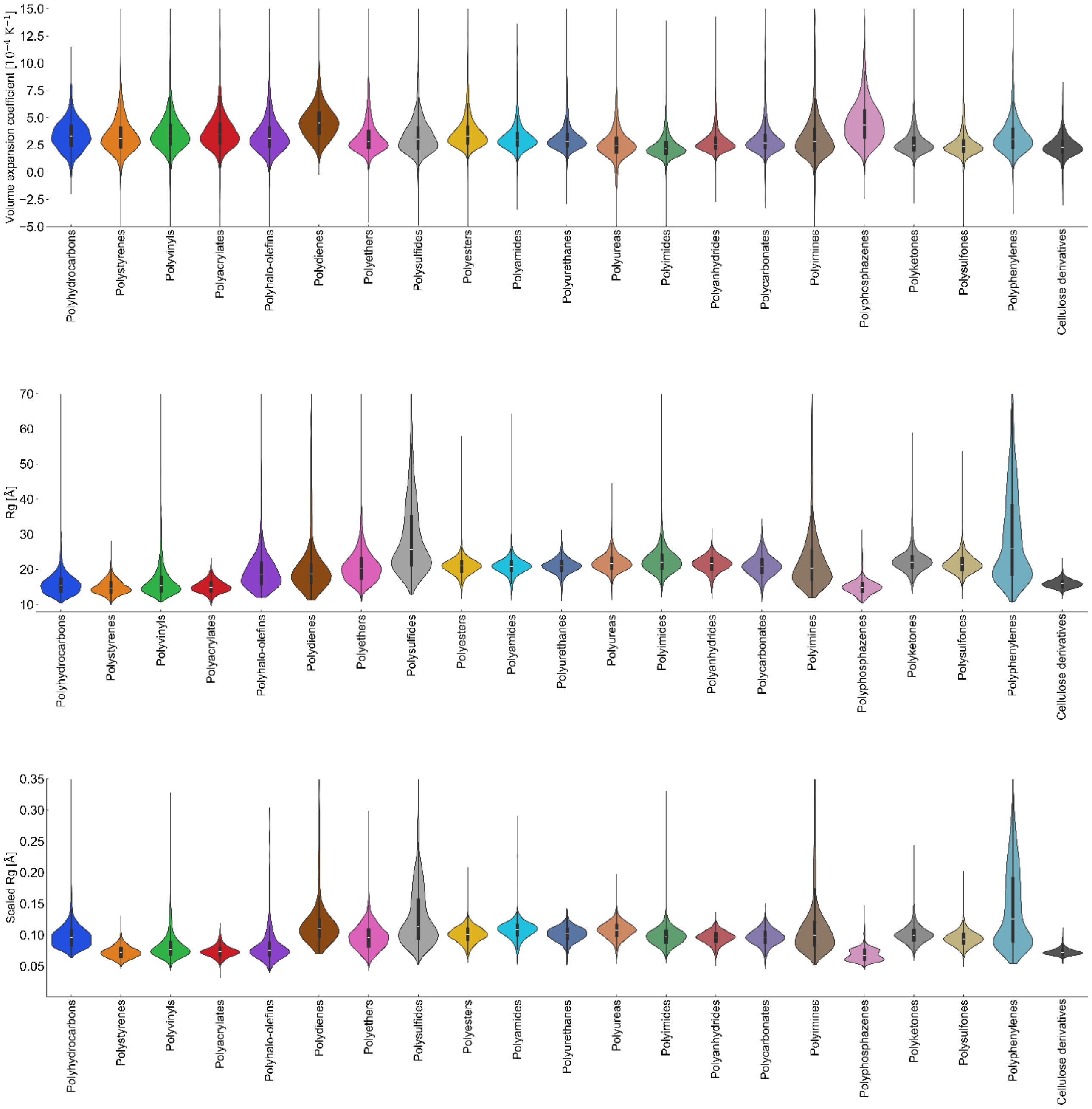

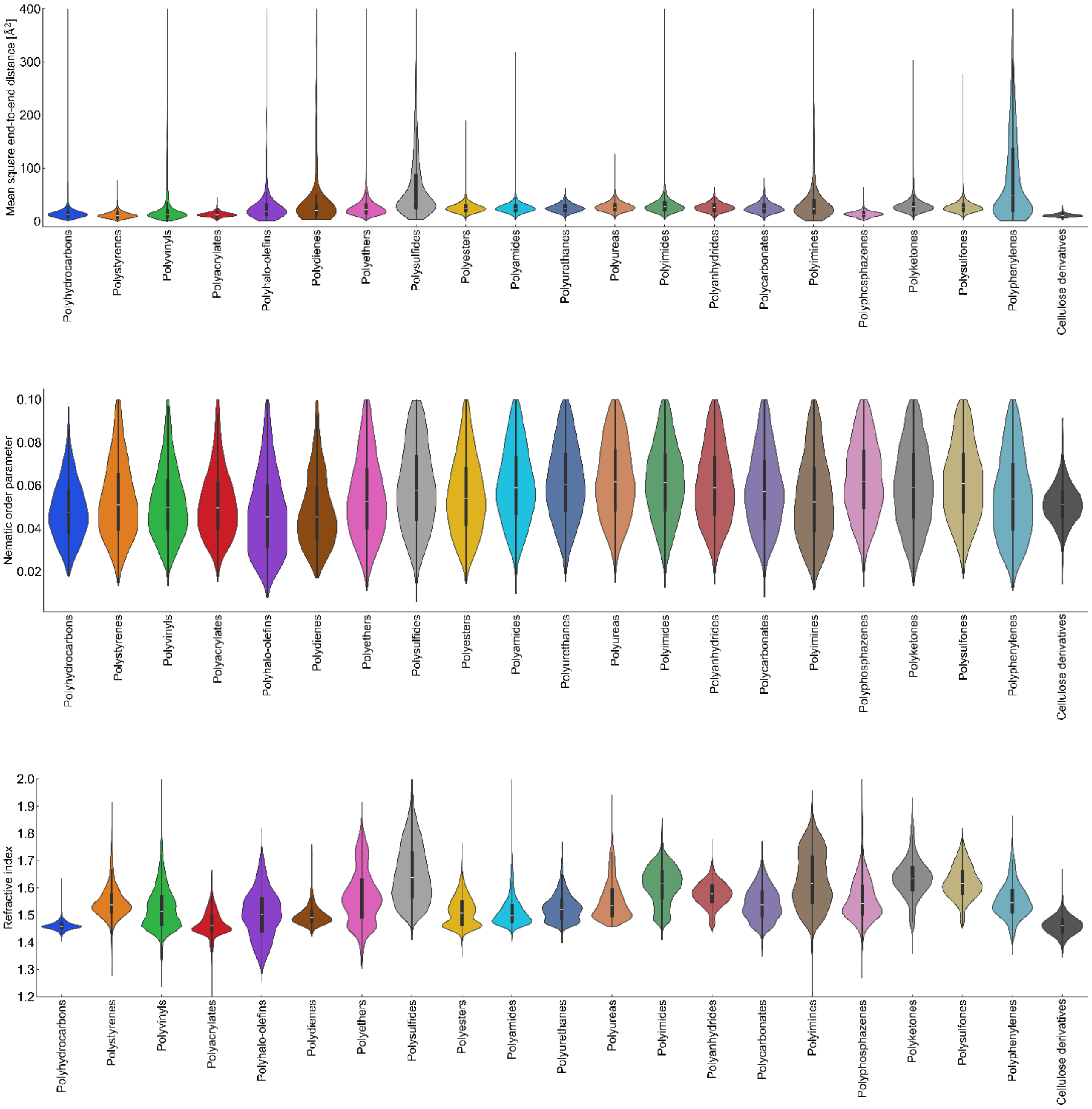

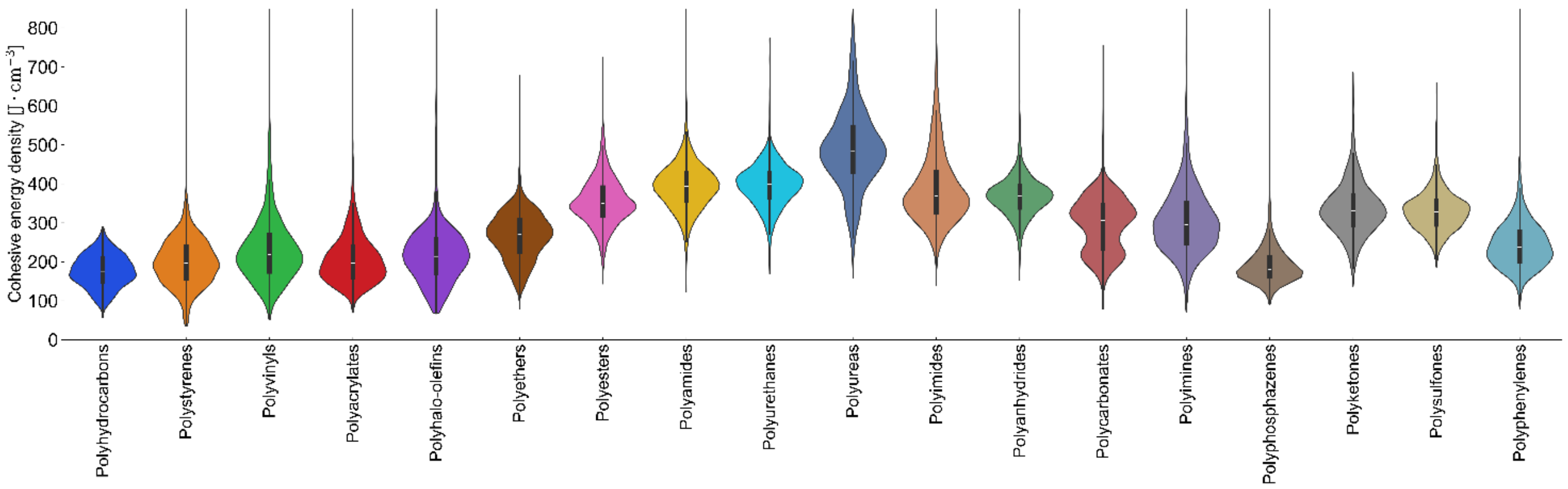

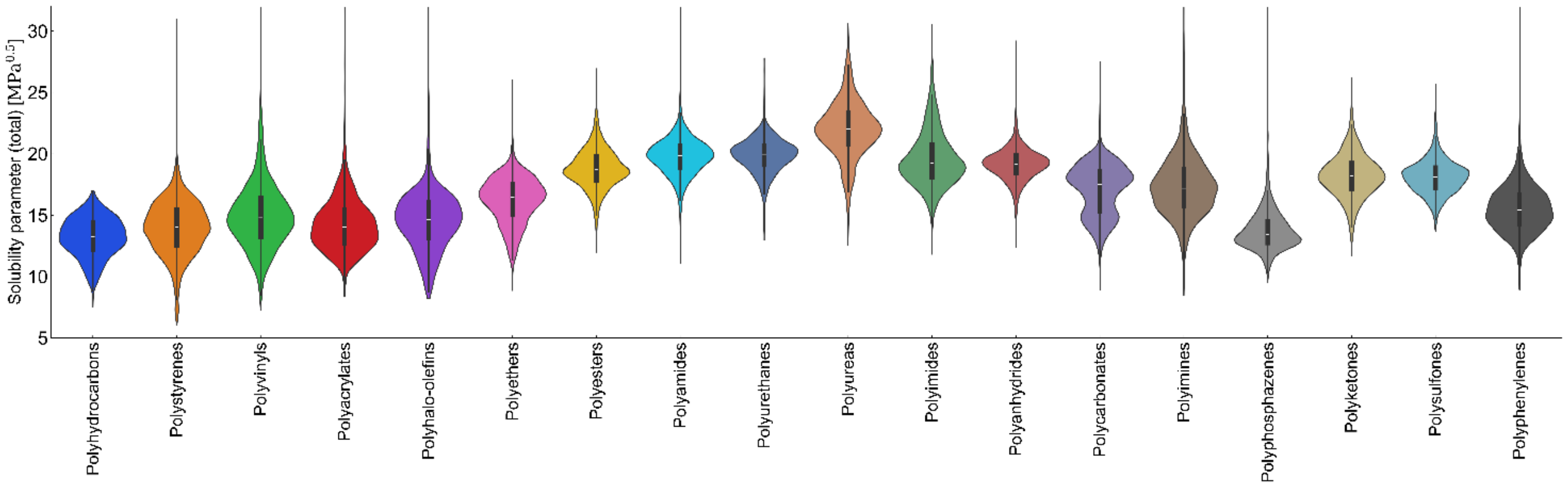

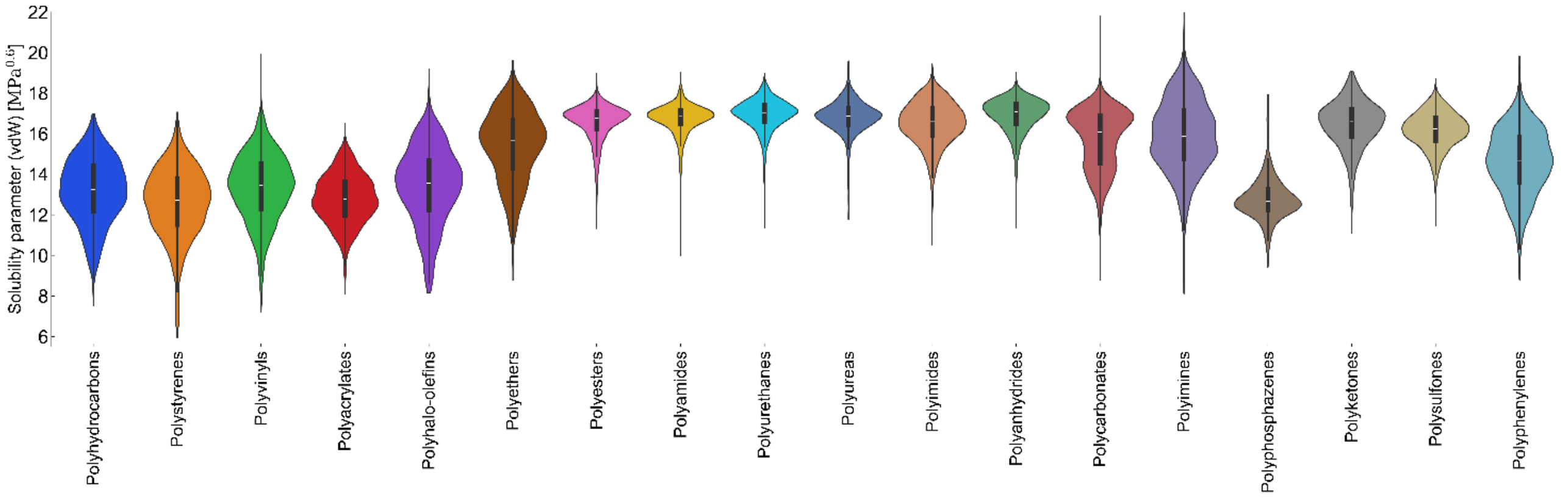

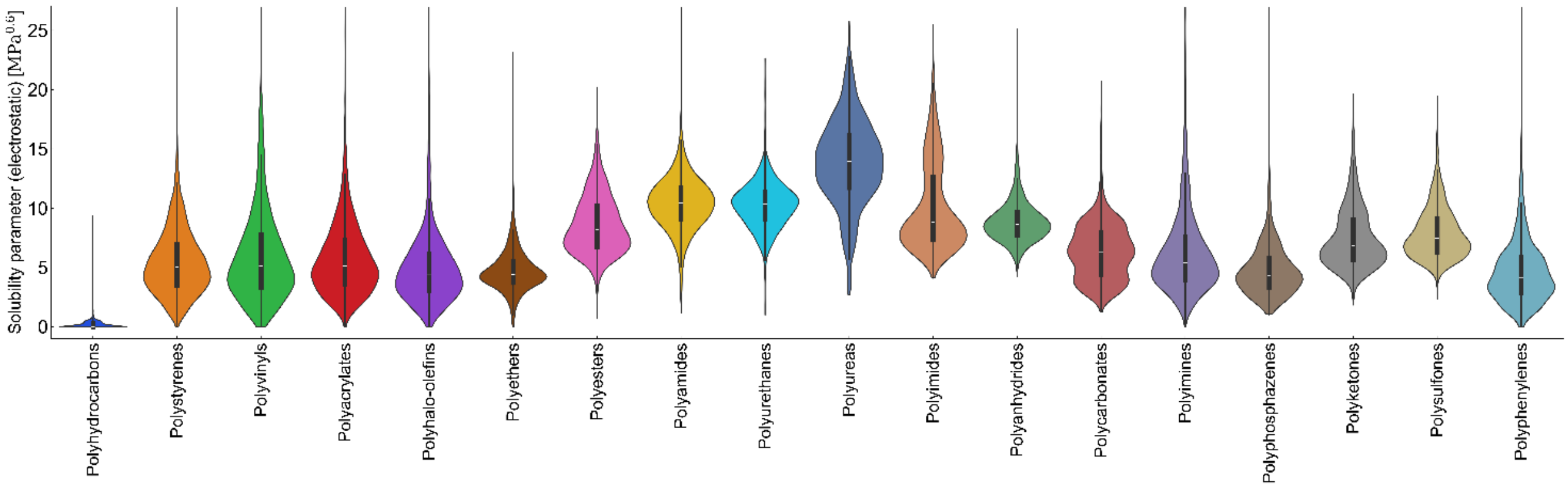

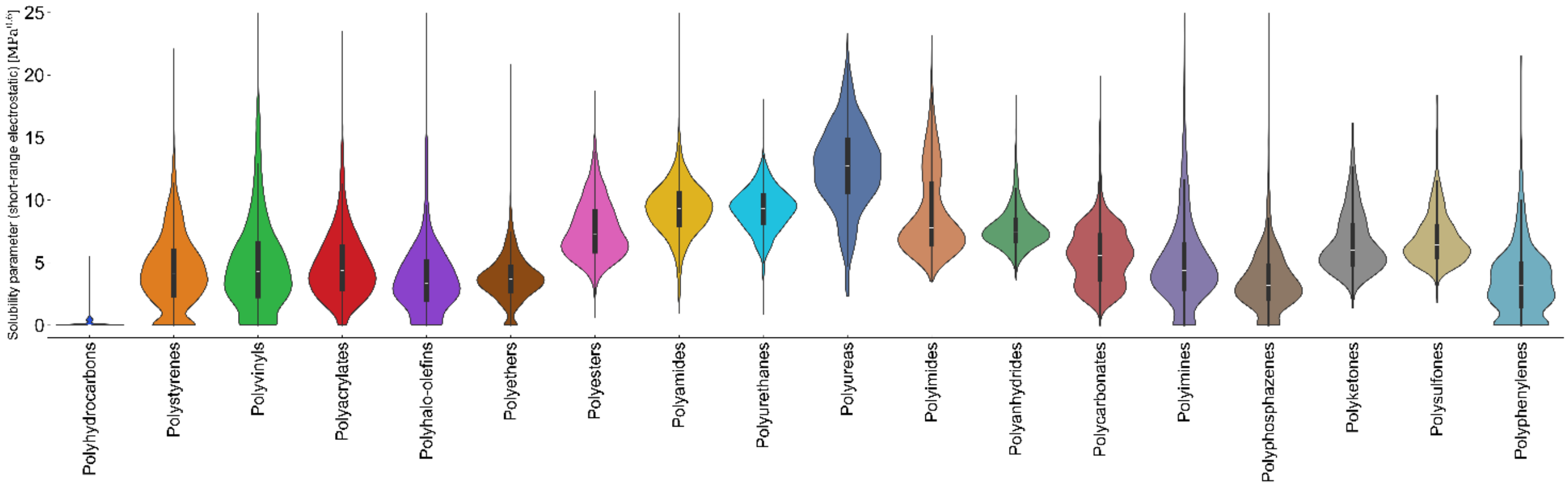

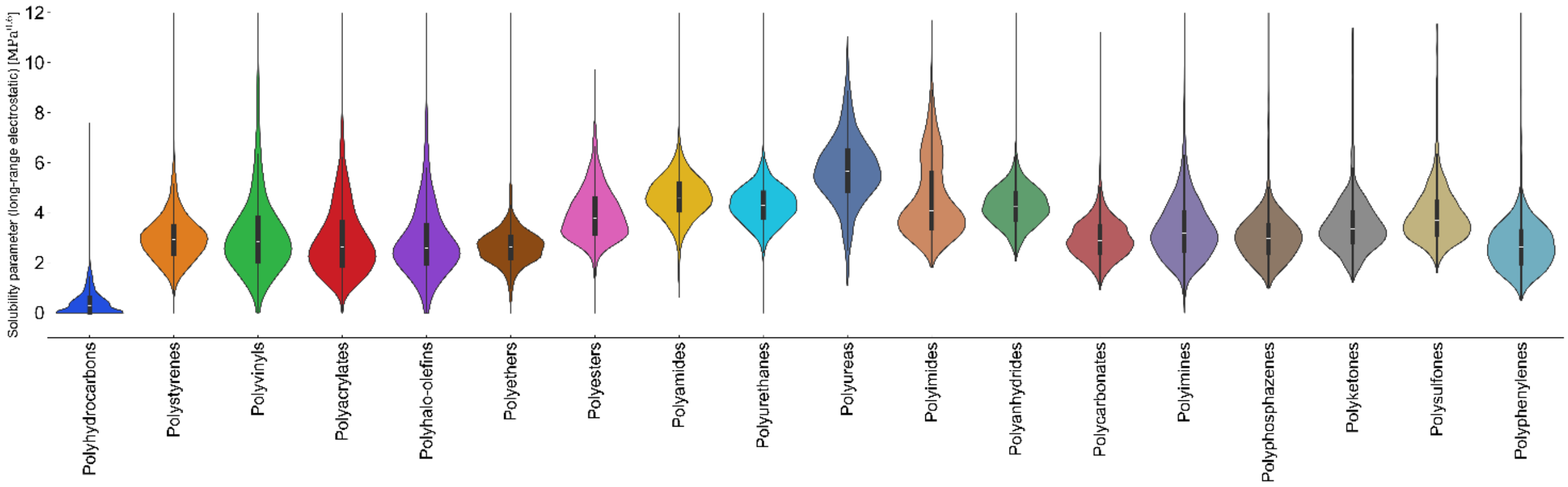

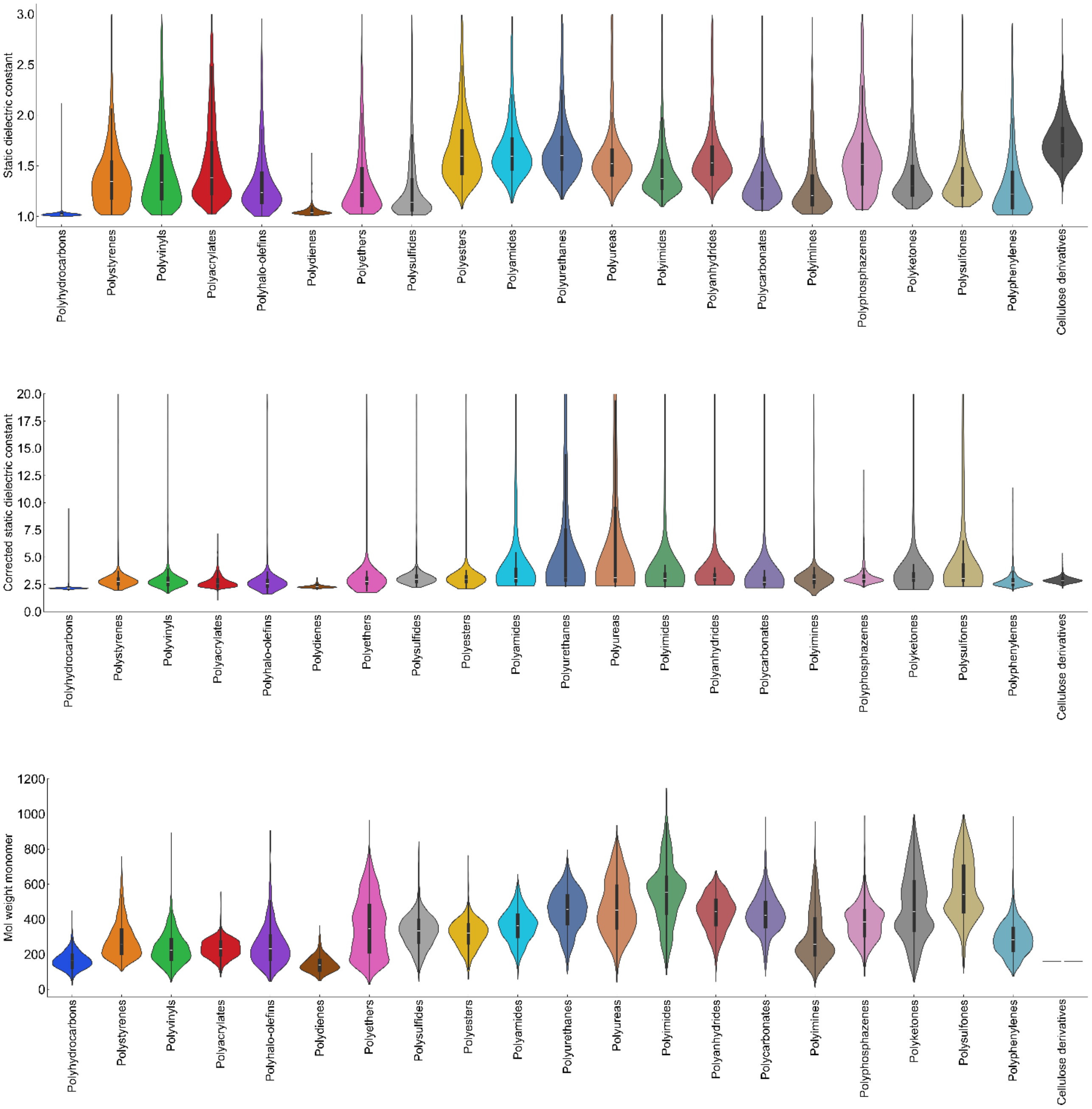

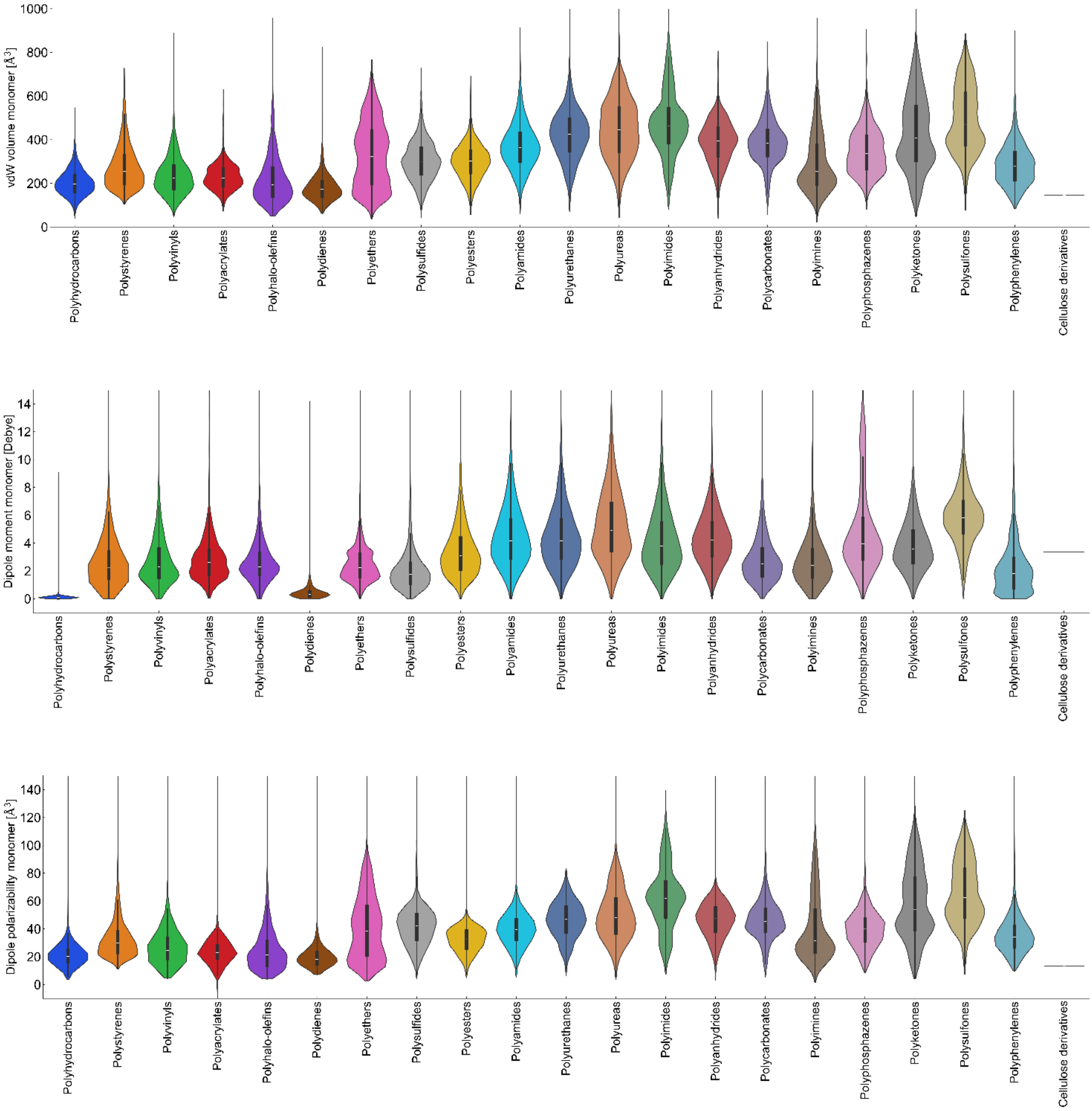

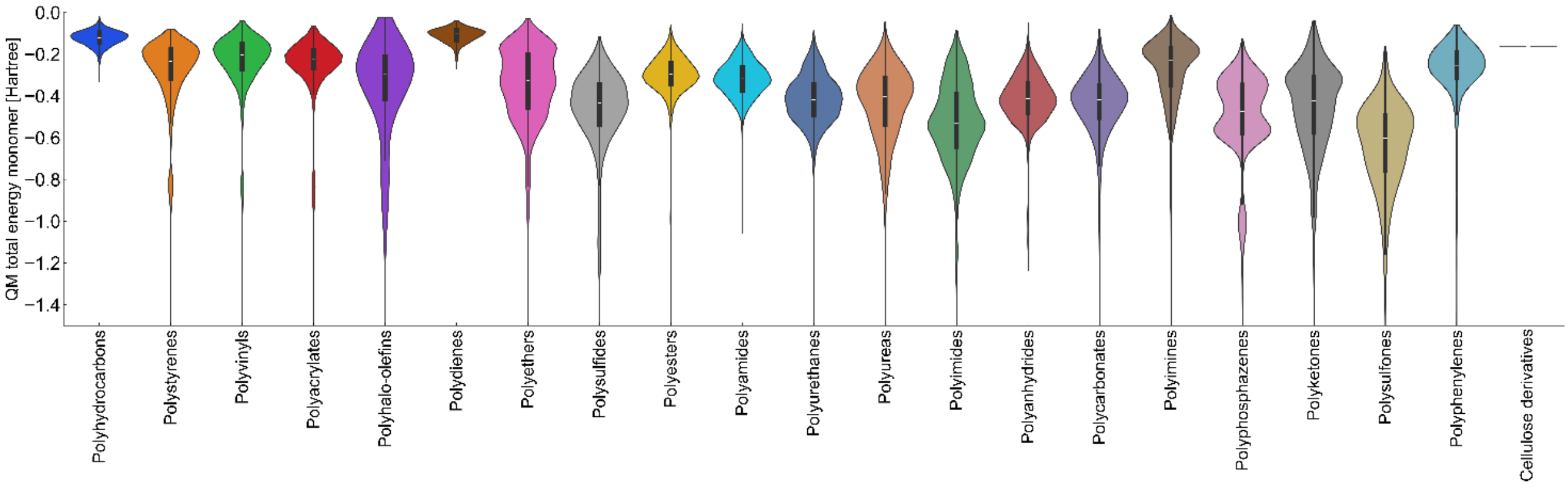

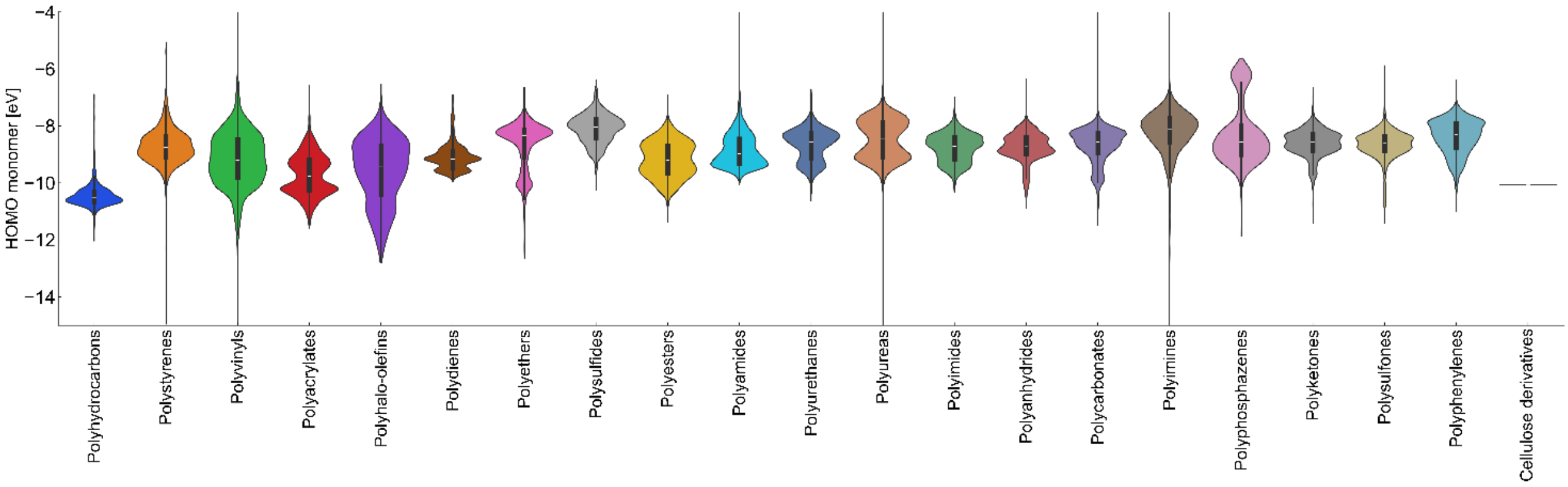

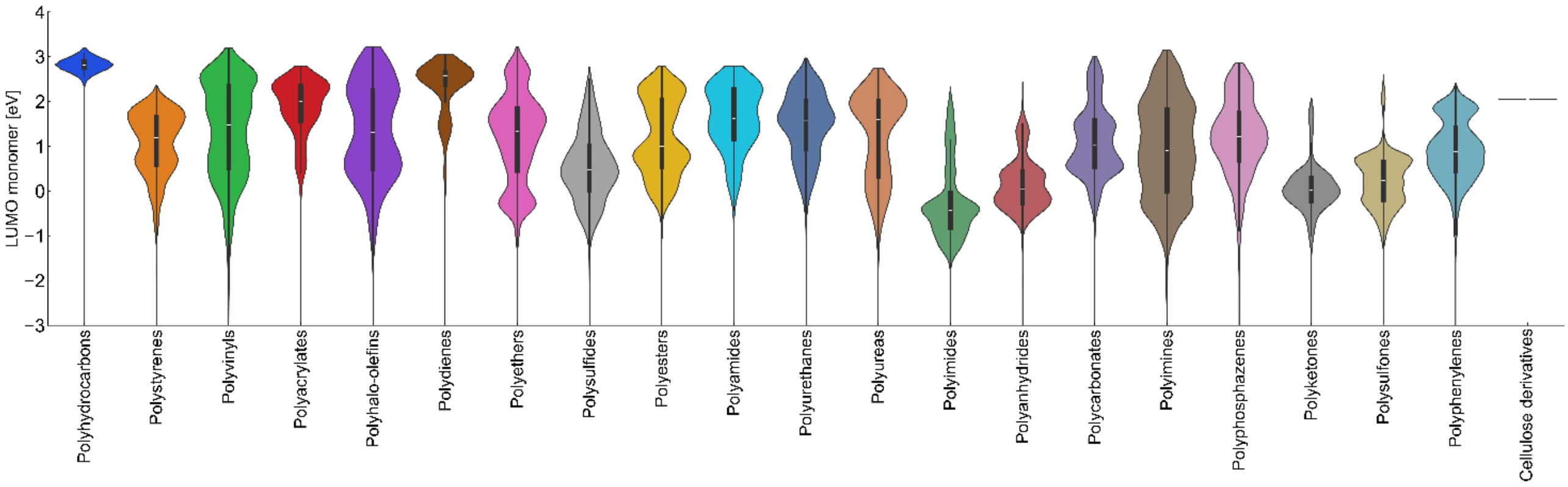

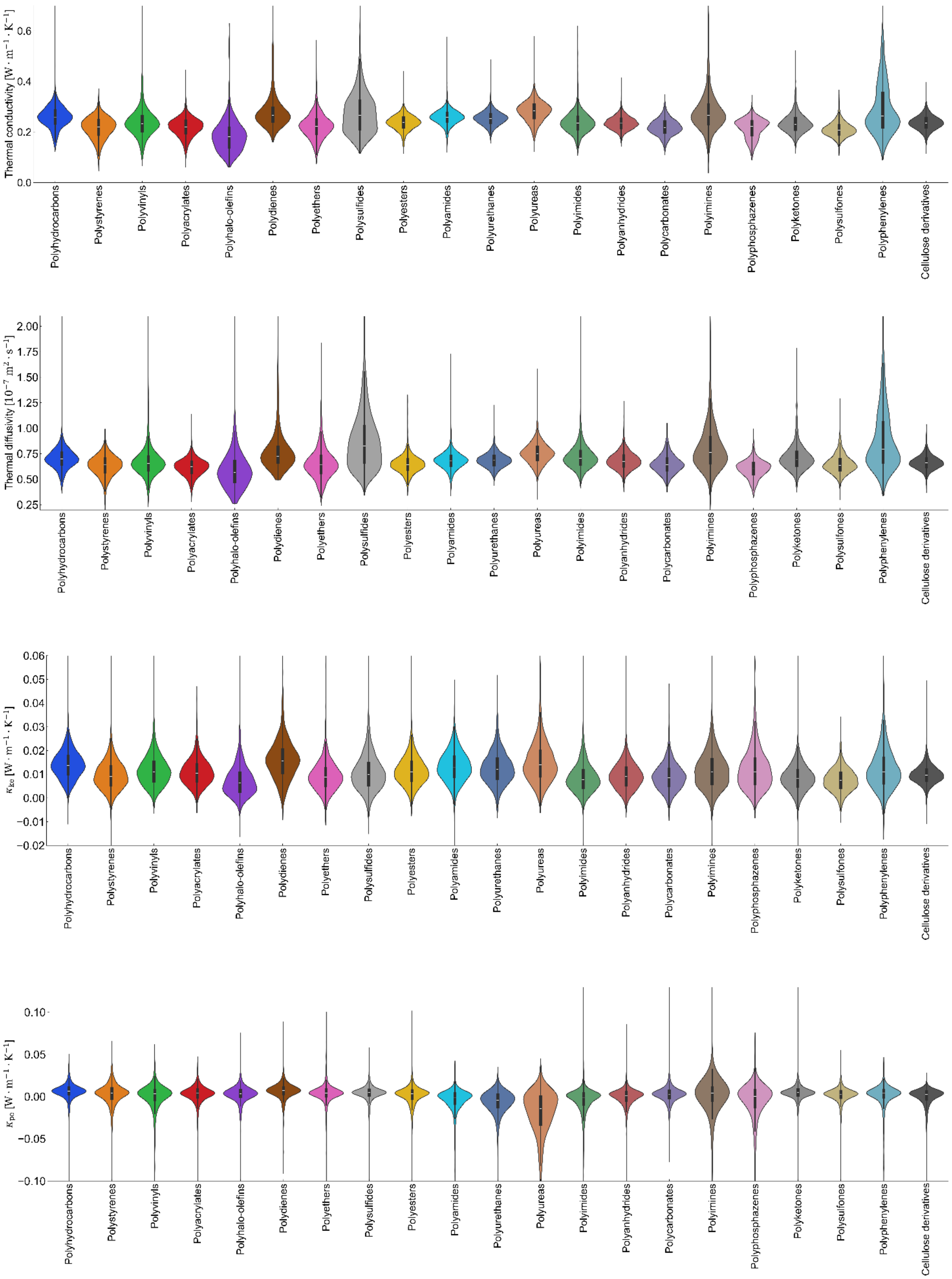

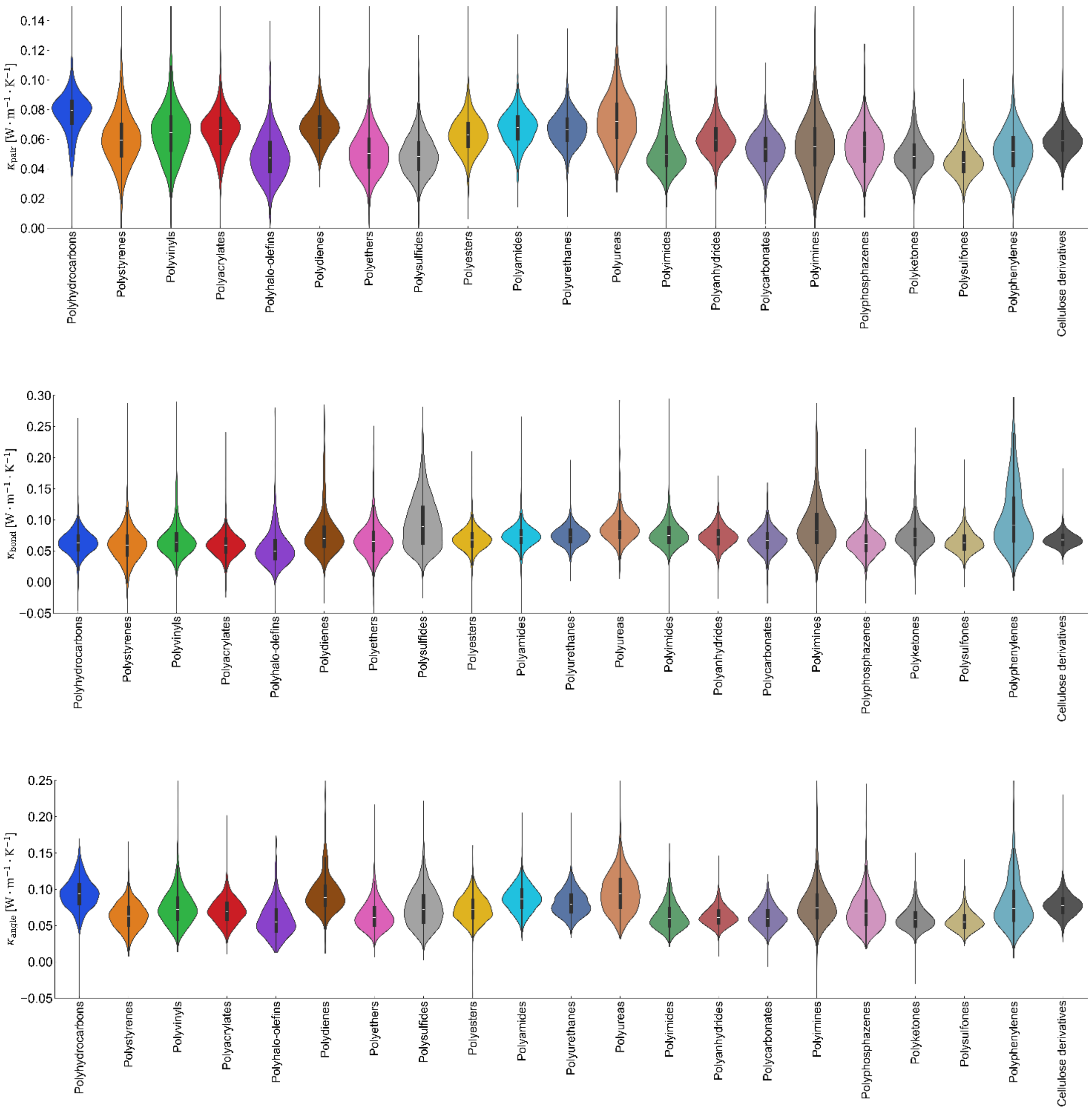

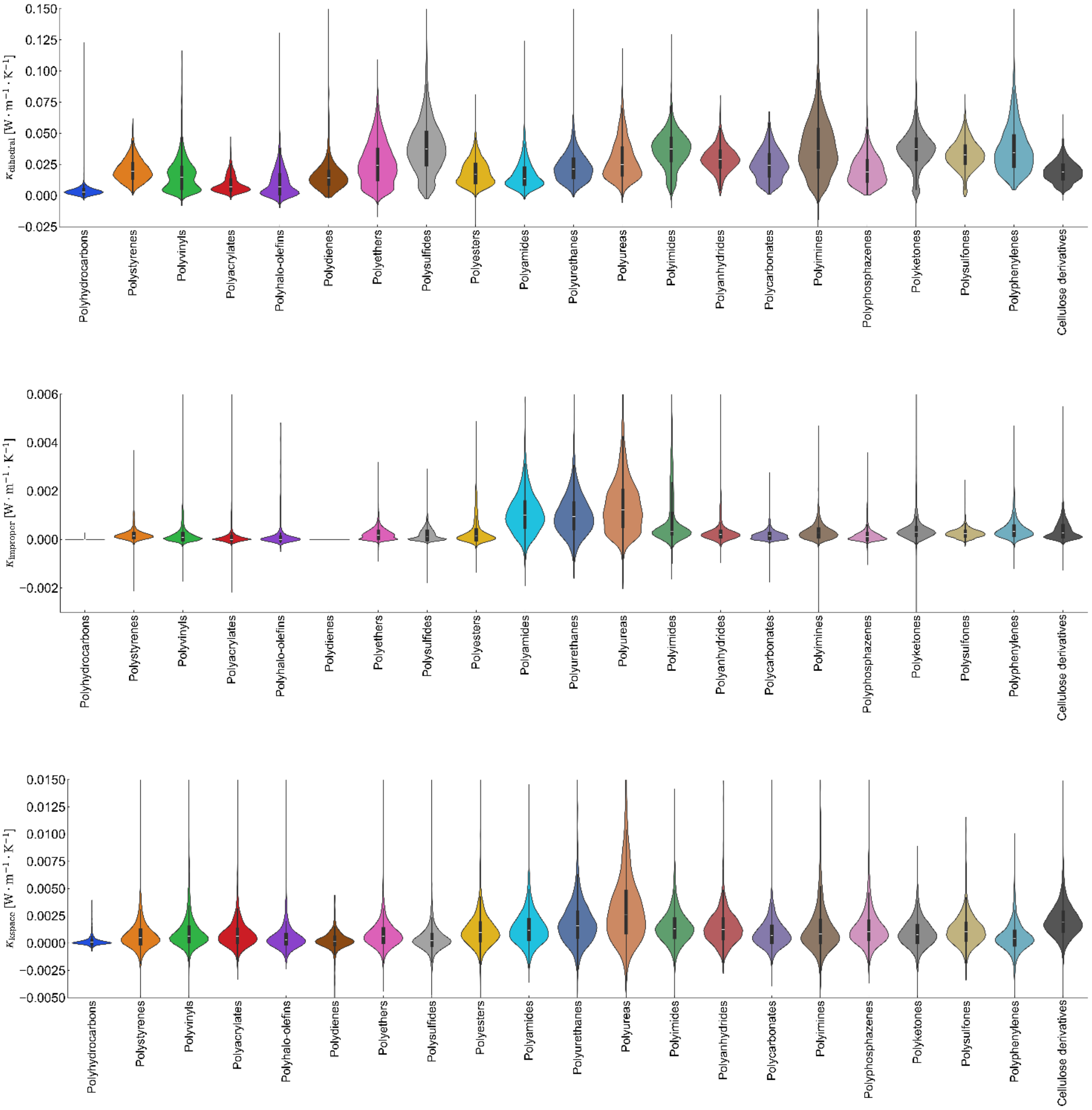

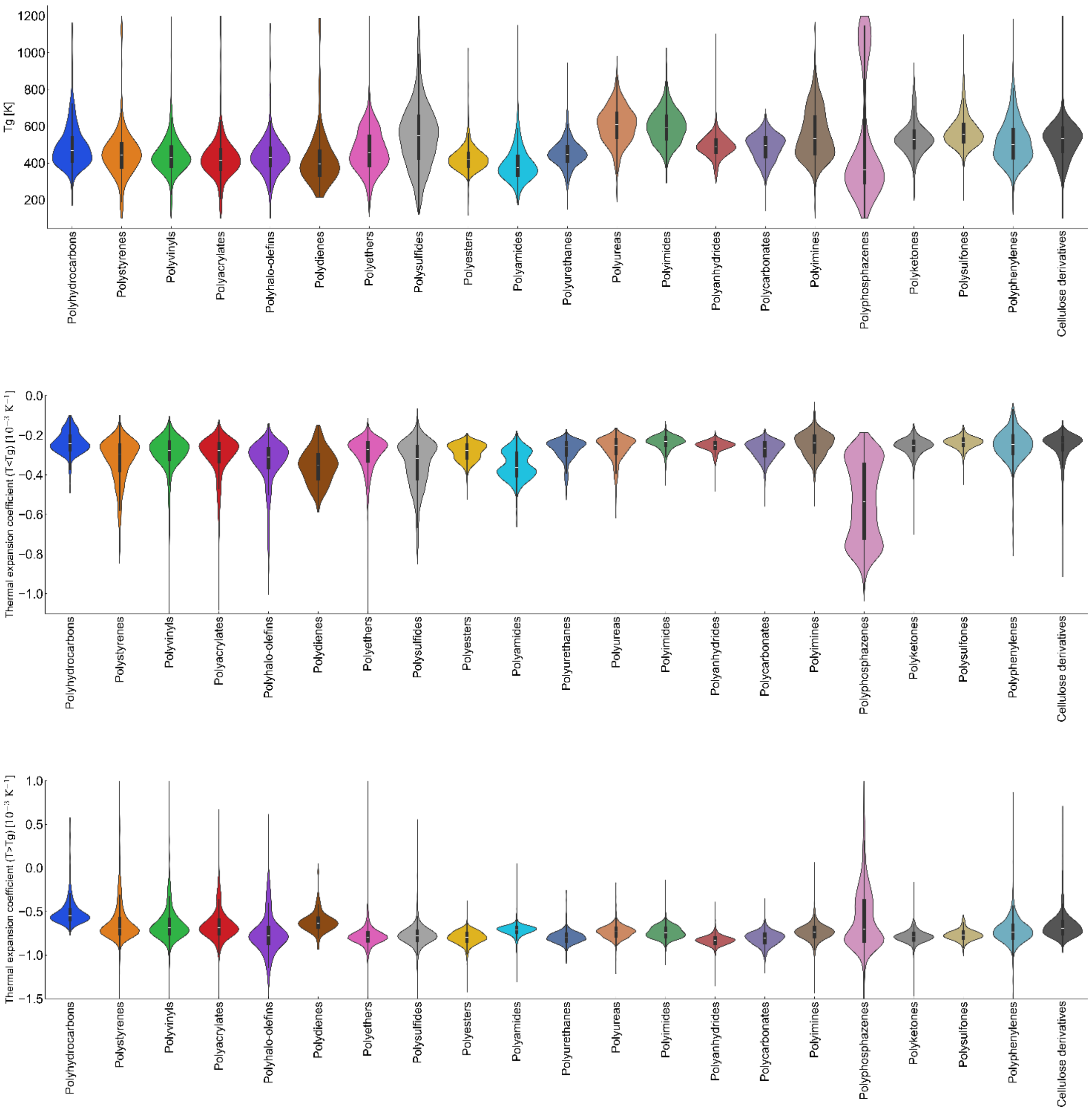

**Fig. S7.** Violin plots showing the distribution of 43 properties across 20 polymer classes of the general polymer dataset and the cellulose derivatives dataset.

# Supplementary Notes

## 1. Automated Molecular Dynamics Simulations using RadonPy

### 1.1 Overview of RadonPy

RadonPy[9], available as an open-source Python library (https://github.com/RadonPy/RadonPy), provides a fully automated pipeline for performing both equilibrium and non-equilibrium molecular dynamics (MD) simulations for a wide range of polymers and their associated physical properties. The simulation engine integrates LAMMPS for classical MD simulations with Psi4 to perform ab initio quantum chemical calculations. The force field parameters for the polymers are assigned using GAFF/GAFF2, and input/output handling is implemented using the RDKit cheminformatics library.

The simulation workflow includes:

- Repeating unit generation
- Automated conformation search
- DFT-based electronic property calculations
- Initial configuration generation
- Force field parameter assignment
- Generation of simulation cells
- Equilibration MD simulations
- Assessment of equilibrium status
- Calculation of a set of properties from the equilibration MD trajectories
- (Optional) Calculation of solubility parameters.
- (Optional) Calculation of glass transition temperature
- (Optional) Calculation of thermal conductivity
- (Optional) Calculation of dielectric properties
- (Optional) Calculation of uniaxially oriented structure properties

This fully automated pipeline enables reproducible high-throughput screening of polymer properties across diverse chemical structures.

### 1.2 Software Dependencies and Installation Guide

Currently, RadonPy operates reliably in environments using Python 3.8–3.12. The system depends not only on the core components, such as LAMMPS, Psi4, and RDKit, but also on several fundamental Python libraries, including SciPy, Pandas, and matplotlib.

Additionally, RadonPy incorporates specialised tools to support key features such as:

- The resp package is a Python implementation of the Restrained Electrostatic Potential (RESP) method for computing atomic partial charges.
- The dftd3 package applies van der Waals dispersion corrections to quantum chemistry calculations.
- The mdtraj library is used for post-processing and analysing molecular dynamics trajectories.

To ensure a stable environmental setup, these dependencies should be installed in advance via Conda before installing RadonPy. Installation guidelines and detailed information on the recommended package versions and configurations are available in the official GitHub repository.

### 1.3 Handling SMILES

The SMILES string, representing the chemical structure of a constitutional repeating unit (CRU), contains two asterisks (*) that designate the polymer connection points. Structural information, including atoms, bonds, and atomic coordinates, is encoded using an RDKit Mol object. In RadonPy, the Mol object is slightly extended to support polymer-specific modelling and MD simulations.

For stereochemical ambiguities, default conventions are applied as:

- If the cis/trans isomerism is unspecified, then the trans isomer is generated by default.
- If R/S chirality is undefined, the S isomer is assumed.

These defaults ensure consistency across automatic structure generation, while allowing users to explicitly specify the required stereochemistry.

## 1.4 Conformational Search for Repeating Units

To explore the initial conformation of a CRU, several thousand conformers (e.g., 1,000) are generated from the input SMILES string using the ETKDG v2 algorithm[11–13] implemented in the RDKit library. The two asterisks (*) in the SMILES notation for each CRU are temporarily capped with hydrogen atoms for conformational generation.

Each generated conformer is optimised geometrically, and its potential energy is calculated by performing molecular mechanics calculations with the GAFF2 force field. The resulting low-energy conformers are clustered using Butina clustering, based on torsion fingerprint deviation[14], to eliminate redundant structures.

From the clustered set, the four most stable conformers are selected for further refinement using density functional theory (DFT). These DFT calculations employ the ωB97M-D3BJ functional[15,16] with the 6-31G(d,p) basis set[17,18]. The final, most stable conformer is determined based on the DFT total energies.

## 1.5 Electronic Property Calculations of Repeating Units

The atomic partial charges of each CRU are computed using the RESP model based on a single-point Hartree–Fock calculation[19] with the 6–31G(d) basis set, applied to the optimised geometry of the most stable conformer.

To evaluate key electronic properties, additional single-point DFT calculations are conducted using the ωB97M-D3BJ functional[15,16] with the 6–311G(d,p) basis set[15,16,20–22] for H, C, N, O, F, P, S, Cl, and Br, and the LanL2DZ basis set[23] for I. These calculations yield the total electronic energy, the highest occupied molecular orbital (HOMO) energy level, the lowest unoccupied molecular orbital (LUMO) energy level, and the dipole moment.

Furthermore, the dipole polarizability tensor is determined using the finite field method under an external electric field of $1.0 \times 10^{-4}$ a.u. Moreover, DFT calculations are conducted with the ωB97M-D3BJ functional in combination with the 6–311+G(2d,p) basis set[15,16,20–22,24,25] for H, C, N, O, F, P, S, and Cl; 6–311G(d,p) for Br; and LanL2DZ for I. The 6–311+G(2d,p) is specifically selected to account for both the diffuse and double polarisation functions, which are essential for accurate polarisability predictions[26].

In this step, various DFT properties of the isolated CRU, such as polarizability and dipole moment, are determined (Section 1.10). Additionally, when evaluating the Abbe number, time-dependent DFT (TD-DFT) is optionally performed. DFT calculations are conducted using the CAM-B3LYP-D3 functional in combination with the 6–311+G(2d,p) basis set.

## 1.6 Polymer Chain Generation

Polymer chains are constructed by connecting repeating units using a self-avoiding random walk (SAW) algorithm. Each repeating unit, specified by a SMILES string with two asterisks representing the attachment points, is capped with hydrogen atoms during chain growth. To prevent undesired chiral inversions and cis/trans isomerisation due to strain accumulation, geometric constraints are enforced. The bond between the head atom of the growing chain and its capping hydrogen, and the bond between the tail atom of the next unit and its capping hydrogen, are aligned coaxially and arranged in antiparallel. Subsequently, the two capping hydrogen atoms are removed, and a new covalent bond between the head and tail atoms is introduced with a fixed bond length of 1.5 Å. The dihedral angle around the new bond is randomly sampled from −180° to +180° during each SAW step. Tacticity is controlled during the chain-building process using RadonPy. For this study, all polymers are generated atactically (random stereochemistry) by default.

To ensure charge neutrality, partial charges of the removed capping hydrogen atoms are redistributed to the bonded atoms. When generating the amorphous polymer dataset, polymer chains containing approximately 1,000 atoms are used, resulting in varying degrees of polymerisation depending on the size of the repeating unit. This approach allows for consistent molecular weights across different polymers, ensuring a fair comparison of the computed physical properties.

## 1.7 Assignment of Force Field Parameters

In RadonPy, GAFF2 parameters can be automatically assigned to each polymer chain, including a comprehensive set of molecular and mechanical parameters.

- Bonded interactions: force constants and equilibrium values for bonds, bond angles, dihedral angles, and improper torsion
- Torsional parameters: multiplicity and phase angles
- Non-bonded interactions: atomic partial charges, vacuum dielectric constant, and Lennard–Jones parameters (defining both the depth of the potential well and the equilibrium interatomic distance)

When GAFF2 lacks predefined parameters, particularly for bond angles involving rare atom groups, RadonPy empirically estimates the values using the same heuristic rules as in the original GAFF2 implementation.

In addition to GAFF2, users can optionally select alternative force fields, such as GAFF or Dreiding, depending on their specific modelling requirements or compatibility constraints.

## 1.8 Construction of Amorphous Polymer Simulation Cells

To simulate amorphous polymers using MD, a simulation cell was constructed by randomly placing and orienting ten polymer chains in a non-overlapping configuration, resulting in a system containing approximately 10,000 atoms. The initial cell density was set to 0.05 $g \cdot cm^{-3}$ and subsequently increased through a multi-step packing simulation.

Owing to its low initial density, a packing protocol was employed to achieve a physically realistic configuration. First, the 1 ns *NVT* simulation was performed using a Nosé–Hoover thermostat while gradually heating the system from 300 to 700 K. Another 1 ns *NVT* simulation was performed at 700 K, during which the simulation box was isotropically compressed to a target density of 0.8 $g \cdot cm^{-3}$.

To prevent the polymer chains from collapsing into globular conformations due to strong intramolecular interactions, Coulombic forces were disabled during this process, and the Lennard–Jones potential was truncated at 3.0 Å. These conditions maintained the polymers in random-coil states while ensuring entanglement through excluded volume effects as chain crossings were prevented.

A time step of 1 fs was considered, and periodic boundary conditions (PBC) were applied in all directions. All the bond lengths and angles, including those involving hydrogen atoms, were constrained using the SHAKE algorithm[27].

## 1.9 Equilibration Simulation

Following the packing process, the amorphous polymer systems were equilibrated using Larsen's 21-step compression–decompression protocol, which alternated the temperature and pressure cycles to remove non-physical configurations and accelerate equilibration. Specifically, the system was repeatedly heated from 300 to 600 K and then cooled back to 300 K while simultaneously undergoing compression up to 50,000 atm, followed by decompression to 1 atm, to construct the amorphous dataset. These cycles were performed over approximately 1.5 ns using a combination of *NVT* and *NPT* ensembles, with the temperature and pressure controlled by the Nosé–Hoover thermostat and barostat.

After the 21-step protocol, the system was subjected to *NPT* simulations at 300 K and 1 atm for over 5 ns. The equilibrium state was evaluated every 5 ns. The equilibration was considered successful when the following relative standard deviation (RSD) thresholds were satisfied:

- Total and kinetic energies: < 0.05%
- Bond and bond angle energies: < 0.1%
- Dihedral energy: < 0.2%
- van der Waals and long-range Coulomb energies: < 0.1%
- Density: < 0.1%
- Radius of gyration: < 1%

Simulations that failed to meet these criteria after 50 ns were deemed unsuccessful and excluded from further analysis.

Technical parameters for the equilibration simulation included:

- Time step: 1 fs
- Periodic boundary conditions (PBC): applied in all directions.
- Bond constraints: enforced using the SHAKE algorithm for all bonds and angles, including those involving hydrogen atoms.
- Non-bonded interactions: handled using the twin-range cutoff method with short- and long-range cutoffs of 8° and 12°, respectively.
- Electrostatics: treated using the particle–particle particle–mesh (PPPM) method.

Finally, the nematic order parameter was evaluated to confirm the amorphous nature of the system (Section 1.10). Structures with values ≥ 0.1 were classified as exhibiting orientational ordering and were excluded from the dataset.

## 1.10 Property Calculations

Various physical properties and structural descriptors were computed from the equilibrium MD trajectories. Following polymer equilibration, NEMD simulations were performed to evaluate thermal conductivity via energy exchange under a temperature gradient (Section 1.12) and to compute the frequency-dependent dielectric properties under an applied alternating electric field. Additionally, the automated simulation module provided functionality for evaluating the material properties under uniaxial stretching, calculating the solubility parameters, and estimating the glass transition temperature.

The **density** $\rho$ (g/cm$^3$) during the *NPT* ensemble simulation was calculated using the total mass ($m$) of all atoms and the volume ($V$) of the simulation box as:

$$\rho = \frac{m}{\langle V \rangle} \tag{1}$$

where the angular brackets $\langle \cdot \rangle$ represent time averaging after the system reaches equilibrium.

The **radius of gyration** $R_g$ (Å) was calculated as:

$$R_g = \sqrt{\frac{1}{N}\sum_{k=1}^{N} \|\boldsymbol{r}_k - \boldsymbol{r}_{\mathrm{mean}}\|^2} \tag{2}$$

where $\boldsymbol{r}_k$ is the centre of a repeating unit $k$, $N$ is the polymerisation degree, $\boldsymbol{r}_{\mathrm{mean}}$ denotes the average positions, and $\|\cdot\|$ represents the Euclidean norm. The centre was calculated as the weighted average of the atomic position vectors, using the atomic masses as weights.

**Scaled $R_g$** (dimensionless) was calculated as:

$$R_{\mathrm{g}}^{\mathrm{scaled}} = \frac{R_g}{M_w^{0.6}} \tag{3}$$

where $M_w$ is the molecular weight of the polymer chains, enabling the comparison between polymers of different lengths.

The **mean square end-to-end distance** $R_e$ (Å) of a polymer chain was calculated as:

$$R_{\mathrm{e}} = \frac{1}{M}\sum_{l=1}^{M} \|\boldsymbol{r}_e(l) - \boldsymbol{r}_0(l)\|^2 \tag{4}$$

where $\boldsymbol{r}_e(l)$ and $\boldsymbol{r}_0(l)$ are the position vectors of the two terminal groups in each polymer chain ($l$). This value was averaged over the simulation trajectory, where $M$ denotes the number of polymer chains in the simulation cell. The end-to-end distance provided a direct measure of the overall chain extension or compactness of a polymer.

**Specific heat capacity at constant pressure** $C_p$ (J·mol$^{-1}$·K$^{-1}$) was calculated from the fluctuations in enthalpy $H$ as:

$$C_{\mathrm{P}} = \frac{\langle \delta H^2 \rangle}{k_{\mathrm{B}} \cdot T^2 \cdot m} \tag{5}$$

where $k_B$ is the Boltzmann constant, $T$ is the temperature, $m$ is the total mass of the system, and $\langle \delta H^2 \rangle$ denotes the variance in enthalpy fluctuations. During *NPT* simulations, the instantaneous pressure exhibited significant fluctuations, which introduced errors in the direct calculation of $C_P$. To mitigate this issue, the enthalpy was calculated at a constant pressure of 1 atm, ensuring a more accurate estimation of the specific heat capacity.

**Isothermal compressibility** $\beta_T$ (Pa$^{-1}$) and **isothermal bulk modulus** $K_T$ (Pa) were calculated from the fluctuations in the volume ($V$) as:

$$\beta_T = \frac{\langle \delta V^2 \rangle}{k_{\mathrm{B}} \cdot T \cdot \langle V \rangle} \tag{6}$$

$$K_T = \frac{1}{\beta_T} \tag{7}$$

where $\langle V \rangle$ is the mean system volume and $\langle \delta V^2 \rangle$ is the variance of the volume fluctuations. The isothermal compressibility quantifies the relative change in volume in response to an infinitesimal change in pressure at a constant temperature, whereas the isothermal bulk modulus represents the resistance of the system to uniform compression.

The **volumetric thermal expansion coefficient** CVTE (K$^{-1}$) was calculated from the covariance $\langle \delta V \delta H \rangle$ between volume ($V$) and enthalpy ($H$) as:

$$\text{CVTE} = \frac{\langle \delta V \cdot \delta H \rangle}{k_{\text{B}} \cdot T^2 \cdot \langle V \rangle} \tag{8}$$

Enthalpy was calculated at a constant pressure of 1 atm to avoid inaccuracies arising from significant pressure fluctuations in the *NPT* simulations.

The **linear thermal expansion coefficient** CLTE ($K^{-1}$) in the isotropic systems was calculated as:

$$\text{CLTE} = \frac{\text{CVTE}}{3} \tag{9}$$

This relationship assumes isotropy, wherein the expansion is considered uniform along all three spatial directions.

The **specific heat capacity at constant volume** $C_V$ ($J \cdot mol^{-1} \cdot K^{-1}$) was calculated by relating $C_V$ to $C_P$, $\alpha_P$, and $\beta_T$ as:

$$C_V = C_P - \frac{\text{CVTE}^2 \cdot T\langle V \rangle}{\beta_T \cdot m} \tag{10}$$

The expression arose from standard thermodynamic relations and accounted for the work performed during thermal expansion at constant pressure. Therefore, $C_V$ was evaluated using the quantities obtained from *NPT* simulations, without requiring direct constant-volume simulations.

The **isentropic compressibility** $\beta_S$ ($Pa^{-1}$) and **isentropic bulk modulus** $K_S$ (Pa) were calculated as:

$$\beta_S = \beta_T \frac{C_V}{C_P} \tag{11}$$

$$K_S = \frac{1}{\beta_S} \tag{12}$$

These relations follow standard thermodynamic identities, where the isentropic compressibility accounts for the change in volume under adiabatic (constant entropy) conditions, and the isentropic bulk modulus represents the corresponding resistance of the system to compression.

The **self-diffusion coefficient** $D$ ($m^2 \cdot s^{-1}$) was calculated using the Einstein equation as:

$$D = \lim_{t \to \infty} \frac{1}{6t} \langle \| \boldsymbol{r}(t + t_0) - \boldsymbol{r}(t_0) \|^2 \rangle \tag{13}$$

where $\boldsymbol{r}(t)$ denotes the atomic position at time ($t$) and the brackets indicate the ensemble average over all particles at the time of origin ($t_0$). The procedure was conducted based on the mean squared displacement (MSD) of atoms in the system. For the long time limit, MSD increased with time, and the proportionality constant was related to the self-diffusion coefficient.

The **refractive index** *RI* (dimensionless) was obtained from the Lorentz–Lorenz equation as:

$$\frac{RI^2 - 1}{RI^2 + 2} = \frac{4\pi}{3} \frac{\rho}{M} \alpha_{\text{polar}} \tag{14}$$

where $\alpha_{\text{polar}}$ is the isotropic dipole polarizability of the repeating unit, obtained from the DFT calculation, and $M$ is the molecular weight of the repeating unit.

The **Abbe number** $\upsilon$ (dimensionless), quantifying the dispersion of the refractive index, was calculated as:

$$v = \frac{RI_\mathrm{D} - 1}{RI_\mathrm{F} - RI_\mathrm{C}} \tag{15}$$

where $RI_\mathrm{D}$, $RI_\mathrm{F}$, and $RI_\mathrm{C}$ represent the refractive indices at wavelengths D (589.3 nm), F (486.1 nm), and C (656.3 nm), respectively. The refractive indices at each wavelength were obtained using the Lorentz–Lorenz equation, with the wavelength-dependent isotropic dipole polarizability given to $\alpha_\mathrm{polar}$.

The **static dielectric constant** $\varepsilon(0)$ (dimensionless) was calculated as:

$$\varepsilon(0) = \frac{\langle\mu^2\rangle - \langle\mu\rangle^2}{3\varepsilon_0 \cdot k_B \cdot T \cdot \langle V\rangle} + 1 \tag{16}$$

where $\mu$ is the total dipole moment of the system and $\varepsilon_0$ is the vacuum permittivity.

The **corrected static dielectric constant** $\varepsilon_\mathrm{corr}(0)$ (dimensionless) was calculated to include the contribution of electronic polarisation as:

$$\varepsilon_\mathrm{corr}(0) = \frac{\langle\mu^2\rangle - \langle\mu\rangle^2}{3\varepsilon_0 \cdot k_B \cdot T \cdot \langle V\rangle} + \varepsilon_{el} \tag{17}$$

where $\varepsilon_\mathrm{el}$ accounts for the contribution of the electronic polarisation to the dielectric constant, which was estimated from the square of the refractive index $RI^2$.

The **nematic order parameter** $S_\mathrm{n}$ (dimensionless) was calculated as the largest eigenvalue of the second-rank ordering tensor $Q_{\alpha\beta}$, defined as:

$$Q_{\alpha\beta} = \frac{1}{N}\sum_{\mathrm{i}=1}^{N} \frac{1}{2}\left(3u_{\mathrm{i}\alpha}u_{\mathrm{i}\beta}^\mathrm{T} - \delta_{\alpha\beta}\right) \tag{18}$$

where $u_{\mathrm{i}\alpha}$ and $u_{\mathrm{i}\beta}$ ($\alpha, \beta = x, y$, or $z$) are the unit vectors of the molecular axis of the $i$-th repeating unit, $\delta_{\alpha\beta}$ is the matrix of the Kronecker delta, and $N$ is the number of repeating units. The molecular axis of each CRU was defined as the principal axis corresponding to the largest moment of inertia obtained from the inertia tensor. The nematic order parameter was set to a value between zero for an isotropic structure and one for a perfectly aligned, fully ordered structure.

The **thermal conductivity** $\kappa$ ($\mathrm{W \cdot m^{-1} \cdot K^{-1}}$) was calculated according to Fourier's law as:

$$\kappa = \frac{J_Q}{(\partial T/\partial x)} = \frac{\Delta E}{2A \cdot \Delta t \cdot (\partial T/\partial x)} \tag{19}$$

where $J_\mathrm{Q}$ is the heat flux and $\partial T/\partial x$ is the temperature gradient obtained from the NEMD simulation. The heat flux was calculated considering the exchanged energy ($\Delta E$) using the Müller–Plathe algorithm, the cross-sectional area ($A$) perpendicular to the heat flux direction, and the simulation time interval ($\Delta t$).

The **thermal diffusivity** $\lambda$ ($\mathrm{m^2 \cdot s^{-1}}$) was obtained from the calculated values of thermal conductivity ($\kappa$), density ($\rho$), and heat capacity at constant pressure ($C_\mathrm{P}$) as:

$$\lambda = \frac{\kappa}{\rho \cdot C_P} \tag{20}$$

The **cohesive energy density** (CED, $\mathrm{J \cdot cm^{-3}}$) of the polymer system was calculated from the potential energy per unit volume as:

$$\mathrm{CED} = \frac{E_{\mathrm{coh}}}{V} = \frac{E_{\mathrm{tot}} - E_{\mathrm{intra}}}{V} \tag{21}$$

where $E_{\mathrm{coh}}$ is the cohesive energy of the system, $E_{\mathrm{tot}}$ is the total potential energy obtained from the MD simulation, and $E_{\mathrm{intra}}$ is the intramolecular contribution (bond stretching, angle bending, and torsional potentials).

The Hansen solubility parameters (HSPs) ($\mathrm{MPa}^{1/2}$) decomposed the cohesive energy density into contributions from dispersion ($\delta_{\mathrm{d}}$), polar ($\delta_{\mathrm{p}}$), and hydrogen-bonding ($\delta_{\mathrm{h}}$) interactions, defined as:

$$\delta_{\mathrm{d}} = \sqrt{\frac{E_{\mathrm{vdw}}}{V}}, \qquad \delta_{\mathrm{p}} = \sqrt{\frac{E_{\mathrm{p}}}{V}}, \qquad \delta_{\mathrm{h}} = \sqrt{\frac{E_{\mathrm{h}}}{V}} \tag{22}$$

where $E_{\mathrm{vdw}}$ is the van der Waals (dispersion) contribution to the potential energy, $E_{\mathrm{p}}$ is the dipole interaction contribution, and $E_{\mathrm{h}}$ is the hydrogen bonding contribution. When the solubility parameters were calculated using the MD simulations, the non-bonded interactions comprised van der Waals and electrostatic interactions. Accordingly, the **solubility parameter** ($\mathrm{MPa}^{1/2}$) comprised three components: $\delta_{\mathrm{vdw}}$ arising from van der Waals interactions, $\delta_{\mathrm{ele}}$ arising from electrostatic interactions, and $\delta_{\mathrm{total}}$ representing the combined contribution of both interactions. For this study, the following correspondence was adopted with the Hansen solubility parameters:

$$\delta_{\mathrm{d}} \equiv \delta_{\mathrm{vdw}}, \qquad \sqrt{\delta_{\mathrm{p}}^2 + \delta_{\mathrm{h}}^2} \equiv \delta_{\mathrm{ele}} \tag{23}$$

The **free volume** $V_{\mathrm{free}}$ (Å$^3$) represents the portion of the total system volume that is not occupied by polymer atoms. It was calculated as:

$$V_{\mathrm{free}} = V_{\mathrm{total}} - V_{\mathrm{occupaied}} \tag{24}$$

where $V_{\mathrm{total}}$ is the total simulation box volume and $V_{\mathrm{occupaied}}$ is the van der Waals volume of all atoms obtained from the sum of the atomic vdW volumes considering overlaps. The **fractional free volume** $V_{\mathrm{FFV}}$ (dimensionless) was calculated as the ratio of the free volume to the total volume, where $V_{\mathrm{FFV}} = V_{\mathrm{free}}/V_{\mathrm{total}}$.

The **tensile (Young's) modulus** $Y$ (GPa) describes the resistance of the polymer to uniaxial deformation. In RadonPy's simulations, the value was estimated from the stress–strain relationship for small deformations as:

$$Y = \frac{\sigma}{\epsilon} \tag{25}$$

where $\sigma$ is the stress along the deformation direction and $\epsilon$ is the applied strain. The tensile modulus was obtained by applying a small deformation to the simulation box and calculating the stress resulting from the virial expression.

The **Poisson ratio** $\nu$ (dimensionless) quantified the transverse contraction relative to longitudinal extension under uniaxial stress as:

$$\nu = -\frac{\epsilon_t}{\epsilon_l} \tag{26}$$

where $\epsilon_t$ is the resulting strain perpendicular to the loading direction and $\epsilon_l$ is the applied strain along the loading direction.
The **bulk modulus** $K_{\mathrm{tensile}}$ (GPa) was calculated using the tensile modulus ($Y$) and Poisson ratio ($\nu$) as:

$$K_{\text{tensile}} = \frac{Y}{3(1-2\nu)} \tag{27}$$

The **shear modulus** $G_{\text{tensile}}$ (GPa), representing the polymer's resistance to shear deformation, was calculated from the tensile modulus ($Y$) and Poisson ratio ($\nu$) as

$$G_{\text{tensile}} = \frac{Y}{2(1+\nu)} \tag{28}$$

The **Lamé constant** ($\Lambda_{\text{tensile}}$) (GPa) was derived from the tensile modulus ($Y$) and Poisson ratio ($\nu$) as:

$$\Lambda_{\text{tensile}} = \frac{Y \cdot \nu}{(1+\nu)(1-2\nu)} \tag{29}$$

The **tensile viscosity** $\eta_{\text{tensile}}$ (Pa·s), which quantifies the polymer's resistance to elongational flow, was obtained from tensile stretching simulations as:

$$\eta_{\text{tensile}} = \frac{\sigma_{zz}}{\dot{\epsilon}} \tag{30}$$

where $\sigma_{zz}$ is the normal stress along the stretching direction and $\dot{\epsilon}$ is the applied strain rate.

The **speed of sound** $c$ ($m \cdot s^{-1}$) in the polymer was calculated using its bulk modulus ($K_{\text{tensile}}$), shear modulus ($G_{\text{tensile}}$), and density as:

$$c = \sqrt{\frac{K_{\text{tensile}} + \frac{3}{4} G_{\text{tensile}}}{\rho}} \tag{31}$$

The speed of sound, which provides information on acoustic propagation, can be used to infer mechanical stiffness.

The properties of the CRU were calculated using DFT, which provides electronic structure information and enables the evaluation of molecular properties relevant to polymer design and material performance. The following properties were evaluated in this study:

The **molecular weight** $M$ ($g \cdot mol^{-1}$) of the CRU was calculated as the sum of the atomic masses of all atoms in the CRU as:

$$M = \sum_i m_i \tag{32}$$

where $m_i$ is the mass of atom $i$.

The **van der Waals volume** $V_{\text{vdw}}$ ($Å^3$) was obtained from the sum of the atomic van der Waals radii $V^{\text{atom}}$, considering their overlap in the optimised molecular geometry, as:

$$V_{\text{vdw}} = \sum_i V_i^{\text{atom}} - \text{overlap correction} \tag{33}$$

The **total electronic energy** $E_{\text{etot}}$ (hartree), **highest occupied molecular orbital (HOMO)** (eV), and **lowest unoccupied molecular orbital (LUMO)** (eV) energies of the CRU were obtained from the converged DFT calculations. The **dipole moment** $\mu$ (debye) was calculated from the electronic charge distribution and nuclear positions as:

$$\mu = \sum_i q_i \cdot \boldsymbol{r}_i \tag{34}$$

where $q_i$ is the partial charge of an atom and $\boldsymbol{r}_i$ is its position vector.

The **polarizability tensor** $a_{ij}$ (Å$^3$) was obtained from the second derivative of the total energy with respect to an external electric field $E$ as:

$$a_{ij} = -\frac{\partial^2 E_{\mathrm{etot}}}{\partial E_i \cdot \partial E_j} \tag{35}$$

The **isotropic polarizability** $a$ (Å$^3$) was set as:

$$a = \frac{1}{3}\left(a_{xx} + a_{yy} + a_{zz}\right) \tag{36}$$

Polarizability was calculated at D (589.3 nm), F (486.1 nm), and C (656.3 nm) spectral lines to evaluate the wavelength-dependent refractive indices and Abbe numbers. Typically, wavelength-dependent polarizability is calculated using the coupled-perturbed Hartree–Fock method; however, it is not implemented for the DFT calculations in Psi4. Therefore, in this study, the wavelength-dependent polarizability $a_{ij}(\omega)$ at frequency ($\omega$) was calculated using the sum-over-states approach as:

$$a_{ij}(\omega) = 2\sum_k \left(\frac{\mu_i^{gk}\mu_j^{kg}}{\hbar\omega_{gk} - (\hbar\omega)^2/(\hbar\omega_{gk})}\right) \tag{37}$$

where $\mu_i^{gk}$ is the transition dipole moment for the $i$-axis ($i \in \{x, y, z\}$) from the ground state ($g$) to the $k$-th excited state, $\hbar\omega_{gk}$ is the excitation energy from the ground state to the $k$-th excited state, and $\hbar$ is the reduced Planck's constant. TD-DFT calculations were performed to determine wavelength-dependent polarizability. In RadonPy, Psi4 is used as the quantum chemistry calculation engine. As TD-DFT calculations are not supported for the ωB97M-D3BJ functional in Psi4, the CAM-B3LYP, a GGA functional incorporating important long-range corrections, was employed to calculate the excited states. The 6-311+G(2d,p) basis set was used for H, C, N, O, F, P, S, and Cl; the 6-311G(d,p) was used for Br, and the LanL2DZ basis set was used for I, respectively.

## 1.11 NEMD Simulations for Thermal Conductivity Calculation

To evaluate the thermal conductivity, we performed reverse NEMD simulations, as proposed by Müller–-Plathe[28]. The simulation box was constructed in triplicate using an equilibrated amorphous cell along the $x$-axis under periodic boundary conditions (PBC).

In the reverse NEMD scheme, the simulation box was divided into $N$ slabs along the direction of heat flux. The heat flux was generated by periodically exchanging the velocities of the coldest atom in slab $N/2$ and the hottest atom in slab 0. This procedure effectively generated slab $N/2$ as the hottest region, while the temperature gradually decreased toward slabs 0 and $N$-1 due to the PBC.

To prevent artificial temperature shifts arising from cell replication, a preheating step using the *NVT* ensemble was performed for 2 ps at 300 K. Subsequently, a reverse NEMD simulation was conducted using the *NVE* ensemble for 1 ns. The simulation parameters were set as: the box was divided into 20 slabs, the velocity exchange frequency was 200 fs, and the time step was 0.2 fs. SHAKE constraints were not applied.

Non-bonded interactions were treated using the twin-range cutoff method with short and long-range cutoffs of 8 and 12 Å, respectively. Long-range Coulomb interactions were computed using the PPPM method. RadonPy was used to validate the simulation and confirm the linearity of the resulting temperature gradient. Any simulation exhibiting poor linearity ($R^2 < 0.95$) was considered unreliable and excluded from the analysis.

## 1.12 Calculation of Frequency-Dependent Dielectric Properties

The dynamic dielectric constant ($\varepsilon'$) and loss tangent ($\tan\delta$) were calculated using MD simulations under the *NPT* ensemble at 1 atm. The system was equilibrated at the target temperature and density prior to applying external fields.

To induce polarisation, an external time-dependent AC electric field $E(t) = E_0 \sin(\omega t)$ was applied to the simulation cell. The polarisation $P(t)$ was defined as

$$P(t) = \frac{\mu(t)}{V} \tag{38}$$

where $\mu(t)$ is the total dipole moment of the system and $V$ is the MD cell volume. An electric field was applied along the $x$, $y$, and $z$ directions, and the corresponding force $F = qE(t)$ was applied to each atom according to its charge $q$. The field amplitude was maintained sufficiently low to avoid non-linear polarisation effects, and the frequency was varied from 10 to 500 GHz to probe the frequency-dependent dielectric behaviour. An AC electric field was implemented using the efield command in LAMMPS.

Simulations were performed under the *NPT* ensemble using a Nosé–Hoover thermostat and barostat to control the temperature and pressure. Each simulation was performed for a sufficiently long duration to capture at least one full period of the applied electric field, ensuring that a steady-state polarisation response was obtained. To improve statistical reliability, simulations were repeated for five independent initial configurations, and the results were averaged.

The time-dependent electric displacement $D(t)$ was calculated as:

$$D(t) = \varepsilon_0 \cdot E(t) + P(t) \tag{39}$$

where $\varepsilon_0$ is the permittivity of free space. The phase difference $(\delta)$ between $D(t)$ and $E(t)$ was determined by fitting the temporal data.

The real part of the dielectric constant ($\varepsilon'$) and the loss tangent ($\tan\delta$) were calculated as:

$$\varepsilon' = 1 + \frac{P_o}{\varepsilon_0 \cdot E_0}\cos\delta + \frac{P_e}{\varepsilon_0 \cdot E} \tag{40}$$

$$\tan\delta = \frac{P_o}{\varepsilon_0 \cdot E_0}\sin\delta \times \frac{1}{\varepsilon'} \tag{41}$$

where $\delta$ represents the phase lag between the applied field and the induced electric displacement, corresponding to energy dissipation. The adopted methodology enables the quantification of frequency-dependent dielectric properties and the correlation of the molecular structure and dipole orientation with the macroscopic dielectric response.

## 1.13 Thermal conductivity calculation of stretched polymer materials

The thermal conductivities of the oriented polymer systems were evaluated using MD simulations. To investigate the effect of chain alignment induced by uniaxial stretching, amorphous polymer cells were subjected to controlled elongation along the $x$-axis and subsequently equilibrated under PBC.

The initial amorphous polymer structures were equilibrated in the *NPT* ensemble at 1 atm and 300 K to achieve realistic densities and thermal equilibria. Uniaxial stretching was applied along the $x$-axis at a constant velocity of 25 m/s and strain of 0.5 and 1.0. After stretching, the system was equilibrated for 500 ps under *NPT* conditions, allowing the lateral dimensions ($y$ and $z$) to relax, while maintaining the stretched $x$-axis length. This step ensured mechanical and thermal equilibration of the oriented polymer structure. The thermal conductivities along the oriented ($x$) and transverse ($y$ and $z$) directions were calculated using the reverse NEMD method.

The relationship between the degree of orientation order and thermal conductivity at strains of 0, 0.5, and 1.0, for 1379 polymers, was calculated. The polymers were distributed in the range of 0 to 0.6 in orientation order and 0.2 to 2.5 W $m^{-1}$ $K^{-1}$ in thermal conductivity. For each polymer class, Figure 3d illustrates the correlation between the average thermal conductivity and average degree of orientation order for each strain. A comparison within the same polymer class revealed that the thermal conductivity tended to increase as the degree of orientation order increased. Conversely, the presence of polymers with high thermal conductivities ranging from 1.5 to 2.0 W $m^{-1}$ $K^{-1}$ was observed, even when the degree of orientation order was less than 0.2. The correlation between the degree of orientation order and thermal conductivity defined the characteristic of each polymer class. In particular, polysulfides, polyphenylenes, and polyimines exhibited high thermal conductivities at a strain of 1.0. Contrarily, polyacrylates, polyphosphazenes, and polystyrenes exhibited minimal change in their thermal conductivity after stretching.

Polymers with rigid main chains containing multiple bonds and aromatic rings exhibited high thermal conductivity owing to their orientation. Conversely, polymers with large side chains demonstrated minimal improvement in thermal conductivity even when their orientation order was increased.

# 2 Polymer Generators

The polymer library comprising the majority of the PolyOmics database was primarily constructed using two virtual polymer generators. The first was a chemical language model trained to replicate the structural patterns of previously synthesised polymers. The second was SMiPoly, a rule-based in silico model that simulated polymerisation reactions. The following sections provide an overview of each model.

## 2.3 Chemical language models

Ikebata et al.[29] introduced a molecular structure generator based on a probabilistic language model, known as an extended n-gram. In this approach, the chemical structure $X$ of each compound in the training dataset is represented using SMILES notation, where $X$ is expressed as a string of length $p$:

$$X = x_1 x_2 \cdots x_p \tag{41}$$

By leveraging the set of SMILES strings corresponding to the molecules synthesised to date, an n-gram language model is trained to construct a structure generator that captures and reproduces the characteristic patterns observed in existing molecules, including frequent fragments and chemically valid bonding rules.

In this framework, the probability of generating a given SMILES string, denoted by $p(X)$, can be factorised as the product of the conditional probabilities according to the n-gram approximation as:

$$p(X) \propto p(x_1) \prod_{i=2}^{p} p(x_\mathrm{i} | x_{\mathrm{i}-1}, \cdots, x_{1,}) \tag{42}$$

where $p(x_\mathrm{i} | x_{\mathrm{i}-1}, \cdots, x_{1,})$ represents the conditional probability of the $i$-th token given the preceding $i$-1 tokens. This formulation enables the generator to produce chemically plausible structures probabilistically by learning the sequential dependencies inherent in the existing molecular SMILES strings.

The synthetic polymers registered in PoLyInfo were classified into 20 distinct classes (Table S2), and class-specific training datasets were constructed. These datasets were used with the XenonPy software to train a class-specific generator for each polymer class. The resulting models were used to construct a virtual library of 1,778,039 polymers covering a broad chemical space. This library was shared within the consortium, enabling the members to generate property datasets collaboratively in a distributed manner. To date, computations have only been performed on a subset of the entire library.

## 2.4 SMiPoly: Polymer Generator based on Rule-Based Polymerisation Reactions

SMiPoly, an open-source Python library, was used for rule-based virtual generation of synthetically accessible polymers. SMiPoly encode a comprehensive set of 22 chemical reaction rules commonly used in polymer synthesis, enabling systematic in silico simulation of polymerisation processes.

The workflow consists of two key modules—monc.py and polg.py:

- monc.py classifies the small molecules into monomer classes based on their reactive functional groups.
- Appropriate polymerisation rules are applied to these monomers to generate the corresponding polymer repeating units.

For this study, 1,083 commercially available monomers were processed using SMiPoly, yielding 169,347 unique linear polymers across seven polymer classes (polyolefins, polyesters, polyethers, polyamides, polyimides, polyurethanes, and polyoxazolidones). These virtual polymers are theoretically synthesizable under practical chemical conditions. To date, computations have only been performed on a subset of the entire library.

## 3. Machine Learning Prediction for Polymer–Solvent Miscibility in PolyOmics

Aoki et al.[30] constructed a machine learning model to predict the Flory–Huggins interaction parameter, known as the $\chi$ parameter, which quantifies polymer–solvent interactions and governs the Gibbs free energy of mixing according to the Flory–Huggins theory.

For model training, three datasets were prepared:

1. **Experimental $\chi$ dataset**:
   Experimental $\chi$ parameters were collected from literature for 1,190 polymer–solvent pairs, consisting of 46 polymers and 140 solvents. This dataset included measurements at various temperatures and polymer–solvent compositions. However, owing to experimental limitations, the data were biased toward miscible pairs and covered a limited region of the chemical space.
2. **Computational $\chi$ dataset**
   To overcome data scarcity, a computational $\chi$ parameter dataset was generated using high-throughput quantum chemical calculations based on the COSMO-RS method[31,32]. This dataset comprised 9,129 polymer–solvent pairs covering a substantially broader chemical space.
3. **Solubility classification dataset**:
   A binary dataset of 429 polymer–solvent pairs was prepared, with each pair labelled as either miscible (good solvent) or immiscible (poor solvent).

A multitask deep neural network was developed to predict the experimental $\chi$ parameter as the main task, while simultaneously learning two auxiliary tasks: the prediction of computationally calculated $\chi$ values and the binary classification of polymer–solvent miscibility. The input variable is a polymer–solvent pair with its chemical structure, in which the compositional, structural, and physicochemical features of the polymer and solvent are encoded into a 397-dimensional descriptor vector using the 190-dimensional force-field kernel mean descriptor from RadonPy and a 207-dimensional descriptor from the Python library RDKit, respectively. The model output comprises three prediction branches corresponding to each task. During training, the losses from all three tasks were combined and optimised collectively to obtain a unified representation of the polymer–solvent interactions. This multitask learning approach enabled the model to leverage relationships among the three related tasks, thereby compensating for the limited size and systematic biases present in the experimental $\chi$ parameter dataset.

Aoki et al. (2023)[30] systematically evaluated the predictive model performance using experimental test data for the main task and compared it with two baseline methods: COSMO-RS quantum chemical calculations and an empirical Hansen solubility parameter (HSP)-based method. The proposed model significantly outperformed both baseline approaches, demonstrating its ability to predict $\chi$ parameters for both miscible and immiscible systems and generalise beyond the scratch learning using only the biased experimental dataset.

Each polymer registered in the PolyOmics database was annotated with machine-learning predicted $\chi$ parameter values for 19 organic solvents and plasticisers commonly used in synthetic processes. These values serve as criteria for selecting candidate solvents and plasticisers for the synthesis of virtual polymers.